\documentclass[aps,prb,twocolumn,showpacs,superscriptaddress,floatfix]{revtex4}

\usepackage{graphicx}
\usepackage{dcolumn}
\usepackage{amsmath}
\usepackage{amsfonts}
\usepackage{amssymb}
\usepackage{bm}
\usepackage{color}
\usepackage{ulem}
\usepackage[squaren]{SIunits}

\begin{document}

\title{Two-resonator circuit~QED: A superconducting quantum switch}

\author{Matteo~Mariantoni\footnote{Authors with equal contributions to this work.}}
\email{Matteo.Mariantoni@wmi.badw.de}
 \affiliation{Walther-Mei{\ss}ner-Institut, Bayerische Akademie der
              Wissenschaften, Walther-Mei{\ss}ner-Strasse 8, D-85748
              Garching, Germany}
 \affiliation{Physics Department, Technische Universit\"at
              M\"{u}nchen, D-85748 Garching, Germany}

\author{Frank~Deppe$^{*}_{}$}
 \affiliation{Walther-Mei{\ss}ner-Institut, Bayerische Akademie der
              Wissenschaften, Walther-Mei{\ss}ner-Strasse 8, D-85748
              Garching, Germany}
 \affiliation{Physics Department, Technische Universit\"at
              M\"{u}nchen, D-85748 Garching, Germany}

\author{A.~Marx}
 \affiliation{Walther-Mei{\ss}ner-Institut, Bayerische Akademie der
              Wissenschaften, Walther-Mei{\ss}ner-Strasse 8, D-85748
              Garching, Germany}

\author{R.~Gross}
 \affiliation{Walther-Mei{\ss}ner-Institut, Bayerische Akademie der
              Wissenschaften, Walther-Mei{\ss}ner-Strasse 8, D-85748
              Garching, Germany}
 \affiliation{Physics Department, Technische Universit\"at
              M\"{u}nchen, D-85748 Garching, Germany}

\author{F.~K.~Wilhelm}
 \affiliation{IQC and Department of Physics and Astronomy,
              University of Waterloo, 200 University Avenue West, Waterloo,
              Ontario, N2L 3G1, Canada}

\author{E.~Solano}
 \affiliation{Physics Department, CeNS and ASC,
              Ludwig-Maximilians-Universit\"at, Theresienstrasse 37, D-80333
              Munich, Germany}
 \affiliation{Departamento de Qu\'{\i}mica F\'{\i}sica,
              Universidad del Pa\'{\i}s Vasco - Euskal Herriko Unibertsitatea,
              Apdo.~644, 48080 Bilbao, Spain}

\begin{abstract}
We introduce a systematic formalism for two-resonator circuit~QED,
where two on-chip microwave resonators are simultaneously coupled
to one superconducting qubit. Within this framework, we
demonstrate that the qubit can function as a quantum switch
between the two resonators, which are assumed to be originally
independent. In this three-circuit network, the qubit mediates a
\textit{geometric second-order circuit interaction} between the
otherwise decoupled resonators. In the dispersive regime, it also
gives rise to a \textit{dynamic second-order perturbative
interaction}. The geometric and dynamic coupling strengths can be
tuned to be equal, thus permitting to switch on and off the
interaction between the two resonators via a qubit population
inversion or a shifting of the qubit operation point. We also show
that our quantum switch represents a flexible architecture for the
manipulation and generation of nonclassical microwave field states
as well as the creation of controlled multipartite entanglement in
circuit~QED. In addition, we clarify the role played by the
geometric interaction, which constitutes a fundamental property
characteristic of superconducting quantum circuits without
counterpart in quantum-optical systems. We develop a detailed
theory of the geometric second-order coupling by means of circuit
transformations for superconducting charge and flux qubits.
Furthermore, we show the robustness of the quantum switch
operation with respect to decoherence mechanisms. Finally, we
propose a realistic design for a two-resonator circuit~QED setup
based on a flux qubit and estimate all the related parameters. In
this manner, we show that this setup can be used to implement a
superconducting quantum switch with available technology.
\end{abstract}

\date{\today}


\maketitle

\section{INTRODUCTION}
 \label{section:introduction}

In the past few years, we have witnessed a tremendous experimental
progress in the flourishing realm of circuit
QED.~\cite{AWallraff:RJSchoelkopf:NatureLett:2004:a,
IChiorescu:JEMooij:NatureLett:2004:a,
JJohansson:HTakayanagi:PhysRevLett:2006:a,
RJSchoelkopf:SMGirvin:NatureHorizons:2008:a} There, different
types of superconducting qubits have been strongly coupled to
on-chip microwave resonators, which act as quantized cavities.
Recently, a quantum state has been stored and coherently
transferred between two superconducting phase qubits via a
microwave
resonator~\cite{MikaASillanpaeae:RaymondWSimmonds:NatureLett:2007:a}
and two transmon qubits have been coupled utilizing an on-chip
cavity as a quantum
bus.~\cite{JMajer:JMChow:RJSchoelkopf:NatureLett:2007:a}
Furthermore, microwave single photons have been generated by
spontaneous
emission~\cite{AAHouck:DISchuster:RJSchoelkopf:NatureLett:2007:a}
and Fock states created in a system based on a phase
qubit.~\cite{MaxHofheinz:JohnMMartinis:ANCleland:NatureLett:2008:a}
In addition, lasing effects have been demonstrated exploiting a
single Cooper-pair box,~\cite{OAstafiev:JSTsai:NatureLett:2007:a}
the nonlinear response of the JC model
observed,~\cite{JMFink:AWallraff:NatureLett:2008:a,
LevSBishop:RJSchoelkopf:arXiv:2008:a} the two-photon driven
Jaynes-Cummings~(JC) dynamics used as a means to probe the
symmetry properties of a flux
qubit,~\cite{FrankDeppe:MatteoMariantoni:RGross:NaturePhysLett:2008:a}
and resonators tuned with high
fidelity.~\cite{APalacios-Laloy:DEsteve:JLowTempPhys:2008:a,
MSandberg:PDelsing:arXiv:2008:a} These formidable advances show
how circuit QED systems are rapidly reaching a level of complexity
comparable to that of the already well-established quantum optical
cavity QED.~\cite{HMabuchi:ACDoherty:ScienceReview:2002:a,
SergeHaroche:Jean-MichelRaimond:Book:2006:a,
HerbertWalther:ThomasBecker:RepProgPhys:2006:a}

Amongst the aims common to these experiments is the possibility to
perform quantum information
processing,~\cite{MichaelANielsen:IsaacLChuang:Book:2000:a} in
particular following the lines of recent proposals, e.g., see
Refs.~\onlinecite{AlexandreBlais:JayGambetta:RJSchoelkopf:PhysRevA:2007:a}
and \onlinecite{FerdinandHelmer:FlorianMarquardt:arXiv:2007:a}.
The latter considers a two-dimensional array of on-chip resonators
coupled to qubits. In this or any other multi-cavity
setup,~\cite{MJStorcz:MMariantoni:ESolano:arXiv:cond-mat:2007:a}
it is highly desirable to switch on and off an interaction between
two resonators or to compensate their spurious crosstalk.
Moreover, investigating the basic properties of two-resonator
circuit~QED, where two resonators are coupled to one qubit,
certainly represents a subject of fundamental relevance. In fact,
when operating such a system in a regime dominated by second-order
(dispersive) interactions, as in the scope of this article, the
requirements on the qubit coherence properties are considerably
relaxed.~\cite{AlexandreBlais:RJSchoelkopf:PhysRevA:2004:a,
MatteoMariantoni:FrankDeppe:unpublished:2008:a} In this manner,
two-resonator architectures constitute an appealing playground for
testing quantum mechanics on a chip. We also notice that
second-order
interactions~\cite{DISchuster:AAHouck:RJSchoelkopf:NatureLett:2007:a,
FrankDeppe:MatteoMariantoni:RGross:NaturePhysLett:2008:a} are
becoming more and more prominent in circuit~QED experiments owing
to the possibility of very large first-order coupling
strengths.~\cite{RHKoch:DPDiVincenzo:PhysRevLett:2006:a,
JJohansson:HTakayanagi:PhysRevLett:2006:a,
DISchuster:AAHouck:RJSchoelkopf:NatureLett:2007:a,
MikaASillanpaeae:RaymondWSimmonds:NatureLett:2007:a,
OAstafiev:JSTsai:NatureLett:2007:a,
RJSchoelkopf:SMGirvin:NatureHorizons:2008:a}

In this article, we theoretically study a three-circuit network
where a superconducting charge or flux
qubit~\cite{YuriyMakhlin:GerdSchoen:RevModPhys:2001:a,
MHDevoret:JMMartinis:arXiv:cond-mat:2004:a,
JQYou:FrancoNori:PhysToday:2005:a,
GoeranWendin:VitalyShumeiko:BookChapter:2006:a} interacts with two
on-chip microwave cavities, a two-resonator circuit~QED setup. In
the absence of the qubit, the resonators are assumed to have
negligible or small geometric first-order (direct) crosstalk. This
scenario is similar to that of quantum optics, where an atom can
interact with two orthogonal cavity
modes.~\cite{ARauschenbeutel:SHaroche:PhysRevARap:2001:a} However,
there are some crucial differences. The nature of the
three-circuit system considered here requires to account for a
\textit{geometric second-order circuit interaction} between the
two resonators. This gives rise to coupling terms in the
interaction Hamiltonian, which are formally equivalent to those
describing a beam splitter. This interaction is mediated by the
circuit part of the qubit and does not depend on the qubit state.
It is noteworthy to mention that this coupling does not exist in
the two-mode Jaynes-Cummings (JC) model studied in quantum optics,
where atoms do not sustain any geometric interaction. This means
that introducing a second resonator causes a departure from the
neat analogy between cavity and one-resonator circuit
QED.~\cite{OlivierBuisson:FrankHekking:BookChapter:2001:a,
Chui-PingYang:SiyuanHan:PhysRevA:2003:a,
FPlastina:GFalci:PhysRevB:2003:a,
JQYou:FrancoNori:PhysRevB:2003:a,
AlexandreBlais:RJSchoelkopf:PhysRevA:2004:a} In the dispersive
regime, where the transition frequency of the qubit is largely
detuned from that of the cavities, also other beam-splitter-type
interaction terms between the two resonators appear. Their
existence is known in quantum
optics~\cite{AMessina:ANapoli:JofModernOpt:2003:a} and results
from a \textit{dynamic second-order perturbative interaction},
which depends on the state of the qubit. The sign of this
interaction can be changed by an inversion of the qubit population
or by shifting the qubit operation point. The latter mechanism can
also be used to change the interaction strength. Notably, for a
suitable set of parameters, the geometric and dynamic second-order
coupling coefficients can be made exactly equal by choosing a
proper qubit-resonator detuning. In this case, the interaction
between the two cavities can be switched on and off, thereby
enabling the implementation of a \textit{discrete quantum switch}
as well as a \textit{tunable coupler}.

In circuit~QED, several other scenarios have been envisioned where
a qubit interacts with different bosonic modes, e.g., those of an
adjacent nanomechanical resonator or similar. It has been proposed
to implement quantum
transducers~\cite{CPSun:FrancoNori:PhysRevA:2005:a} as well as
Jahn-Teller models and Kerr
nonlinearities,~\cite{FLSemitao:GJMilburn:arXiv:2008:a} to
generate nontrivial nonclassical states of the microwave
radiation,~\cite{MJStorcz:MMariantoni:ESolano:arXiv:cond-mat:2007:a,
AlexandreBlais:RobertJSchoelkopf:APSDenver:2007:a} to create
entanglement via Landau-Zener
sweeps,~\cite{MartijnWubs:PeterHaenggi:PhysicaE:2007:a} and to
carry out high fidelity measurements of microwave quantum
fields.~\cite{AlexandreBlais:RobertJSchoelkopf:APSDenver:2007:a,
FerdinandHelmer:FlorianMarquardt:arXiv:2007:b} Moreover,
multi-resonator setups might serve to probe quantum
walks~\cite{PengXue:KevinLalumiere:arXiv:2008:a} and to study the
scattering process of single microwave
photons.~\cite{LanZhou:FrancoNori:arXiv:2008:a} All of these
proposals, however, do not develop a systematic theory of a
realistic architecture based on two on-chip microwave resonators
and do take into account the fundamental geometric second-order
coupling between them. Also, our quantum switch is inherently
different from the quantum switches investigated in atomic
systems.~\cite{LDavidovich:SHaroche:PhysRevLett:1993:a} First, we
consider a qubit simultaneously coupled to two resonators, which
are not positioned one after the other in a cascade configuration
as in Ref.~\onlinecite{LDavidovich:SHaroche:PhysRevLett:1993:a}.
Second, our switch behaves as a tunable quantum coupler between
the two resonators. Last, atomic systems naturally lack a
geometric second-order coupling. Furthermore, it is important to
stress that the dynamic interaction studied here cannot be cast
within the framework of the quantum reactance theory (capacitance
or inductance, depending on the specific
implementation).~\cite{DVAverin:CBruder:PhysRevLett:2003:a,
EIlapichev:AMZagoskin:PhysRevLett:2003:a,
BLTPlourde:JohnClarkePhysRevBRap:2004:a,
ALupacscu:JEMooij:PhysRevLett:2004:a,
MASillanpaeae:PJHakonen:PhysRevLett:2005:a,
TDuty:PDelsing:PhysRevLett:2005:a,
GJohansson:GWendin:JPhysCondensMatter:2006:a,
JKoenemann:ABZorin:PhysRevB:2007:a} The main hypothesis for a
quantum reactance to be defined is a resonator characterized by a
transition frequency extremely different from that of the qubit.
Typically, the resonator frequency is considered to be very low
(practically zero) compared to the qubit one. Such a scenario is
undesirable for the purposes of this work, where a truly quantized
high-frequency cavity initialized in the vacuum state has to be
used. Also, to our knowledge, the quantum reactance works
mentioned above do not directly exploit a geometric coupling
between two resonators to compensate a dynamic one. Nevertheless,
we believe that a circuit theory
approach~\cite{LeonOChua:CharlesADesoer:ErnestSKuh:Book:1987:a,
BernardYurke:JohnSDenker:PhysRevA:1984:a,
MHDevoret:JMMartinis:arXiv:cond-mat:2004:a,
GuidoBurkard:DavidPDiVincenzo:PhysRevB:2004:a,
GuidoBurkard:PhysRevB:2005:a,GuidoBurkard:BookChapter:2005:a,
GWendin:VSShumeiko:LowTempPhys:2007:a} to two-resonator
circuit~QED, which we pursue throughout this manuscript, allows
for a deep comprehension of the matter discussed here. Finally, we
point out that the geometric first-order coupling between two
resonators can be reduced or erased by simple engineering, whereas
the second-order coupling due to the presence of a qubit circuit
is a fundamental issue. As we show later, whenever the coupling
between qubit and resonators is wanted to be large, an appreciable
geometric second-order coupling inevitably appears, especially for
resonators perfectly isolated in first order. In summary, our
quantum switch is based on a combination of geometric and dynamic
interactions competing against each other and constitutes a
promising candidate to perform nontrivial quantum operations
between different resonators.

The paper is organized as follows. In
Sec.~\ref{section:two:resonator:circuit:qed}, we develop a
systematic formalism for two-resonator circuit~QED employing
second-order circuit theory. In
Sec.~\ref{section:derivation:of:the:quantum:switch:hamiltonian},
we focus on the dispersive regime of two-resonator circuit~QED and
derive the quantum switch Hamiltonian. In
Sec.~\ref{section:treatment:of:decoherence}, we discuss the main
limitations to the quantum switch operation due to decoherence
processes of qubit and cavities. In
Sec.~\ref{section:an:example:of:two:resonator:circuit:qed:with:a:flux:qubit},
we propose a realistic implementation of a two-resonator
circuit~QED architecture, which is suitable for the realization of
a superconducting quantum switch. Finally, in
Sec.~\ref{section:summary:and:conclusions}, we summarize our main
results, draw our conclusions, and give a brief outlook.

\section{TWO-RESONATOR CIRCUIT~QED}
 \label{section:two:resonator:circuit:qed}

\begin{figure*}[t!]
\centering{%
 \includegraphics[width=0.70\textwidth]{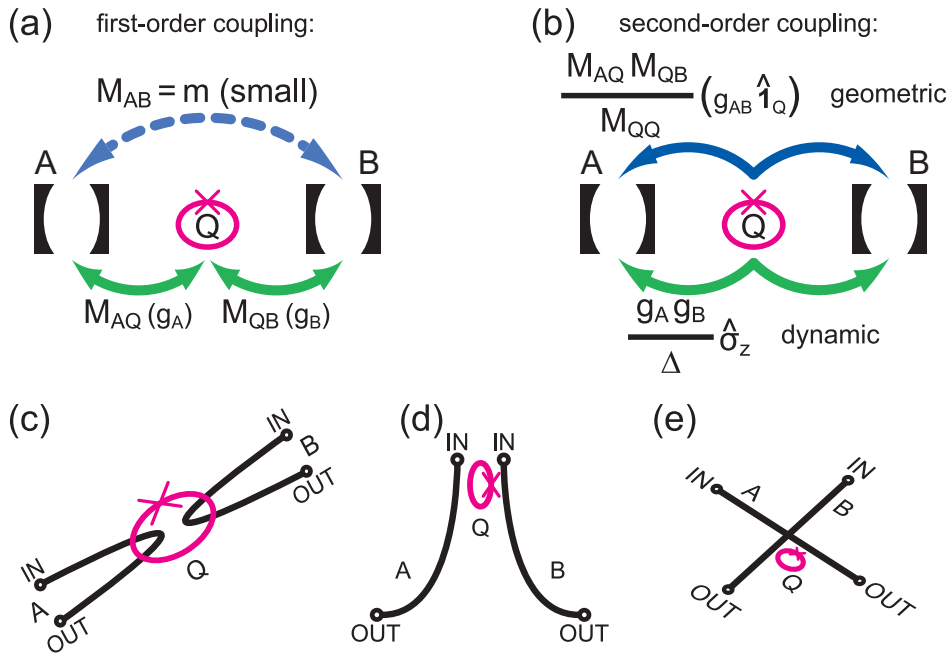}}
\caption{(Color online). Sketch of the system under analysis. All
constants are defined in the main body of the paper. Only
inductive couplings are considered. (a)~Schematic representation
of the first-order coupling Hamiltonian of our three-node network.
Two cavities (resonators) A and B interact with a generic
superconducting qubit Q. A and B can have a weak geometric
first-order coupling $M^{}_{\rm A B} {} = {} m$ [broken blue (dark
grey) arrow], as in the Hamiltonian $\widehat{H}^{\left( 1
\right)}_{\rm A B}$ of Eq.~(\ref{H:op:(1):AB}). The two solid
green (light grey) arrows represent a two-mode Jaynes-Cummings
dynamics with coupling coefficients $g^{}_{\rm A} {} \propto {}
M^{}_{\rm A Q}$ and $g^{}_{\rm B} {} \propto {} M^{}_{\rm Q B}$,
respectively. (b)~Visualization of the effective second-order
coupling Hamiltonian $\widehat{H}^{}_{\rm eff}$ of
Eq.~(\ref{H:op:eff}). The solid blue (dark grey) arrows show the
second-order \textit{geometric} coupling channel mediated by a
virtual excitation of the circuit associated with Q, as in the
Hamiltonian $\widehat{H}^{\left( 2 \right)}_{\rm A B}$ of
Eq.~(\ref{H:op:(2):AB}). This channel is characterized by a
constant $g^{}_{\rm A B} {} \propto {} M^{}_{\rm A Q} M^{}_{\rm Q
B} / M^{}_{\rm Q Q}$ (the small contribution from $m$ is
neglected) and is qubit-state independent. The solid green (light
grey) arrows show the second-order \textit{dynamic} channel
mediated by a virtual excitation of the qubit Q. This channel is
characterized by a constant $g^{}_{\rm A} g^{}_{\rm B} / \Delta$
and is qubit-state dependent. Three generic sketches of a possible
setup. (c)~A flux qubit (Q) sits at the current antinode of, e.g.,
the first mode of two $\lambda / 2$ resonators (solid black lines,
only inner conductor shown). The open circles at the ``IN" and
``OUT" ports denote the position of the coupling capacitors to be
used in real implementations [e.g.,
cf.~Sec.~\ref{section:an:example:of:two:resonator:circuit:qed:with:a:flux:qubit}
and Fig.~\ref{QS:Figure:6:abcdef:Matteo:Mariantoni:2008}(a)].
(d)~A charge qubit (Q) sits at the voltage antinode of, e.g., the
first mode of two $\lambda / 2$ resonators. (e)~A charge or flux
qubit sits at the voltage (e.g., second mode, $\lambda$
resonators) or current (e.g., first mode, $\lambda / 2$
resonators) antinode of two orthogonal
resonators.~\cite{FerdinandHelmer:FlorianMarquardt:arXiv:2007:a}}
 \label{QS:Figure:1:abcde:Matteo:Mariantoni:2008}
\end{figure*}

In this section, we take the perspective of classical circuit
theory~\cite{LeonOChua:CharlesADesoer:ErnestSKuh:Book:1987:a} and
extend it to the quantum regime to derive the Hamiltonian of a
quantized three-circuit network. In general, the latter is
composed of two on-chip microwave resonators and a superconducting
qubit. Our approach is similar to that of
Refs.~\onlinecite{BernardYurke:JohnSDenker:PhysRevA:1984:a,
MHDevoret:JMMartinis:arXiv:cond-mat:2004:a,
GuidoBurkard:DavidPDiVincenzo:PhysRevB:2004:a,
GuidoBurkard:PhysRevB:2005:a,GuidoBurkard:BookChapter:2005:a}, and
\onlinecite{GWendin:VSShumeiko:LowTempPhys:2007:a}. In addition,
we account for second-order circuit elements linking different
parts of the network, which is considered to be closed and
nondissipative. Here, closed means that we assume no energy flow
between the network under analysis and other possible adjacent
networks. These could be additional circuitry used to access the
three-circuit network from outside and where excitations could
possibly decay. Nondissipative means that we consider capacitive
and inductive circuit elements only, more in general, reactive
elements. We neglect resistors, which could represent dissipation
processes of qubit and resonators. In summary, the network of our
model is altogether a conservative system. The detailed role of
decoherence mechanisms is studied later in
Sec.~\ref{section:treatment:of:decoherence}.

The first step of our derivation is to demonstrate a geometric
second-order coupling between the circuit elements of a simple
\textit{three-node network}. This means that we assume the various
circuit elements to be concentrated in three confined regions of
space (nodes). Any topologically complex \textit{three-circuit
network} can be reduced to such a three-node network, where each
node is fully characterized by its capacitance matrix $\mathbf{C}$
and/or inductance matrix $\mathbf{M}$. The topology of the
different circuits (e.g., two microstrip or coplanar waveguide
resonators coupled to a superconducting qubit) is thus absorbed in
the definition of $\mathbf{C}$ and $\mathbf{M}$, simplifying the
analysis significantly. The system Hamiltonian can then be
straightforwardly obtained. In fact, the classical energy of a
conservative network can be expressed as $E = (
\vec{V}^T_{}\,\mathbf{C}\,\vec{V}^{}_{} +
\vec{I}^{{\,}T}_{}\,\mathbf{M}\,\vec{I}^{}_{} ) / 2$, where the
vectors $\vec{V}^{}_{}$ and $\vec{I}^{}_{}$ represent the voltages
and currents on the various capacitors and
inductors.~\cite{LeonOChua:CharlesADesoer:ErnestSKuh:Book:1987:a}
The usual quantization of voltages and
currents~\cite{AlexandreBlais:RJSchoelkopf:PhysRevA:2004:a} and
the addition of the qubit Hamiltonian allows us to obtain the
fully quantized Hamiltonian of the three-node network
(cf.~Subsec.~\ref{subsection:the:hamiltonian:of:a:generic:three:node:network}).
Special attention is then reserved to compute contributions to the
matrices $\mathbf{C}$ and $\mathbf{M}$ up to second order. These
are consequently redefined as $\mathbf{C^{\left( 2 \right)}_{}}$
and $\mathbf{M^{\left( 2 \right)}_{}}$, respectively
(cf.~Subsec.~\ref{subsection:the:capacitance:and:inductance:matrices:up:to:second:order}).
Corrections of third or higher order to the capacitance and
inductance matrices are discussed in
Appendix~\ref{appendix:a:higher:order:corrections:to:the:capacitance:and:inductance:matrices},
where we show that they are not relevant for this work.

We finally consider two examples of possible implementations of
two-resonator circuit QED
(cf.~Subsec.~\ref{subsection:the:role:of:circuit:topology:two:examples}).
These examples account for two superconducting resonators coupled
to a charge quantum circuit (e.g., a Cooper-pair box or a
transmon) or a flux quantum circuit (e.g., a superconducting one-
or three-Josephson-junction loop). Before moving to a two-level
approximation, the Hamiltonians of these devices can be used to
deduce the geometric second-order circuit interaction between the
two resonators. This result is better understood considering the
lumped-element equivalent circuits of the entire systems. In this
way, also the conceptual step from a three-circuit to a three-node
network is clarified and the role played by the topology of the
different circuits becomes more evident. We show that special care
must be taken when quantizing the interaction Hamiltonian between
charge or flux quantum circuits and microwave fields by the simple
promotion of an AC classical field to a quantum one.
Interestingly, comparing the standard Hamiltonian of charge and
flux quantum circuits coupled to quantized fields with \textit{ab
initio} models based on lumped-element equivalent circuits, we
prove that the latter are better suited to describe circuit~QED
systems.

\subsection{The Hamiltonian of a generic three-node network}
 \label{subsection:the:hamiltonian:of:a:generic:three:node:network}

The system to be studied is sketched in
Figs.~\ref{QS:Figure:1:abcde:Matteo:Mariantoni:2008}(a) and
\ref{QS:Figure:1:abcde:Matteo:Mariantoni:2008}(b), where the
microwave resonators are represented by symbolic mirrors. A more
realistic setup is discussed in
Sec.~\ref{section:an:example:of:two:resonator:circuit:qed:with:a:flux:qubit}
and is drawn in
Fig.~\ref{QS:Figure:6:abcdef:Matteo:Mariantoni:2008}(a). A and B
represent the two cavities and Q a superconducting qubit, making
altogether a three-node network. The coupling channels between the
three nodes are assumed to be capacitive and/or inductive. We also
hypothesize the first-order interaction between A and B to be weak
and that between A or B and Q to be strong by design. In other
words, the first-order capacitance and inductance matrices are
$\mathbf{C} = C^{}_{k l}$ and $\mathbf{M} = M^{}_{k l}$, with $k ,
l \in \left\{ {\rm A} , {\rm B} , {\rm Q} \right\}$, where
$C^{}_{k l} = C^{}_{l k}$ and $M^{}_{k l} = M^{}_{l k}$ because of
symmetry reasons. In addition, we assume $C^{}_{\rm A B} \equiv c
\ll C^{}_{k l \neq {\rm A B}}$ and $M^{}_{\rm A B} \equiv m \ll
M^{}_{k l \neq {\rm A B}}$. The elements $c$ and $m$ represent a
first-order crosstalk between A and B, which can be either
spurious or engineered and, here, is considered to be small. In
Sec.~\ref{section:an:example:of:two:resonator:circuit:qed:with:a:flux:qubit},
we delve into a more detailed analysis of the geometric
first-order coupling between two microstrip resonators.
Restricting the cavities to a single relevant mode, the total
Hamiltonian of the system is given by
\begin{equation}
\widehat{H}^{}_{\rm T} {} = {} \frac{1}{2} \vec{\hat{V}}^{T}_{} \,
\mathbf{C^{( n )}_{}}\,\vec{\hat{V}}^{}_{} + \frac{1}{2}
\vec{\hat{I}}^{T}_{} \, \mathbf{M^{( n )}_{}} \,
\vec{\hat{I}}^{}_{} + \frac{1}{2} G \left( E^{}_{\rm c} ,
E^{}_{\rm J} \right) \hat{\bar{\sigma}}^{}_x \, ,
 \label{H:op:T}
\end{equation}
where $\mathbf{C^{( n )}_{}}$ and $\mathbf{M^{( n )}_{}}$ are the
renormalized capacitance and inductance matrices up to the $n$-th
order, with $\mathbf{C^{( 1 )}_{}} {} \equiv {} \mathbf{C}$ and
$\mathbf{M^{( 1 )}_{}} {} \equiv {} \mathbf{M}$. Also,
$\vec{\hat{V}} \equiv [ \hat{V}^{}_{\rm A} , \hat{V}^{}_{\rm B} ,
\hat{V}^{}_{\rm Q} ]^{T}_{}$ and $\vec{\hat{I}} \equiv [
\hat{I}^{}_{\rm A} , \hat{I}^{}_{\rm B} , \hat{I}^{}_{\rm Q}
]^{T}_{}$. In general, $G$ is a function of the charging energy
$E^{}_{\rm c}$ and/or coupling energy $E^{}_{\rm J}$ of the
Josephson tunnel junctions in the qubit. For instance, $G {} = {}
E^{}_{\rm J}$ for a charge qubit and $G {} \propto {}
\sqrt{E^{}_{\rm c} E^{}_{\rm J}} \exp ( - \mu \sqrt{E^{}_{\rm c} /
E^{}_{\rm J}} )$ for a flux qubit ($\mu {} \equiv {} {\rm
const}$). Furthermore, $\hat{V}^{}_{\rm A} {} \equiv {} v^{}_{\rm
DC} + v^{}_{{\rm A} 0} ( \hat{a}^{\dag}_{} + \hat{a}^{}_{} )$,
$\hat{V}^{}_{\rm B} \equiv v^{}_{{\rm B} 0} ( \hat{b}^{\dag}_{} +
\hat{b}^{}_{} )$, $\hat{V}^{}_{\rm Q} \equiv v^{}_{\rm Q}
\hat{\bar{\sigma}}^{}_z$, $\hat{I}^{}_{\rm A} \equiv i^{}_{\rm DC}
+ i^{}_{{\rm A} 0}\,j ( \hat{a}^{\dag}_{} - \hat{a}^{}_{} )$,
$\hat{I}^{}_{\rm B} \equiv i^{}_{{\rm B} 0}\,j ( \hat{b}^{\dag}_{}
- \hat{b}^{}_{} )$, and $\hat{I}^{}_{\rm Q} \equiv i^{}_{\rm Q}
\hat{\bar{\sigma}}^{}_z$. In these expressions,
$\hat{\bar{\sigma}}^{}_x$ and $\hat{\bar{\sigma}}^{}_z$ are the
usual Pauli operators for a spin-$1 / 2$ system in the diabatic
basis, which consists of the eigenstates $\left| - \right\rangle$
and $\left| + \right\rangle$ of $C^{}_{\rm A Q} v^{}_{\rm DC}
v^{}_{\rm Q} \hat{\bar{\sigma}}^{}_z$ (charge case) or $M^{}_{\rm
A Q} i^{}_{\rm DC} i^{}_{\rm Q} \hat{\bar{\sigma}}^{}_z$ (flux
case). Additionally, $\hat{a}^{\dag}_{}$, $\hat{b}^{\dag}_{}$,
$\hat{a}^{}_{}$, and $\hat{b}^{}_{}$ are bosonic creation and
annihilation operators for the fields of cavities A and B,
respectively, and $j \equiv \sqrt{- 1}$. The DC voltage $v^{}_{\rm
DC}$ and current $i^{}_{\rm DC}$ account for the quasi-static
polarization of the qubit and can be applied through any suitable
bias circuit. For definiteness, we have chosen here cavity A to
perform this function. This is the standard approach followed by
the charge qubit circuit QED
community.~\cite{AWallraff:RJSchoelkopf:NatureLett:2004:a}
However, for flux qubits the current $i^{}_{\rm DC}$ is more
easily applied via an external
coil.~\cite{RHKoch:DPDiVincenzo:PhysRevLett:2006:a,
JJohansson:HTakayanagi:PhysRevLett:2006:a,
FDeppe:RGross:PhysRevB:2007:a,
FrankDeppe:MatteoMariantoni:RGross:NaturePhysLett:2008:a} In the
latter case, we impose $i^{}_{\rm DC} {} = {} 0$ and add to the
Hamiltonian of Eq.~(\ref{H:op:T}) the term $( \Phi^{\rm DC}_{\rm
x} - \Phi^{}_0 / 2 ) \hat{I}^{}_{\rm Q}$, where $\Phi^{\rm
DC}_{\rm x}$ is an externally applied flux bias and $\Phi^{}_0 {}
\equiv {} h / 2 e {} = {} 2.07 \times 10^{- 15}_{}$\,Wb is the
flux quantum. The results of our derivation are not affected by
this particular choice. The vacuum (zero point) fluctuations of
the voltage and current of each resonator are given by $v^{}_{{\rm
A} 0} {} \equiv {} \sqrt{\hbar \omega^{}_{\rm A} / 2 C^{}_{\rm A
A}}$, $v^{}_{{\rm B} 0} {} \equiv {} \sqrt{\hbar \omega^{}_{\rm B}
/ 2 C^{}_{\rm B B}}$, $i^{}_{{\rm A} 0} {} \equiv {} \sqrt{\hbar
\omega^{}_{\rm A} / 2 M^{}_{\rm A A}}$, and $i^{}_{{\rm B} 0} {}
\equiv {} \sqrt{\hbar \omega^{}_{\rm B} / 2 M^{}_{\rm B B}}$,
respectively. Here, $\omega^{}_{\rm A}$ and $\omega^{}_{\rm B}$
are the transition angular frequencies of the two cavities.
Finally, $v^{}_{\rm Q}$ and $i^{}_{\rm Q}$ represent the voltage
of the superconducting island(s) and the current through the loop
of the qubit circuit. Depending on the specific qubit
implementation, either $v^{}_{\rm Q}$ or $i^{}_{\rm Q}$ dominates,
thus defining the charge and flux regimes.

\subsection{The capacitance and inductance matrices up to second order}
 \label{subsection:the:capacitance:and:inductance:matrices:up:to:second:order}

The matrices $\mathbf{C^{( n )}_{}}$ and $\mathbf{M^{( n )}_{}}$
account for corrections up to the $n$-th order interaction process
between the elements of the network. In fact, in order to write
the exact Hamiltonian of the circuit, all possible electromagnetic
paths connecting its nodes must be considered. A consequence of
this approach to circuit theory is that the direct coupling
\begin{equation}
\widehat{H}^{( 1 )}_{\rm A B} {} = {} \hat{V}^{}_{\rm A} \, c \,
\hat{V}^{}_{\rm B} + \hat{I}^{}_{\rm A} \, m \, \hat{I}^{}_{\rm B}
 \label{H:op:(1):AB}
\end{equation}
between resonators A and B
[cf.~Fig.~\ref{QS:Figure:1:abcde:Matteo:Mariantoni:2008}(a)], here
assumed to be small, is not the only interaction mechanism to be
considered. In fact, an indirect coupling mediated by the circuit
associated with the qubit Q has also to be included in the
Hamiltonian. The dominating term for the A-Q-B excitation pathway
can be derived from its second-order electromagnetic energy
[cf.~Fig.~\ref{QS:Figure:1:abcde:Matteo:Mariantoni:2008}(b)],
which gives
 \setlength\arraycolsep{0pt}
\begin{eqnarray}
\widehat{H}^{( 2 )}_{\rm A B} & {} = {} & \widehat{H}^{( 1 )}_{\rm
A B} \nonumber\\
& & {} + \hat{V}^{}_{\rm A} \, C^{}_{\rm A Q} \frac{1}{C^{}_{\rm Q
Q}} C^{}_{\rm Q B} \, \hat{V}^{}_{\rm B} \nonumber\\
& & {} + \hat{I}^{}_{\rm A} \, M^{}_{\rm A Q} \frac{1}{M^{}_{\rm Q
Q}} M^{}_{\rm Q B} \, \hat{I}^{}_{\rm B} \, .
 \label{H:op:(2):AB}
\end{eqnarray}
Note that the inverse path (B-Q-A) is already included in this
equation. In our work, we assume $0 {} \lesssim {} c {} \lesssim
{} C^{}_{\rm A Q} C^{}_{\rm Q B} / C^{}_{\rm Q Q}$ and $0 {}
\lesssim {} m {} \lesssim {} M^{}_{\rm A Q} M^{}_{\rm Q B} /
M^{}_{\rm Q Q}$
(cf.~Sec.~\ref{section:an:example:of:two:resonator:circuit:qed:with:a:flux:qubit}).
When $c , m {} \simeq {} 0$, the direct coupling between A and B
is negligible, i.e., the contribution of $\widehat{H}^{( 1 )}_{\rm
A B}$ can be omitted. On the other hand, when $c {} > {} 0$ and/or
$m {} > {} 0$, both first- and second-order circuit theory
contributions are relevant. In this case, $c$ and $m$ can
represent a spurious or an engineered crosstalk. The latter can
deliberately be exploited to increase the strength of the
geometric second-order coupling. However, $c$ and $m$ should be
small enough to leave the mode structure and quality factors of A
and B unaffected.

From the knowledge of $\widehat{H}^{( 2 )}_{\rm A B}$, the
capacitance matrix up to second order is readily obtained
 {\setlength\arraycolsep{2.0mm}
\begin{equation}
 {\mathbf C}^{\left( 2 \right)}_{} {} = {}
  \begin{bmatrix}
   \raisebox{0mm}[6.5mm][0mm]{$C^{}_{\rm A A}$}
   &
   c + \dfrac{C^{}_{\rm A Q} C^{}_{\rm Q B}}{C^{}_{\rm Q Q}}
   &
   C^{}_{\rm A Q}
   \\
   c + \dfrac{C^{}_{\rm B Q} C^{}_{\rm Q A}}{C^{}_{\rm Q Q}}
   &
   C^{}_{\rm B B}
   &
   C^{}_{\rm B Q}
   \\
   \raisebox{0mm}[0mm][3.0mm]{$C^{}_{\rm Q A}$}
   &
   C^{}_{\rm Q B}
   &
   C^{}_{\rm Q Q}
  \end{bmatrix} \, .
 \label{C:(2):matrix}
\end{equation}
The second-order corrections to the self-capacitances, i.e., the
diagonal elements $C^{}_{k k}$ are absorbed in their
definitions~\cite{capacitances:values}
(cf.~Subsec.~\ref{subsection:the:role:of:circuit:topology:two:examples}).
In analogy, the corrected inductance matrix $\mathbf{M^{( 2
)}_{}}$ is found substituting $C^{}_{k l}$ with $M^{}_{k l}$ and
$c$ with $m$ in matrix~(\ref{C:(2):matrix}) yielding
\begin{equation}
 {\mathbf M}^{\left( 2 \right)}_{} {} = {}
  \begin{bmatrix}
   \raisebox{0mm}[6.5mm][0mm]{$M^{}_{\rm A A}$}
   &
   m + \dfrac{M^{}_{\rm A Q} M^{}_{\rm Q B}}{M^{}_{\rm Q Q}}
   &
   M^{}_{\rm A Q}
   \\
   m + \dfrac{M^{}_{\rm B Q} M^{}_{\rm Q A}}{M^{}_{\rm Q Q}}
   &
   M^{}_{\rm B B}
   &
   M^{}_{\rm B Q}
   \\
   \raisebox{0mm}[0mm][3.0mm]{$M^{}_{\rm Q A}$}
   &
   M^{}_{\rm Q B}
   &
   M^{}_{\rm Q Q}
  \end{bmatrix} \, .
 \label{M:(2):matrix}
\end{equation}
Again, second-order corrections to the self-inductances are
absorbed in the definition of $M^{}_{k k}$. The matrices
$\mathbf{C^{( 2 )}_{}}$ and $\mathbf{M^{( 2 )}_{}}$ constitute the
\textit{first main result} of this work. They show that, if a
large qubit-resonator coupling (i.e., a vacuum Rabi coupling
$\propto C^{}_{\rm A Q} \, , \, C^{}_{\rm Q B}$ for charge quantum
circuits and $\propto M^{}_{\rm A Q} \, , \, M^{}_{\rm Q B}$ for
flux quantum circuits) is present, as in most circuit~QED
implementations,~\cite{AWallraff:RJSchoelkopf:NatureLett:2004:a,
JJohansson:HTakayanagi:PhysRevLett:2006:a,
RJSchoelkopf:SMGirvin:NatureHorizons:2008:a,
MikaASillanpaeae:RaymondWSimmonds:NatureLett:2007:a,
JMajer:JMChow:RJSchoelkopf:NatureLett:2007:a,
AAHouck:DISchuster:RJSchoelkopf:NatureLett:2007:a,
MaxHofheinz:JohnMMartinis:ANCleland:NatureLett:2008:a,
OAstafiev:JSTsai:NatureLett:2007:a,
JMFink:AWallraff:NatureLett:2008:a,
LevSBishop:RJSchoelkopf:arXiv:2008:a,
FrankDeppe:MatteoMariantoni:RGross:NaturePhysLett:2008:a,
RHKoch:DPDiVincenzo:PhysRevLett:2006:a,
DISchuster:AAHouck:RJSchoelkopf:NatureLett:2007:a} a relevant
geometric second-order coupling ($\propto C^{}_{\rm A Q} C^{}_{\rm
Q B} / C^{}_{\rm Q Q}$ or $\propto M^{}_{\rm A Q} M^{}_{\rm Q B} /
M^{}_{\rm Q Q}$ for charge and flux quantum circuits,
respectively) has to be expected. This coupling becomes relevant
in the dispersive
regime,~\cite{AlexandreBlais:RJSchoelkopf:PhysRevA:2004:a,
DISchuster:AAHouck:RJSchoelkopf:NatureLett:2007:a} where a dynamic
second-order coupling, whose magnitude can be comparable to that
of the geometric one, is also present
(cf.~Subsec.~\ref{subsection:balancing:the:geometric:and:dynamic:coupling}).
We study in more detail the relationship between $m$ and
$M^{}_{\rm A Q} M^{}_{\rm Q B} / M^{}_{\rm Q Q}$ in
Sec.~\ref{section:an:example:of:two:resonator:circuit:qed:with:a:flux:qubit}.
There, we show that for a realistic design engineered for a flux
qubit, which is our experimental
expertise,~\cite{FDeppe:RGross:PhysRevB:2007:a,
FrankDeppe:MatteoMariantoni:RGross:NaturePhysLett:2008:a} the
geometric second-order interaction dominates over the first-order
one.}

Figures~\ref{QS:Figure:1:abcde:Matteo:Mariantoni:2008}(c)-(e) show
three generic sketches, where the coupling of two on-chip
resonators to one superconducting qubit is illustrated. In
particular, the sketch drawn in
Fig.~\ref{QS:Figure:1:abcde:Matteo:Mariantoni:2008}(c) is suitable
when a flux qubit is intended to be utilized. In this case, the
qubit is positioned at the current antinode of the first
mode~\cite{MMariantoni:ESolano:arXiv:cond-mat:2005:a} of two
$\lambda / 2$ resonators. Moreover, this design clearly allows for
engineering a strong coupling between the qubit and each
resonator, while reducing the geometric first-order coupling
between resonators A and B. This is due to the fact that the two
cavities are close to each other only in the restricted region
where the qubit is located and then develop abruptly towards
opposite directions. The sketch in
Fig.~\ref{QS:Figure:1:abcde:Matteo:Mariantoni:2008}(d), instead,
is more suitable for charge qubit applications. The qubit can
easily be fabricated near a voltage
antinode.~\cite{AlexandreBlais:RJSchoelkopf:PhysRevA:2004:a,
AWallraff:RJSchoelkopf:NatureLett:2004:a} Similar arguments as in
the previous case apply for the qubit-resonator couplings and the
geometric first-order coupling between A and B. Finally, the
sketch of Fig.~\ref{QS:Figure:1:abcde:Matteo:Mariantoni:2008}(e)
relies on an orthogonal-cavity design, which can be used for both
charge and flux qubits. The main properties of such a setup have
already been presented in two of our previous
works,~\cite{MJStorcz:MMariantoni:ESolano:arXiv:cond-mat:2007:a,
FerdinandHelmer:FlorianMarquardt:arXiv:2007:a} where orthogonal
cavities have been exploited for different purposes. In
conclusion, we want to stress that based on the general sketches
of Figs.~\ref{QS:Figure:1:abcde:Matteo:Mariantoni:2008}(c)-(e), a
large variety of specific experimental implementations can be
envisioned.

\subsection{The role of circuit topology: Two examples}
 \label{subsection:the:role:of:circuit:topology:two:examples}

\begin{figure*}[t!]
\centering{%
 \includegraphics[width=0.99\textwidth]{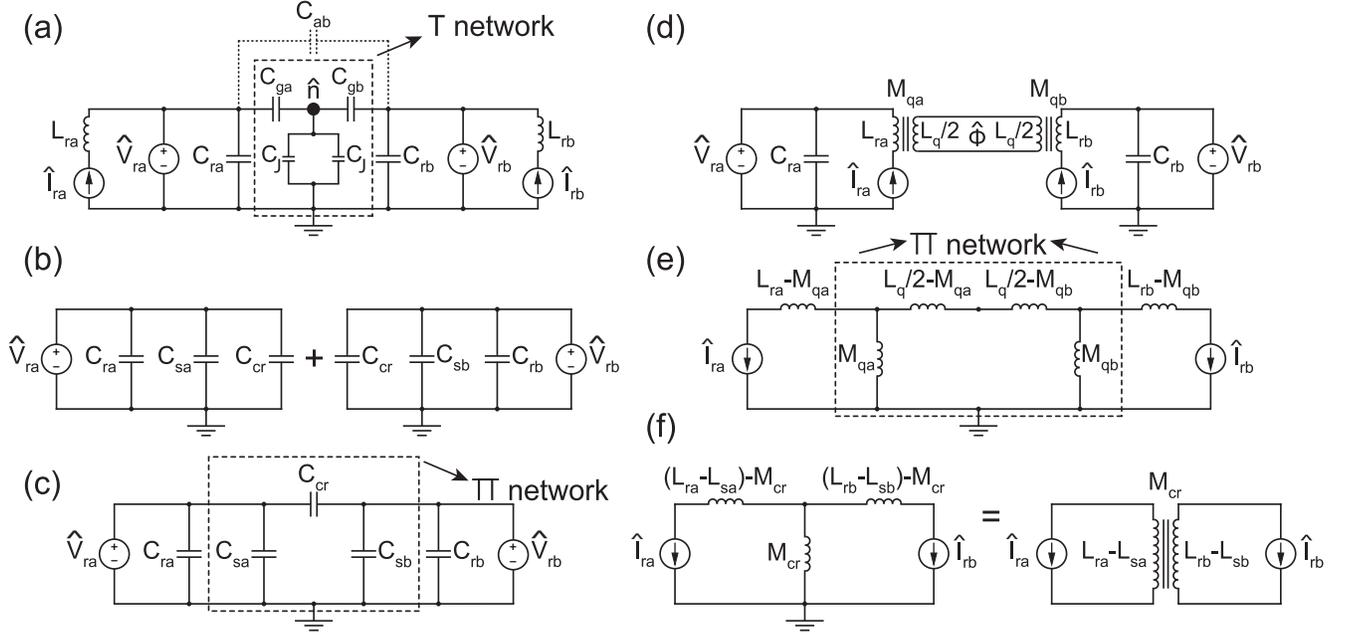}}
\caption{Equivalent circuit diagrams for two different
implementations of two-resonator circuit~QED based on either a
charge qubit [(a)-(c)] or a flux qubit [(d)-(f)].
Cf.~Subsec.~\ref{subsection:the:role:of:circuit:topology:two:examples}
for details. (a)~$\widehat{V}^{}_{\rm r a}$ and
$\widehat{V}^{}_{\rm r b}$: Quantized voltage sources associated
with resonators A and B in parallel to the self-capacitances
$C^{}_{\rm r a}$ and $C^{}_{\rm r b}$ of the resonators.
$\hat{I}^{}_{\rm r a}$ and $\hat{I}^{}_{\rm r b}$: Quantized
current sources associated with resonators A and B in series to
the self-inductances $L^{}_{\rm r a}$ and $L^{}_{\rm r b}$ of the
resonators. The number of excess Cooper pairs on the charge qubit
island (big dot) is $\left\langle n \right| \hat{n} \left| n
\right\rangle$. $C^{}_{\rm J}$: Capacitance of each of the two
Josephson tunnel junctions connecting the island to ground.
$C^{}_{\rm g a}$ and $C^{}_{\rm g b}$: Coupling capacitances
between the qubit and the two resonators. $C^{}_{\rm a b}$:
First-order cross-capacitance between A and B (typically small,
dotted line). The dashed box marks a T-network composed of
$C^{}_{\rm g a}$, $2 C^{}_{\rm J}$, and $C^{}_{\rm g b}$.
(b)~$C^{}_{\rm cr} {} \equiv {} C^{}_{\rm g a} C^{}_{\rm g b} /
C^{}_{\Sigma}$: Second-order cross-capacitance. $C^{}_{\rm s a} {}
\equiv {} 2 C^{}_{\rm J} C^{}_{\rm g a} / C^{}_{\Sigma}$ and
$C^{}_{\rm s b} {} \equiv {} 2 C^{}_{\rm J} C^{}_{\rm g b} /
C^{}_{\Sigma}$: Resonator shift capacitances. $C^{}_{\rm a b}$ is
neglected for simplicity. (c)~The circuits of (b) rearranged as a
single $\Pi$-network (dashed box). The latter is equivalent to the
T-network of (a). The magnitudes of $C^{}_{\rm r a}$ and
$C^{}_{\rm r b}$ are increased by the presence of the shift
capacitances $C^{}_{\rm s a}$ and $C^{}_{\rm s b}$. (d)~Two
resonators A and B inductively coupled via $M^{}_{\rm q a}$ and
$M^{}_{\rm q b}$ to a flux qubit with total self-inductance
$L^{}_{\rm q} {} = {} L^{}_{\rm q} / 2 + L^{}_{\rm q} / 2$ and
flux operator $\widehat{\Phi}$. The first-order mutual inductance
$m$ between the two resonators is neglected to simplify the
notation. (e)~The disconnected circuit of (d) is transformed into
a connected
circuit.~\cite{LeonOChua:CharlesADesoer:ErnestSKuh:Book:1987:a}
Again, we can identify a $\Pi$-network (dashed box). (f)~Left
side: T-network obtained from the $\Pi$-network of (e). We
identify the second-order mutual inductance $M^{}_{\rm cr} {}
\equiv {} M^{}_{\rm q a} M^{}_{\rm q b} / L^{}_{\rm q}$ and the
shift inductances $L^{}_{\rm s a} {} \equiv {} M^2_{\rm q a} /
L^{}_{\rm q}$ and $L^{}_{\rm s b} {} \equiv {} M^2_{\rm q b} /
L^{}_{\rm q}$. Right side: The connected circuit on the left side
is transformed into a disconnected
circuit.~\cite{LeonOChua:CharlesADesoer:ErnestSKuh:Book:1987:a}}
 \label{QS:Figure:2:abcdef:Matteo:Mariantoni:2008}
\end{figure*}

All results of
Subsecs.~\ref{subsection:the:hamiltonian:of:a:generic:three:node:network}
and
\ref{subsection:the:capacitance:and:inductance:matrices:up:to:second:order}
are general and do not rely \textit{a priori} on the knowledge of
the three-circuit network topology. Here, we explain with the aid
of two easy examples how to obtain a reduced three-node network
starting from a three-circuit one. The examples are based on the
coupling of two superconducting coplanar waveguide or microstrip
resonators to a single Cooper-pair
box~\cite{AlexandreBlais:RJSchoelkopf:PhysRevA:2004:a,
AWallraff:RJSchoelkopf:NatureLett:2004:a} (or a
transmon~\cite{JensKoch:TerriMYu:JayGambetta:AlexandreBlais:RJSchoelkopf:PhysRevA:2007:a,
JASchreier:JensKoch:RJSchoelkopf:PhysRevBRap:2008:a,
AAHouck:JensKoch:RJSchoelkopf:arXiv:2008:a}) or to a
superconducting loop interrupted by one (or three)
Josephson-tunnel
junctions.~\cite{Chui-PingYang:SiyuanHan:PhysRevA:2003:a,
MMariantoni:ESolano:arXiv:cond-mat:2005:a,
RHKoch:DPDiVincenzo:PhysRevLett:2006:a,
TLindstroem:AYaTzalenchuk:JPhysConfSer:2008:a}

\textit{The first example} is the case of a single Cooper-pair box
(a charge quantum circuit), which is formally equivalent to the
more appealing case of the transmon. A single Cooper-pair
box~\cite{AlexandreBlais:RJSchoelkopf:PhysRevA:2004:a,
AWallraff:RJSchoelkopf:NatureLett:2004:a} is made of a
superconducting island connected to a large reservoir via two
Josephson tunnel junctions with Josephson energy $E^{}_{\rm J}$
and capacitance $C^{}_{\rm J}$. The box is capacitevely coupled to
two resonators A and B by the gate capacitors $C^{}_{\rm g a}$ and
$C^{}_{\rm g b}$, respectively. In the charge basis, the
Hamiltonian of a single Cooper-pair box can be written
as~\cite{AlexandreBlais:RJSchoelkopf:PhysRevA:2004:a}
\begin{eqnarray}
\widehat{H}^{}_{\rm c} & {} = {} & 4 E^{}_{\rm C} \sum_n^{} (
\hat{n}
- n^{}_{\rm g} )^2_{} \left| n \rangle \langle n \right| \nonumber\\[1.5mm]
& & {} - \frac{E^{}_{\rm J}}{2} \sum_n^{} ( \left| n \rangle
\langle n + 1 \right| + \left| n + 1 \rangle \langle n \right| )
\, ,
 \label{H:op:c:I}
\end{eqnarray}
where $E^{}_{\rm C} {} = {} e^2_{} / 2 C^{}_{\Sigma}$ is the box
electrostatic energy ($e$ is the electron charge), $C^{}_{\Sigma}
{} = {} C^{}_{\rm g a} + 2 C^{}_{\rm J} + C^{}_{\rm g b}$ is its
total capacitance,~\cite{single:Cooper-pair:box:self-capacitance}
$\left\langle n \right| \hat{n} \left| n \right\rangle$ represents
the number of excess Cooper pairs on the island, and $n^{}_{\rm
g}$ is the global dimensionless gate charge applied to it. The
latter is the sum of a DC signal $n^{\rm DC}_{\rm g}$ (here,
considered to be applied through cavity A) and a high-frequency
excitation $\delta n^{}_{\rm g}$ applied through cavities A and/or
B, $n^{}_{\rm g} {} \equiv {} n^{\rm DC}_{\rm g} + \delta
n^{}_{\rm g}$. In particular, $\delta n^{}_{\rm g}$ can represent
the quantized electric fields (equivalent to the voltages) of the
two cavities acting as quantum harmonic oscillators. Restricting
ourselves to the two lowest charge states $n {} = {} 0 , 1$, we
can rewrite the Hamiltonian of Eq.~(\ref{H:op:c:I}) as
\begin{eqnarray}
\widehat{H}^{}_{\rm c} & {} = {} & 2 E^{}_{\rm C} \left( 1 - 2
n^{}_{\rm g} + 2 n^2_{\rm g} + \hat{\bar{\sigma}}^{}_z - 2
n^{}_{\rm g} \hat{\bar{\sigma}}^{}_z \right) - \frac{E^{}_{\rm
J}}{2} \hat{\bar{\sigma}}^{}_x \nonumber\\[1.5mm]
& {} = {} & 2 E^{}_{\rm C} \left( 1 - 2 n^{\rm DC}_{\rm g} \right)
\hat{\bar{\sigma}}^{}_z - \frac{E^{}_{\rm J}}{2}
\hat{\bar{\sigma}}^{}_x \nonumber\\[1.5mm]
& & {} - 4 E^{}_{\rm C} \delta n^{}_{\rm g} \left( 1 - 2 n^{\rm
DC}_{\rm g} - \delta n^{}_{\rm g} + \hat{\bar{\sigma}}^{}_z
\right) \, .
 \label{H:op:c:II}
\end{eqnarray}
The second line of the above equation forms the standard charge
qubit, which can be controlled by the quasi-static bias $n^{\rm
DC}_{\rm g} {} \equiv {} C^{}_{\rm g a} v^{}_{\rm DC} / 2 e$. The
third line contains four high-frequency interaction terms. Among
those, two of them are particularly
interesting.~\cite{displacement-type:operators} These are $4
E^{}_{\rm C} \delta n^2_{\rm g}$ and $- 4 E^{}_{\rm C} \delta
n^{}_{\rm g} \hat{\bar{\sigma}}^{}_z$. We now quantize the
high-frequency excitations $\delta n^{}_{\rm g} {} \rightarrow {}
\delta \hat{n}^{}_{\rm g} {} \equiv {} C^{}_{\rm g a} v^{}_{{\rm
A} 0} ( \hat{a}^{\dag}_{} + \hat{a}^{}_{} ) / 2 e + C^{}_{\rm g b}
v^{}_{{\rm B} 0} ( \hat{b}^{\dag}_{} + \hat{b}^{}_{} ) / 2 e$,
using the fact that they are the quantized voltages of the two
resonators. We subsequently perform a rotating-wave
approximation~(RWA) and, finally, write the interaction
Hamiltonian
\begin{eqnarray}
\widehat{H}^{\rm int}_{\rm c} & {} = {} & \hbar G^{}_{\rm A B} (
\hat{a}^{\dag}_{} + \hat{a}^{}_{} )
( \hat{b}^{\dag}_{} + \hat{b}^{}_{} ) \nonumber\\[1.5mm]
& & {} - \hbar G^{}_{\rm A} \hat{\bar{\sigma}}^{}_z (
\hat{a}^{\dag}_{} + \hat{a}^{}_{} ) - \hbar G^{}_{\rm B}
\hat{\bar{\sigma}}^{}_z ( \hat{b}^{\dag}_{} + \hat{b}^{}_{} ) {}
\nonumber\\[1.5mm]
& & {} + \hbar \tilde{\omega}^{}_{\rm A} \hat{a}^{\dag}_{}
\hat{a}^{}_{} + \hbar \tilde{\omega}^{}_{\rm B} \hat{b}^{\dag}_{}
\hat{b}^{}_{} \, ,
 \label{H:op:c:int}
\end{eqnarray}
where all constant energy offsets, e.g., the Lamb shifts, have
been neglected. Remarkably, in the first line of the above
equation we identify a geometric resonator-resonator interaction
term with second-order coupling coefficient $G^{}_{\rm A B} {}
\equiv {} v^{}_{{\rm A} 0} v^{}_{{\rm B} 0} C^{}_{\rm g a}
C^{}_{\rm g b} / C^{}_{\Sigma} \hbar$. Furthermore, the two terms
of the second line of this equation represent the expected
first-order qubit-resonator interactions with coupling
coefficients $G^{}_{\rm A} {} \equiv {} e ( C^{}_{\rm g a} /
C^{}_{\Sigma} ) v^{}_{{\rm A} 0} / \hbar$ and $G^{}_{\rm B} {}
\equiv {} e ( C^{}_{\rm g b} / C^{}_{\Sigma} ) v^{}_{{\rm B} 0} /
\hbar$, respectively. In the third line, the two small
renormalizations $\widetilde{\omega}^{}_{\rm A} {} \equiv {} (
C^{}_{\rm g a} v^{}_{{\rm A} 0} )^2_{} / C^{}_{\Sigma} \hbar$ and
$\widetilde{\omega}^{}_{\rm B} {} \equiv {} ( C^{}_{\rm g b}
v^{}_{{\rm B} 0} )^2_{} / C^{}_{\Sigma} \hbar$ of the resonator
angular frequencies are artifacts due to the simple model behind
the Hamiltonian of Eq.~(\ref{H:op:c:I}). A more advanced model
based on a realistic circuit topology yields similar
renormalization terms, which, however, are governed by different
topology-dependent constants. Among the possible ways to find the
correct constants, we choose the circuit transformations of
Figs.~\ref{QS:Figure:2:abcdef:Matteo:Mariantoni:2008}(a)-(c). This
approach also allows us to better understand the geometric
second-order interaction term.

In Fig.~\ref{QS:Figure:2:abcdef:Matteo:Mariantoni:2008}(a), the
two cavities are represented as $L C$-resonators with total
capacitances and inductances $C^{}_{\rm r a}$, $C^{}_{\rm r b}$,
$L^{}_{\rm r a}$, and $L^{}_{\rm r b}$, respectively. The
quantized voltages and currents of the two resonators are
$\widehat{V}^{}_{\rm r a} {} \equiv {} v^{}_{{\rm A} 0} (
\hat{a}^{\dag}_{} + \hat{a}^{}_{} )$, $\widehat{V}^{}_{\rm r b} {}
\equiv {} v^{}_{{\rm B} 0} ( \hat{b}^{\dag}_{} + \hat{b}^{}_{} )$,
$\hat{I}^{}_{\rm r a} {} \equiv {} i^{}_{{\rm A} 0}\,j (
\hat{a}^{\dag}_{} - \hat{a}^{}_{} )$, and $\hat{I}^{}_{\rm r b} {}
\equiv {} i^{}_{{\rm B} 0}\,j ( \hat{b}^{\dag}_{} - \hat{b}^{}_{}
)$, respectively. Also, $C^{}_{\rm a b}$ accounts for a
first-order cross-capacitance between resonators A and B, which,
for simplicity, is neglected in Eqs.~(\ref{H:op:c:II}) and
(\ref{H:op:c:int}). In addition, here we are only interested in
the geometric properties of the charge quantum circuit. The
dynamic properties of this circuit are studied following a more
canonical approach within a two-level approximation in
Sec.~\ref{section:derivation:of:the:quantum:switch:hamiltonian}.
The dynamic properties are governed by the two Josephson tunnel
junctions and by the number of excess Cooper pairs on the island,
$\left\langle n \right| \hat{n} \left| n \right\rangle$. To
simplify our derivations, we can then assume $\hat{n} = 0$ and
consider only the capacitance $C^{}_{\rm J}$ of the two Josephson
tunnel junctions, but not their Josephson energy.

We now derive in three steps the geometric part of the interaction
Hamiltonian by means of circuit theory. The procedure is
visualized in
Figs.~\ref{QS:Figure:2:abcdef:Matteo:Mariantoni:2008}(a)-(c). The
steps are:

\textit{(i)} - First, we assume that the circuit associated to the
charge qubit is positioned at a voltage
antinode~\cite{AlexandreBlais:RJSchoelkopf:PhysRevA:2004:a} of
both resonators. Consequently, we can replace the two current
sources of Fig.~\ref{QS:Figure:2:abcdef:Matteo:Mariantoni:2008}(a)
with open circuits, $\hat{I}^{}_{\rm r a} {} = {} \hat{I}^{}_{\rm
r b} {} = {} 0$. Thus, we can eliminate both $L^{}_{\rm r a}$ and
$L^{}_{\rm r b}$ from the circuit diagram because they are in
series to open circuits.

\textit{(ii)} - Second, we apply the superposition principle of
circuit
theory.~\cite{LeonOChua:CharlesADesoer:ErnestSKuh:Book:1987:a} One
at the time, we replace each of the two voltage sources with short
circuits, $\hat{V}^{}_{\rm r a} {} = {} 0$ or $\hat{V}^{}_{\rm r
b} {} = {} 0$. This allows us to split up the circuit of
Fig.~\ref{QS:Figure:2:abcdef:Matteo:Mariantoni:2008}(a) into the
two subcircuits of
Fig.~\ref{QS:Figure:2:abcdef:Matteo:Mariantoni:2008}(b), which are
topologically less complex. As a consequence, in the respective
subcircuits, $C^{}_{\rm r b}$ or $C^{}_{\rm r a}$ can be
substituted by short circuits and all other capacitors opportunely
rearranged. In this way, for the case of cavity A, we find the
small shift capacitance $C^{}_{\rm s a} {} \equiv {} 2 C^{}_{\rm
J} C^{}_{\rm g a} / C^{}_{\Sigma}$, which gives the correct
angular frequency renormalization of the resonator,
$\widetilde{\omega}^{\rm corr}_{\rm A} {} \equiv {} 2 C^{}_{\rm J}
C^{}_{\rm g a} v^2_{{\rm A} 0} / C^{}_{\Sigma} \hbar$. Remarkably,
we also find the second-order cross-capacitance $C^{}_{\rm cr} {}
\equiv {} C^{}_{\rm g a} C^{}_{\rm g b} / C^{}_{\Sigma}$,
corresponding to the geometric second-order coupling between the
resonators. This coincides with our result obtained in
Eq.~(\ref{H:op:(2):AB}) of
Subsec.~\ref{subsection:the:capacitance:and:inductance:matrices:up:to:second:order}
and is consistent with the simple model of
Eqs.~(\ref{H:op:c:I})-(\ref{H:op:c:int}). We notice that
$C^{}_{\rm cr}$ deviates from the simple series of the two gate
capacitances $C^{}_{\rm g a}$ and $C^{}_{\rm g b}$ because of the
presence of $C^{}_{\rm J}$ in $C^{}_{\Sigma}$. For the case of
cavity B, $C^{}_{\rm s b} {} \equiv {} 2 C^{}_{\rm J} C^{}_{\rm g
b} / C^{}_{\Sigma}$ and $\widetilde{\omega}^{\rm corr}_{\rm B} {}
\equiv {} 2 C^{}_{\rm J} C^{}_{\rm g b} v^2_{{\rm B} 0} /
C^{}_{\Sigma} \hbar$ can be derived in an analogous manner. In
Subsec.~\ref{subsection:the:capacitance:and:inductance:matrices:up:to:second:order},
the two renormalization constants as well as $C^{}_{\rm J}$ are
absorbed in the definitions of $C^{}_{\rm A A}$, $C^{}_{\rm B B}$,
and $C^{}_{\rm Q Q}$, respectively.

\textit{(iii)} - Third, we notice that the cross-capacitance
$C^{}_{\rm cr}$, which is responsible for the geometric
second-order interaction between A and B, is subjected to both
quantum voltages $\widehat{V}^{}_{\rm r a}$ and
$\widehat{V}^{}_{\rm r b}$. Hence, we can finally draw the circuit
diagram of
Fig.~\ref{QS:Figure:2:abcdef:Matteo:Mariantoni:2008}(c). Indeed,
we could have identified the T-network of
Fig.~\ref{QS:Figure:2:abcdef:Matteo:Mariantoni:2008}(a) (indicated
by a dashed box) and transformed it into the equivalent
$\Pi$-network of
Fig.~\ref{QS:Figure:2:abcdef:Matteo:Mariantoni:2008}(c) (also
indicated by a dashed box) in one single
step,~\cite{LeonOChua:CharlesADesoer:ErnestSKuh:Book:1987:a}
obtaining the same results. We prefer to explicitly show the steps
of Fig.~\ref{QS:Figure:2:abcdef:Matteo:Mariantoni:2008}(b) for
pedagogical reasons.

\textit{The second example} is based on a superconducting loop
interrupted by one Josephson tunnel junction (a flux quantum
circuit). Such a device is also known as radio-frequency~(RF)
superconducting quantum interference device~(SQUID). We choose the
RF~SQUID here for pure pedagogical reasons. In fact, our treatment
could be extended to the more common case of three
junctions.~\cite{TPOrlando:JEMooij:PhysRevB:1999:a,
MatteoMariantoni:PhDThesis:2008:a} The Hamiltonian of an RF~SQUID
can be expressed
as~\cite{YuriyMakhlin:GerdSchoen:RevModPhys:2001:a,
MHDevoret:JMMartinis:arXiv:cond-mat:2004:a,
Chui-PingYang:SiyuanHan:PhysRevA:2003:a,
MMariantoni:ESolano:arXiv:cond-mat:2005:a}
\begin{equation}
\widehat{H}^{}_{\rm f} {} = {} \frac{\widehat{Q}^2}{2 C^{}_{\rm
J}} + \frac{\left( \widehat{\Phi} - \Phi^{}_{\rm x}
\right)^2_{}}{2 L^{}_{\rm q}} - E^{}_{\rm J} \cos \left( 2 \pi
\frac{\widehat{\Phi}}{\Phi^{}_0} \right) \, ,
 \label{H:op:f}
\end{equation}
where $\widehat{Q}$ is the operator for the charge accumulated on
the capacitor $C^{}_{\rm J}$ associated with the Josephson tunnel
junction. The flux operator $\widehat{\Phi}$ is the conjugated
variable of $\widehat{Q}$, i.e., $[ \widehat{\Phi} , \widehat{Q} ]
= j \hbar$. In analogy to the dimensionless gate charge $n^{}_{\rm
g}$ of the previous example, the flux bias $\Phi^{}_{\rm x} \equiv
\Phi^{\rm DC}_{\rm x} + \delta \Phi^{}_{\rm x}$ consists of a DC
and an AC component. The self-inductance of the superconducting
loop is defined as $L^{}_{\rm q}$. When the RF~SQUID is coupled to
two quantized resonators, we can quantize the high-frequency
excitations performing the transformations $\delta \Phi^{}_{\rm x}
{} \rightarrow {} \delta \widehat{\Phi}^{}_{\rm x} {} \equiv {}
M^{}_{\rm q a} i^{}_{{\rm A} 0}\,j ( \hat{a}^{\dag}_{} -
\hat{a}^{}_{} ) + M^{}_{\rm q b} i^{}_{{\rm B} 0}\,j (
\hat{b}^{\dag}_{} - \hat{b}^{}_{} )$. Here, $M^{}_{\rm q a}$ and
$M^{}_{\rm q b}$ are the mutual inductances between the loop and
each resonator. We can then assume $\widehat{\Phi} {} = {} 0$,
perform a two-level approximation and a RWA, and finally obtain
the same interaction Hamiltonian as in Eq.~(\ref{H:op:c:int}).
However, in this case the coefficients are redefined as
$\widetilde{\omega}^{}_{\rm A} {} \equiv {} ( M^{}_{\rm q a}
i^{}_{{\rm A} 0} )^2_{} / L^{}_{\rm q} \hbar$,
$\widetilde{\omega}^{}_{\rm B} {} \equiv {} ( M^{}_{\rm q b}
i^{}_{{\rm B} 0} )^2_{} / L^{}_{\rm q} \hbar$, and $G^{}_{\rm A B}
\equiv i^{}_{{\rm A} 0} i^{}_{{\rm B} 0} M^{}_{\rm q a} M^{}_{\rm
q b} / L^{}_{\rm q} \hbar$. The term with coupling coefficient
$G^{}_{\rm A B}$ constitutes the geometric second-order
interaction between A and B. As it appears clear from the
discussion below, once again the renormalization terms
$\widetilde{\omega}^{}_{\rm A}$ and $\widetilde{\omega}^{}_{\rm
B}$ do not catch the circuit topology properly. This issue can be
clarified analyzing the circuit diagram drawn in
Fig.~\ref{QS:Figure:2:abcdef:Matteo:Mariantoni:2008}(d), where all
the geometric elements for this example are shown. The geometric
first-order mutual inductance $m$ between the two resonators is
neglected to simplify the notation. Again, the Josephson tunnel
junctions responsible for the dynamic behaviour are not included.

We now study the geometric part of the interaction Hamiltonian
between the RF~SQUID and the two resonators following a similar
path as for the case of the single Cooper-pair box
[cf.~Figs.~\ref{QS:Figure:2:abcdef:Matteo:Mariantoni:2008}(d)-(f)].
The four main transformation steps are:

\textit{(i)} - First, we assume the circuit corresponding to the
flux qubit to be positioned at a current antinode. Thus, in
Fig.~\ref{QS:Figure:2:abcdef:Matteo:Mariantoni:2008}(e), we
replace all voltage sources and capacitors of
Fig.~\ref{QS:Figure:2:abcdef:Matteo:Mariantoni:2008}(d) with short
circuits. The self-inductance of the qubit loop is split up into
two $L^{}_{\rm q} / 2$ inductances to facilitate the following
transformation steps.

\textit{(ii)} - Second, a well-known theorem of circuit
theory~\cite{LeonOChua:CharlesADesoer:ErnestSKuh:Book:1987:a}
allows us to transform the three disconnected circuits of
Fig.~\ref{QS:Figure:2:abcdef:Matteo:Mariantoni:2008}(d) into the
connected circuit of
Fig.~\ref{QS:Figure:2:abcdef:Matteo:Mariantoni:2008}(e). Here, the
region indicated by the dashed box evidently forms a
$\Pi$-network.

\textit{(iii)} - Third, a $\Pi$-to-T-network
transformation~\cite{LeonOChua:CharlesADesoer:ErnestSKuh:Book:1987:a}
results in the circuit on the left side of
Fig.~\ref{QS:Figure:2:abcdef:Matteo:Mariantoni:2008}(f).

\textit{(iv)} - Fourth, applying the inverse theorem of that used
in step \textit{(ii)} finally allows us to draw the equivalent
circuit on the right side of
Fig.~\ref{QS:Figure:2:abcdef:Matteo:Mariantoni:2008}(f). Here,
$M^{}_{\rm cr} {} \equiv {} M^{}_{\rm q a} M^{}_{\rm q b} /
L^{}_{\rm q}$ represents the second-order mutual inductance
between resonators A and B, corresponding to the geometric
second-order coupling between them. Remarkably, this coincides
with our result obtained in Eq.~(\ref{H:op:(2):AB}) of
Subsec.~\ref{subsection:the:capacitance:and:inductance:matrices:up:to:second:order}
and is consistent with the simple model of Eq.~(\ref{H:op:f}).
However, in this model the small shift inductances $L^{}_{\rm s a}
{} \equiv {} M^2_{\rm q a} / L^{}_{\rm q}$ and $L^{}_{\rm s b} {}
\equiv {} M^2_{\rm q b} / L^{}_{\rm q}$ (here defined to be
strictly positive) acquire the wrong sign. Our circuit approach
reveals that the correct renormalization constants of the
resonators angular frequency are $\widetilde{\omega}^{}_{\rm A} {}
= {} - L^{}_{\rm s a} i^2_{{\rm A} 0} / \hbar$ and
$\widetilde{\omega}^{}_{\rm B} {} = {} - L^{}_{\rm s b} i^2_{{\rm
B} 0} / \hbar$. This result is also confirmed by our numerical
simulations
(cf.~Sec.~\ref{section:an:example:of:two:resonator:circuit:qed:with:a:flux:qubit}
and Table~\ref{QS:Table:1:Matteo:Mariantoni:2008}). In
Subsec.~\ref{subsection:the:capacitance:and:inductance:matrices:up:to:second:order},
these renormalization constants are absorbed in the definitions of
$M^{}_{\rm A A}$ and $M^{}_{\rm B B}$.

\section{DERIVATION OF THE QUANTUM SWITCH HAMILTONIAN}
 \label{section:derivation:of:the:quantum:switch:hamiltonian}

In this section, we analyze the Hamiltonian of a three-node
quantum network as found in
Subsec.~\ref{subsection:the:capacitance:and:inductance:matrices:up:to:second:order}.
In particular, we focus on the relevant case of large
qubit-resonator detuning, i.e., the dispersive regime of
two-resonator circuit~QED. Under this assumption, we are able to
derive an effective Hamiltonian describing a quantum switch
between two resonators. We compare the analytical results to those
of extensive simulations
(cf.~Subsec.~\ref{subsection:balancing:the:geometric:and:dynamic:coupling}).
We also propose a protocol for the quantum switch operation
stressing two possible variants. One is based on a qubit
population inversion and the other on an adiabatic shift pulse
with the qubit in the energy groundstate
(cf.~Subsec.~\ref{subsection:a:quantum:switch:protocol}). Finally,
we give a few examples of advanced applications of the quantum
switch and, more in general, of dispersive two-resonator
circuit~QED
(cf.~Subsec.~\ref{subsection:advanced:applications:nonclassical:states:and:entanglement}).

\begin{figure*}[t!]
\centering{%
 \includegraphics[width=0.85\textwidth]{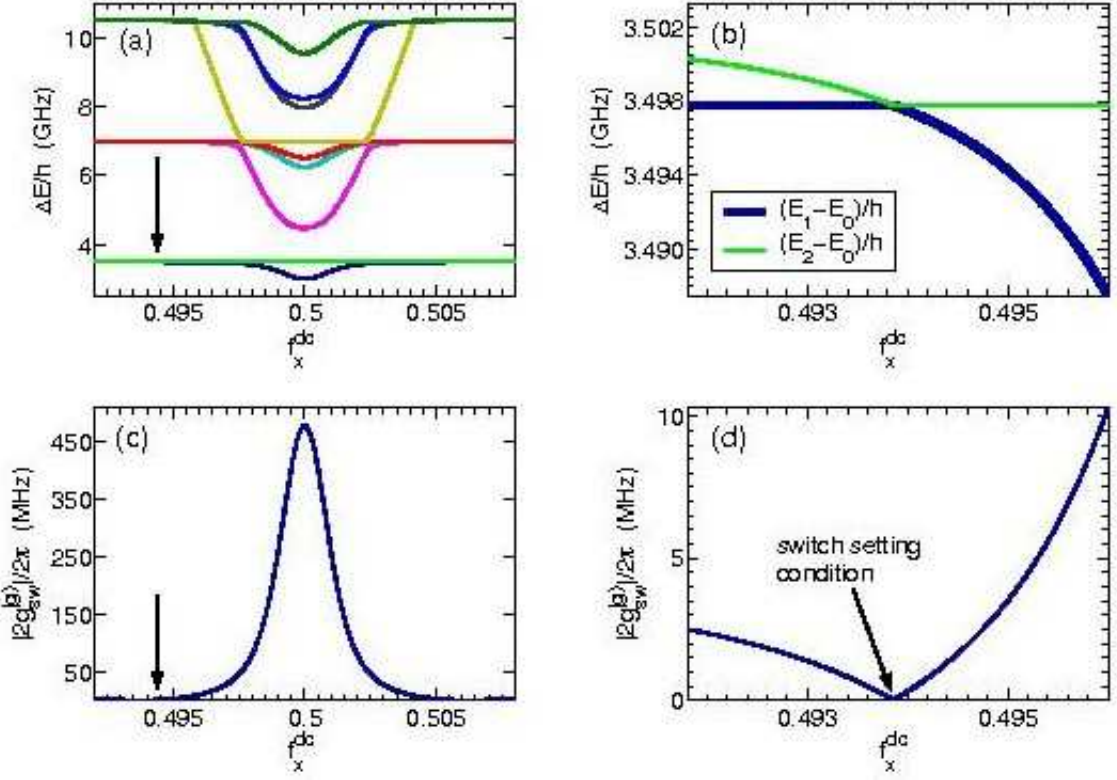}}
\caption{(Color online) Simulation of the Hamiltonian of
Eq.~(\ref{H:op:prime}) in the dispersive regime
(cf.~Subsec.~\ref{subsection:balancing:the:geometric:and:dynamic:coupling}
for a detailed description of the system parameters). (a)~The
differences between the first nine excited energy levels of the
quantum switch Hamiltonian and the groundstate energy level,
$\Delta E$, as a function of the frustration $f^{\rm DC}_{\rm x}
{} \equiv {} \Phi^{\rm DC}_{\rm x} / \Phi^{}_0$. The two lowest
lines [blue (dark grey) and green (light grey), respectively] are
associated with resonators A and B. The dispersive action of the
qubit, which modifies the shape of the resonator lines, is clearly
noticeable in the vicinity of the qubit degeneracy point. In this
region, the third energy difference (hyperbolic shape, red line)
represents the modified transition frequency of the qubit.
(b)~Close-up of the area indicated by the black arrow in (a).
Here, the two modified resonator lines [thick blue (dark grey) and
thin green (light grey), respectively] cross each other.
(c)~Quantum switch coupling coefficient $| 2 g^{\left| {\rm g}
\right\rangle}_{\rm sw} |$ extrapolated from the energy spectrum
of (a) plotted versus $f^{\rm DC}_{\rm x}$. (d)~Close-up of the
area indicated by the black arrow of (c). The switch setting
condition $| 2 g^{\left| {\rm g} \right\rangle}_{\rm sw} | {} = {}
0$ is reached at $f^{\rm DC}_{\rm x} {} \simeq {} 0.4938$.}
 \label{QS:Figure:3:abcd:Matteo:Mariantoni:2008}
\end{figure*}

\subsection{Balancing the geometric and dynamic coupling}
 \label{subsection:balancing:the:geometric:and:dynamic:coupling}

We now give the total Hamiltonian of the three-node quantum
network of Figs.~\ref{QS:Figure:1:abcde:Matteo:Mariantoni:2008}(a)
and \ref{QS:Figure:1:abcde:Matteo:Mariantoni:2008}(b). In order to
avoid unnecessarily cumbersome calculations, we restrict ourselves
to purely inductive interactions up to geometric second-order
corrections. In this framework, the most suitable quantum circuit
to be used is a flux qubit. Hereafter, all specific parameters and
corresponding simulations refer to this particular case.
Nevertheless, the formalism which we develop remains general and
can be extended to purely capacitive interactions (charge qubits)
straightforwardly.

The flux qubit is assumed to be positioned at a current antinode.
As a consequence, the vacuum fluctuations $i^{}_{{\rm A} 0}$ and
$i^{}_{{\rm B} 0}$ have maximum values $i^{\max}_{{\rm A} 0}$ and
$i^{\max}_{{\rm B} 0}$ at the qubit position and we can impose
$v^{}_{{\rm A} 0} {} = {} v^{}_{{\rm B} 0} {} = {} 0$. Also, in
the standard operation of a flux qubit no DC voltages are applied,
i.e., $v^{}_{\rm DC} {} = {} 0$, and the quasi-static flux bias is
usually controlled by an external coil and not by the cavities
(cf.~Subsec.~\ref{subsection:the:hamiltonian:of:a:generic:three:node:network}).
Again, we impose $i^{}_{\rm DC} {} = {} 0$ and add to the
Hamiltonian of Eq.~(\ref{H:op:T}) the term $( \Phi^{\rm DC}_{\rm
x} - \Phi^{}_0 / 2 ) \hat{I}^{}_{\rm Q}$. Under all these
assumptions and substituting $\mathbf{M}^{\left( n \right)}_{}$ of
Eq.~(\ref{H:op:T}) by $\mathbf{M}^{\left( 2 \right)}_{}$ of
matrix~(\ref{M:(2):matrix}), we readily obtain
\begin{eqnarray}
\widehat{H}^{\prime}_{} & {} = {} & \frac{1}{2} \hbar \epsilon
\hat{\bar{\sigma}}^{}_z + \frac{1}{2} \hbar \delta^{}_{\rm Q}
\hat{\bar{\sigma}}^{}_x + \hbar \omega^{}_{\rm A}
\hat{a}^{\dag}_{} \hat{a}^{}_{} + \hbar \omega^{}_{\rm B}
\hat{b}^{\dag}_{} \hat{b}^{}_{} \nonumber\\[1.5mm]
& & {} + \hbar g^{}_{\rm A} \hat{\bar{\sigma}}^{}_z (
\hat{a}^{\dag}_{} + \hat{a}^{}_{} ) + \hbar g^{}_{\rm B}
\hat{\bar{\sigma}}^{}_z ( \hat{b}^{\dag}_{} + \hat{b}^{}_{} )
\nonumber\\[1.5mm]
& & {} + \hbar g^{}_{\rm A B} ( \hat{a}^{\dag}_{} + \hat{a}^{}_{}
) ( \hat{b}^{\dag}_{} + \hat{b}^{}_{} ) \, .
 \label{H:op:prime}
\end{eqnarray}
Here, all global energy offsets have been neglected and we have
included both first- and second-order circuit theory
contributions. In this equation, $\hbar \epsilon {} \equiv {} 2
i^{}_{\rm Q} ( \Phi^{\rm DC}_{\rm x} - \Phi^{}_0 / 2 )$ is the
qubit energy bias, $\delta^{}_{\rm Q} {} \equiv {} \delta^{}_{\rm
Q} \left( E^{}_{\rm c} , E^{}_{\rm J} \right)$ is the qubit
gap,~\cite{YuriyMakhlin:GerdSchoen:RevModPhys:2001:a,
TPOrlando:JEMooij:PhysRevB:1999:a} $\omega^{}_{\rm A} {} \equiv {}
1 / \sqrt{M^{}_{\rm A A} C^{}_{\rm A A}}$ and $\omega^{}_{\rm B}
{} \equiv {} 1 / \sqrt{M^{}_{\rm B B} C^{}_{\rm B B}}$ are the
angular frequencies of resonators A and B, respectively,
$g^{}_{\rm A} {} \equiv {} i^{}_{\rm Q} i^{}_{{\rm A} 0} M^{}_{\rm
A Q} / \hbar$ and $g^{}_{\rm B} {} \equiv {} i^{}_{\rm Q}
i^{}_{{\rm B} 0} M^{}_{\rm B Q} / \hbar$ are the qubit-resonator
coupling coefficients, and, finally, the second-order coupling
coefficient $g^{}_{\rm A B} {} \equiv {} i^{}_{{\rm A} 0}
i^{}_{{\rm B} 0} \left( m + M^{}_{\rm A Q} M^{}_{\rm Q B} /
M^{}_{\rm Q Q} \right) / \hbar$. In general, $g^{}_{\rm A}$ and
$g^{}_{\rm B}$ can be different due to parameter spread during the
sample fabrication. Later, we show that the architecture proposed
here is robust with respect to such imperfections. We now rotate
the system Hamiltonian of Eq.~(\ref{H:op:prime}) into the qubit
energy eigenbasis $\left\{ \left| {\rm g} \right\rangle , \left|
{\rm e} \right\rangle \right\}$ obtaining
\begin{eqnarray}
\widehat{H}^{}_{} & {} = {} & \hbar \frac{\Omega^{}_{\rm Q}}{2}
\hat{\sigma}^{}_z + \hbar \omega^{}_{\rm A} \hat{a}^{\dag}_{}
\hat{a}^{}_{} + \hbar \omega^{}_{\rm B} \hat{b}^{\dag}_{} \hat{b}^{}_{}
\nonumber\\[1.5mm]
& & {} + \hbar g^{}_{\rm A} \cos \theta \hat{\sigma}^{}_z (
\hat{a}^{\dag}_{} + \hat{a}^{}_{} ) + \hbar g^{}_{\rm B} \cos
\theta \hat{\sigma}^{}_z ( \hat{b}^{\dag}_{} + \hat{b}^{}_{} )
\nonumber\\[1.5mm]
& & {} - \hbar g^{}_{\rm A} \sin \theta \hat{\sigma}^{}_x (
\hat{a}^{\dag}_{} + \hat{a}^{}_{} ) - \hbar g^{}_{\rm B}
\cos \theta \hat{\sigma}^{}_x ( \hat{b}^{\dag}_{} + \hat{b}^{}_{} )
\nonumber\\[1.5mm]
& & {} + \hbar g^{}_{\rm A B} ( \hat{a}^{\dag}_{} + \hat{a}^{}_{}
) ( \hat{b}^{\dag}_{} + \hat{b}^{}_{} ) \, .
 \label{H:op}
\end{eqnarray}
Here, $\Omega^{}_{\rm Q} {} = {} \sqrt{\epsilon^2_{} +
\delta^2_{\rm Q}}$ is the $\Phi^{\rm DC}_{\rm x}$-dependent
transition frequency of the qubit and $\theta {} = {} \arctan
\left( \delta^{}_{\rm Q} / \epsilon \right)$ is the usual mixing
angle. Hereafter, we use the redefined Pauli operators
$\hat{\sigma}^{}_x$ and $\hat{\sigma}^{}_z$, where
$\hat{\sigma}^{}_x {} = {} \hat{\sigma}^+_{} + \hat{\sigma}^-_{}$
and $\hat{\sigma}^+_{}$ and $\hat{\sigma}^-_{}$ are the lowering
and raising operators between the qubit energy groundstate $\left|
{\rm g} \right\rangle$ and excited state $\left| {\rm e}
\right\rangle$, respectively. Expressing $\widehat{H}^{}_{}$ in an
interaction picture with respect to the qubit and both resonators,
$\hat{a}^{\dag}_{} {} \rightarrow {} \hat{a}^{\dag}_{} \exp \left(
+ j \omega^{}_{\rm A} t \right)$, $\hat{a}^{}_{} {} \rightarrow {}
\hat{a}^{}_{} \exp \left( - j \omega^{}_{\rm A} t \right)$,
$\hat{b}^{\dag}_{} {} \rightarrow {} \hat{b}^{\dag}_{} \exp \left(
+ j \omega^{}_{\rm B} t \right)$, $\hat{b}^{}_{} {} \rightarrow {}
\hat{b}^{}_{} \exp \left( - j \omega^{}_{\rm B} t \right)$,
$\hat{\sigma}^{\mp}_{} {} \rightarrow {} \hat{\sigma}^{\mp}_{}
\exp \left( \mp j \Omega^{}_{\rm Q} t \right)$, assuming
$\omega^{}_{\rm A} {} = {} \omega^{}_{\rm B} \equiv \omega \equiv
2 \pi f$, and performing a RWA yields
\begin{eqnarray}
\widehat{\widetilde{H}} & {} \! = \! {} & \hbar \sin \theta \!\!
\left[ \hat{\sigma}^-_{} \!\! \left( g^{}_{\rm A}
\hat{a}^{\dag}_{} \! + g^{}_{\rm B} \hat{b}^{\dag}_{} \right) \!\!
e^{- j \Delta t}_{} \! + \hat{\sigma}^+_{} \!\! \left( g^{}_{\rm
A} \hat{a}^{}_{} \! + g^{}_{\rm B} \hat{b}^{}_{} \right) \!\! e^{j
\Delta t}_{} \right]
\nonumber\\[1.5mm]
& & {} + \hbar g^{}_{\rm A B} \left( \hat{a}^{\dag}_{}
\hat{b}^{}_{} \! + \hat{a}^{}_{} \hat{b}^{\dag}_{} \right) \, .
 \label{H:op:ind:secord:int:pic}
\end{eqnarray}
Here, $\Delta {} \equiv {} \Omega^{}_{\rm Q} - \omega$ is the
qubit-resonator detuning. The first two terms of
Eq.~(\ref{H:op:ind:secord:int:pic}) represent a standard two-mode
JC dynamics.~\cite{ARauschenbeutel:SHaroche:PhysRevARap:2001:a,
CWildfeuer:DHSchiller:PhysRevA:2003:a} The last term, instead,
constitutes a beam-splitter-type interaction specific to
two-resonator circuit~QED. This interaction is not present in the
quantum optical
version.~\cite{ARauschenbeutel:SHaroche:PhysRevARap:2001:a,
CWildfeuer:DHSchiller:PhysRevA:2003:a} The coupling coefficient
$g^{}_{\rm A B}$ is typically much smaller than $g^{}_{\rm A}$ and
$g^{}_{\rm B}$ (see below). However, in the dispersive regime
($\left| \Delta \right| \gg \max \left\{ g^{}_{\rm A} , g^{}_{\rm
B} , g^{}_{\rm A B} \right\}$), $g^{}_{\rm A B}$ becomes
comparable to all other dispersive coupling strengths. To gain
further insight into this matter, we can define two superoperators
$\hat\Xi^{\dag}_{} {} \equiv {} \hat{\sigma}^+_{} \left( g^{}_{\rm
A} \hat{a}^{}_{} + g^{}_{\rm B} \hat{b}^{}_{} \right)$ and
$\hat\Xi^{}_{} {} \equiv {} \hat{\sigma}^-_{} \left( g^{}_{\rm A}
\hat{a}^{\dag}_{} + g^{}_{\rm B} \hat{b}^{\dag}_{} \right)$. It
can be shown that the Dyson series for the evolution operator
associated with the time-dependent Hamiltonian of
Eq.~(\ref{H:op:ind:secord:int:pic}) can be rewritten in the
exponential form $\widehat{U} {} = {} \exp \left( - j
\widehat{\widetilde{H}}^{}_{\rm eff} t / \hbar \right)$, where
$\widehat{\widetilde{H}}^{}_{\rm eff} {} = {} \hbar \left[
\hat{\Xi}^{\dag}_{} , \hat{\Xi}^{}_{} \right] / \Delta + \hbar
g^{}_{\rm A B} \left( \hat{a}^{\dag}_{} \hat{b}^{}_{} +
\hat{a}^{}_{} \hat{b}^{\dag}_{} \right)$. Thus
\begin{eqnarray}
\!\!\!\!\!\!\!\! \widehat{\widetilde{H}}^{}_{\rm eff} & {} = {} &
\hbar \frac{\left( g^{}_{\rm A} \sin \theta \right)^2_{}}{\Delta}
\hat{\sigma}^{}_z \left( \hat{a}^{\dag}_{} \hat{a}^{}_{} + \frac{1}{2} \right) \nonumber\\[1.5mm]
& & {} + \hbar \frac{\left( g^{}_{\rm B} \sin \theta
\right)^2_{}}{\Delta} \hat{\sigma}^{}_z \left( \hat{b}^{\dag}_{}
\hat{b}^{}_{} + \frac{1}{2} \right) \nonumber\\[1.5mm]
& & {} + \hbar \left( \frac{g^{}_{\rm A} g^{}_{\rm B} \sin^2_{}
\theta}{\Delta} \hat{\sigma}^{}_z + g^{}_{\rm A B} \right) \left(
\hat{a}^{\dag}_{} \hat{b}^{}_{} + \hat{a}^{}_{} \hat{b}^{\dag}_{}
\right) \, .
 \label{H:tilde:op:eff}
\end{eqnarray}
In this Hamiltonian, the first two terms represent dynamic
(AC-Zeeman) shifts (AC-Stark shifts in the case of charge qubits)
of the transition angular frequency of resonators A and B,
respectively. If $g^{}_{\rm A} {} = {} g^{}_{\rm B} {} \equiv {}
g$ and we only use eigenstates of $\hat{\sigma}^{}_z$, the first
two terms of Eq.~(\ref{H:tilde:op:eff}) equally renormalize
$\omega^{}_{\rm A}$ and $\omega^{}_{\rm B}$, respectively. The
Hamiltonian of Eq.~(\ref{H:tilde:op:eff}) can be further
simplified through an additional unitary transformation described
by $\widehat{U}^{}_0 {} = {} \exp ( j \widehat{H}^{}_0 t / \hbar
)$, where $\widehat{H}^{}_0 {} \equiv {} \hbar ( g^2_{\rm A}
\sin^2_{} \theta / \Delta ) \hat{\sigma}^{}_z ( \hat{a}^{\dag}_{}
\hat{a}^{}_{} + 1 / 2 ) + \hbar ( g^2_{\rm B} \sin^2_{} \theta /
\Delta ) \hat{\sigma}^{}_z ( \hat{b}^{\dag}_{} \hat{b}^{}_{} + 1 /
2 )$. When $g^{}_{\rm A} {} = {} g^{}_{\rm B} {} \equiv {} g$,
this transformation yields the final effective Hamiltonian
\begin{equation}
\widehat{H}^{}_{\rm eff} {} = {} \hbar \left( \frac{g^{2}
\sin^2_{} \theta}{\Delta} \hat{\sigma}^{}_z + g^{}_{\rm A B}
\right) \left( \hat{a}^{\dag}_{} \hat{b}^{}_{} + \hat{a}^{}_{}
\hat{b}^{\dag}_{} \right) \, ,
 \label{H:op:eff}
\end{equation}
which constitutes the \textit{second main result} of this work.
This Hamiltonian is the key ingredient for the implementation of a
quantum switch between the two resonators. In fact, it clearly
represents a tunable interaction between A and B characterized by
an effective coupling coefficient
\begin{equation}
\begin{array}{r c l}
g^{\left| {\rm g} \right\rangle}_{\rm sw} & {} \equiv {} &
g^{}_{\rm A B} - \dfrac{g^2_{} \sin^2_{} \theta}{\Delta} \\[2mm]
g^{\left| {\rm e} \right\rangle}_{\rm sw} & {} \equiv {} &
g^{}_{\rm A B} + \dfrac{g^2_{} \sin^2_{} \theta}{\Delta}
\end{array} \quad ,
 \label{g:ge:sw}
\end{equation}
for $\left| {\rm g} \right\rangle$ and $\left| {\rm e}
\right\rangle$, respectively. The switching of such an interaction
triggers, or prevents, the exchange of quantum information between
A and B. On the one hand, the first part of this interaction is a
purely geometric coupling, which is constant and qubit-state
independent. On the other hand, the second part is a dynamic
coupling, which depends on the state of the qubit. The
\textit{switch setting condition}
\begin{equation}
\frac{g^2_{} \sin^2_{} \theta}{\left| \Delta \right|} {} = {}
\left| g^{}_{\rm A B} \right|
 \label{sw:sett:cond}
\end{equation}
can easily be fulfilled varying $\Delta$, changing $\sin \theta$,
or inducing AC-Zeeman or -Stark
shifts.~\cite{JMajer:JMChow:RJSchoelkopf:NatureLett:2007:a} In the
special case of a charge qubit, not treated here in detail, this
task can also be accomplished modifying the qubit transition
frequency $\Omega^{}_{\rm Q}$ via a suitable quasi-static magnetic
field.~\cite{AWallraff:RJSchoelkopf:NatureLett:2004:a} This allows
one to keep the qubit at the degeneracy point. Here, we focus on
the first option, i.e., finding a suitable qubit bias for which
the detuning $\Delta$ fulfills the relation of
Eq.~(\ref{sw:sett:cond}). For a flux qubit, this can be realized
polarizing the qubit by means of an external flux.

To better understand the switch setting condition, we numerically
diagonalize the entire system Hamiltonian of
Eq.~(\ref{H:op:prime}), without performing any approximation. The
results are presented in
Fig.~\ref{QS:Figure:3:abcd:Matteo:Mariantoni:2008}, which shows
the energy spectrum of the quantum switch Hamiltonian and the
effective coupling coefficient $| 2 g^{\left| {\rm g}
\right\rangle}_{\rm sw} |$ for a flux qubit with $i^{}_{\rm Q} {}
= {} 370$\,nA, $\delta^{}_{\rm Q} / 2 \pi {} = {} 4$\,GHz, $f {} =
{} 3.5$\,GHz, $g^{}_{} / 2 \pi {} = {} 472$\,MHz, and $g^{}_{\rm A
B} / 2 \pi {} = {} 2.2$\,MHz. The parameters for the flux qubit
are chosen from our previous experimental
works,~\cite{FDeppe:RGross:PhysRevB:2007:a,
FrankDeppe:MatteoMariantoni:RGross:NaturePhysLett:2008:a} whereas
the three coupling coefficients are the result of detailed
simulations
(cf.~Sec.~\ref{section:an:example:of:two:resonator:circuit:qed:with:a:flux:qubit}).
It is noteworthy to mention that large vacuum Rabi couplings
$g^{}_{} / 2 \pi$ on the order of $500$\,MHz have already been
achieved both for flux and charge
qubits.~\cite{RHKoch:DPDiVincenzo:PhysRevLett:2006:a,
RJSchoelkopf:SMGirvin:NatureHorizons:2008:a} We have chosen the
qubit to be already detuned from both resonators by $0.5$\,GHz
when biased at the flux degeneracy point. Moving sufficiently far
from the degeneracy point enables us to increase the
qubit-resonator detuning such that the system can be modeled by
the Hamiltonian of Eq.~(\ref{H:op:eff}).
\begin{figure*}[t!]
\centering{%
 \includegraphics[width=0.85\textwidth]{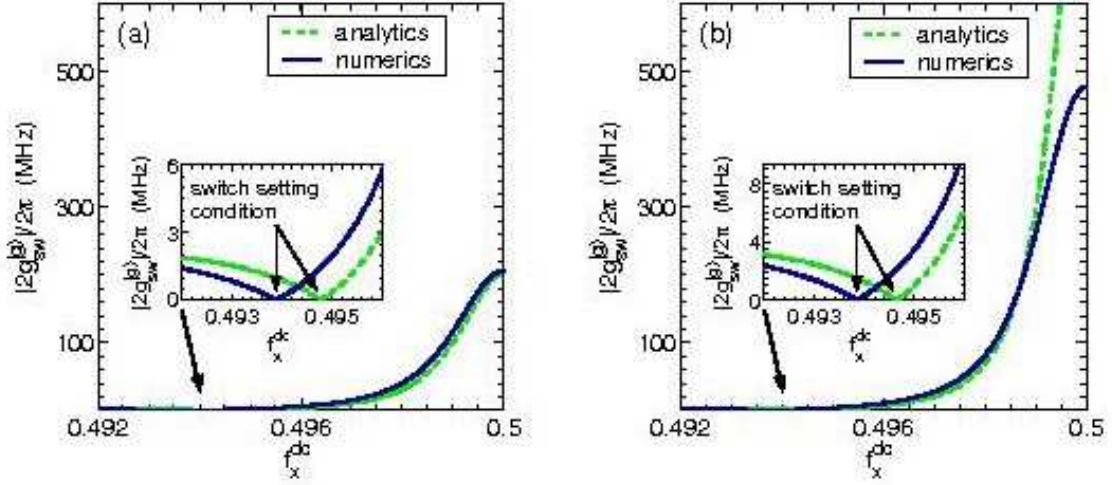}}
\caption{(Color online) Comparison between the $f^{\rm DC}_{\rm
x}$-dependence of the analytical expression for the coupling
coefficient $| 2 g^{\left| {\rm g} \right\rangle}_{\rm sw} |$
obtained from Eq.~(\ref{g:ge:sw}) and the one found by numerically
diagonalizing the Hamiltonian of Eq.~(\ref{H:op:prime}). (a)~We
choose a center frequency $f^{}_{\rm A} {} = {} f^{}_{\rm B} {} =
{} f = 2.7$\,GHz for the two resonators. All the other parameters
are the same as those used to obtain the results of
Fig.~\ref{QS:Figure:3:abcd:Matteo:Mariantoni:2008}. The analytical
[dashed green (light grey) line] and the numerical [solid blue
(dark grey) line] results are in excellent agreement. In the large
detuning limit far away from the qubit degeneracy point, $| 2
g^{\left| {\rm g} \right\rangle}_{\rm sw} |$ saturates to the
value $| 2 g^{}_{\rm A B} | {} \simeq {} 2.6$\,MHz. Inset:
Close-up of the region near the switch setting condition.
(b)~Here, we choose a center frequency $f^{}_{\rm A} {} = {}
f^{}_{\rm B} {} = {} f {} = {} 3.5$\,GHz for the two resonators.
The analytical [dashed green (light grey) line] and numerical
[solid blue (dark grey) line] results are in good agreement away
from the qubit degeneracy point. Closer to it they diverge
(cf.~Subsec.~\ref{subsection:balancing:the:geometric:and:dynamic:coupling}
for more details). In the large detuning limit far away from the
qubit degeneracy point, $| 2 g^{\left| {\rm g} \right\rangle}_{\rm
sw} |$ saturates to the value $| 2 g^{}_{\rm A B} | {} \simeq {}
4.4$\,MHz. Inset: Close-up of the region near the switch setting
condition.}
 \label{QS:Figure:4:ab:insab:Matteo:Mariantoni:2008}
\end{figure*}
Figure~\ref{QS:Figure:3:abcd:Matteo:Mariantoni:2008}(a) shows the
differences between the first nine excited energy levels of the
quantum switch Hamiltonian and the groundstate energy level,
$\Delta E^{}_i {} \equiv {} E^{}_i - E^{}_0$ with $i {} = {}
\left\{ 1 , \dots , 10 \right\}$, as a function of the frustration
$f^{}_{\rm x} {} \equiv {} \Phi^{\rm DC}_{\rm x} / \Phi^{}_0$.
Here, $E^{}_i$ is the energy level of the $i$-th excited state and
$E^{}_0$ that of the groundstate. Due to the qubit-resonator
detuning, the two lowest energy differences [blue (dark grey) and
green (light grey) lines, respectively] correspond to the modified
transition frequencies of the two resonators. Owing to the
interaction with the qubit these lines are not flat. This effect
becomes particularly evident in the region close to the qubit
degeneracy point, where dispersivity is reduced. In this region,
the third energy difference (hyperbolic shape, red line)
represents the modified transition frequency of the qubit. When
moving away from the qubit degeneracy point, a crossing between
the modified resonator lines is encountered, as clearly shown in
Fig.~\ref{QS:Figure:3:abcd:Matteo:Mariantoni:2008}(b) [see, thick
blue (dark grey) and thin green (light grey) lines]. This crossing
represents the switch setting condition of
Eq.~(\ref{sw:sett:cond}).
Figures~\ref{QS:Figure:3:abcd:Matteo:Mariantoni:2008}(c) and
\ref{QS:Figure:3:abcd:Matteo:Mariantoni:2008}(d) show the absolute
value of the flux-dependent coupling coefficient $| 2 g^{\left|
{\rm g} \right\rangle}_{\rm sw} |$ in the flux windows of
Figs.~\ref{QS:Figure:3:abcd:Matteo:Mariantoni:2008}(a) and
\ref{QS:Figure:3:abcd:Matteo:Mariantoni:2008}(b), respectively.
The switch setting condition $| 2 g^{\left| {\rm g}
\right\rangle}_{\rm sw} | {} = {} 0$ is reached at $f^{\rm
DC}_{\rm x} {} \simeq {} 0.4938$.

A comparison between the analytic expression of
Eq.~(\ref{g:ge:sw}) with the qubit in $\left| {\rm g}
\right\rangle$ [dashed green (light grey) lines] and a numerical
estimate of the effective coupling coefficient $| 2 g^{\left| {\rm
g} \right\rangle}_{\rm sw} |$ [solid blue (dark grey) lines] is
shown in Figure~\ref{QS:Figure:4:ab:insab:Matteo:Mariantoni:2008}.
To clarify similarities and differences between analytical and
numerical calculations, we choose two different sets of
parameters. In
Fig.~\ref{QS:Figure:4:ab:insab:Matteo:Mariantoni:2008}(a), the
center frequencies of the two resonators are set to be $f^{}_{\rm
A} {} = {} f^{}_{\rm B} {} = {} f {} = {} 2.7$\,GHz, whereas in
Fig.~\ref{QS:Figure:4:ab:insab:Matteo:Mariantoni:2008}(b) we
choose $f^{}_{\rm A} {} = {} f^{}_{\rm B} {} = {} f {} = {}
3.5$\,GHz. All the other parameters are equal to those used to
obtain the results of
Fig.~\ref{QS:Figure:3:abcd:Matteo:Mariantoni:2008}. In
Fig.~\ref{QS:Figure:4:ab:insab:Matteo:Mariantoni:2008}(a),
analytics and numerics agree over the entire frustration window.
The inset shows that the switch setting condition obtained from
the numerical simulation is only slightly shifted with respect to
the analytical prediction. Also in
Fig.~\ref{QS:Figure:4:ab:insab:Matteo:Mariantoni:2008}(b), the
agreement between analytical and numerical estimates is good far
away from the qubit degeneracy point. However, closer to it the
qubit and the two resonators are not detuned enough to guarantee
dispersivity. Therefore, analytics and numerics start to deviate,
as expected. Again, the inset shows that the switch setting
condition can be fulfilled. It is noteworthy to point out that
both analytical and numerical estimates converge to the value $| 2
g^{}_{\rm A B} |$ in the limit of large detuning. We find $| 2
g^{}_{\rm A B} | / 2 \pi {} \simeq {} 2.6$\,MHz and $| 2 g^{}_{\rm
A B} | / 2 \pi {} \simeq {} 4.4$\,MHz from the simulations that
produce Figs.~\ref{QS:Figure:4:ab:insab:Matteo:Mariantoni:2008}(a)
and \ref{QS:Figure:4:ab:insab:Matteo:Mariantoni:2008}(b),
respectively.

\begin{figure}[t!]
\centering{%
 \includegraphics[width=0.90\columnwidth,clip=]{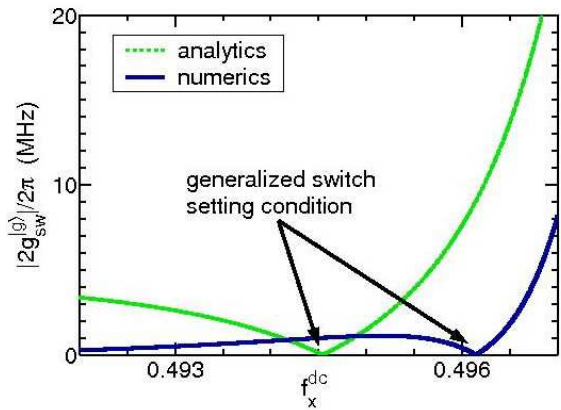}}
\caption{(Color online) Robustness of quantum switch to
fabrication imperfections. Solid blue (dark grey) line: Numerical
simulation of the quantum switch coupling coefficient $| 2
g^{\left| {\rm g} \right\rangle}_{\rm sw} |$ as a function of the
frustration $f^{\rm DC}_{\rm x}$. Here, we assume a relatively
large spread of $1 \%$ for the resonators center
frequencies~\cite{ThomasNiemczyk:PrivateComm:2008:a} and a
difference of approximately $15 \%$ between $g^{}_{\rm A}$ and
$g^{}_{\rm B}$. Dashed green (light grey) line: Plot of $| 2
g^{\left| {\rm g} \right\rangle}_{\rm sw} |$ extracted from the
generalized switch setting condition of
Eq.~(\ref{gen:sw:sett:cond}) for the same parameter spread as in
the numerical simulations. For both the analytical and numerical
result the switch setting condition is fulfilled (see black
arrows).}
 \label{QS:Figure:5:Matteo:Mariantoni:2008}
\end{figure}

Finally, we demonstrate that the quantum switch Hamiltonian is
robust to parameter spread due to fabrication inaccuracies.
Typically, for a center frequency of $5$\,GHz the expected spread
around this value is
approximately~\cite{ThomasNiemczyk:PrivateComm:2008:a} $\mp
10$\,MHz for two resonators fabricated on the same chip. Also, the
coupling coefficients $g^{}_{\rm A}$ and $g^{}_{\rm B}$ can differ
slightly. In this case, a generalized effective Hamiltonian for
the quantum switch can be
derived~\cite{MatteoMariantoni:FrankDeppe:unpublished:2008:a}
 \newlength{\myplus}
 \settowidth\myplus{${}+{}$}
\begin{eqnarray}
\widehat{H}^{\rm gen}_{\rm eff} & {} = {} & \hbar \frac{\left(
g^{}_{\rm A} \sin \theta \right)^2_{}}{\Delta^{}_{\rm A}}
\hat{\sigma}^{}_z \hat{a}^{\dag}_{} \hat{a}^{}_{} + \hbar
\frac{\left( g^{}_{\rm B} \sin \theta \right)^2_{}}{\Delta^{}_{\rm
B}} \hat{\sigma}^{}_z \hat{b}^{\dag}_{} \hat{b}^{}_{}
\nonumber\\[1.5mm]
& & {} + \hbar \left[ \frac{g^{}_{\rm A} g^{}_{\rm B} \sin^2_{}
\theta}{2} \left( \frac{1}{\Delta^{}_{\rm A}} +
\frac{1}{\Delta^{}_{\rm B}} \right) \hat{\sigma}^{}_z + g^{}_{\rm A B} \right]
\nonumber\\[1.5mm]
& & {} \hspace{\myplus} \times \left( \hat{a}^{\dag}_{}
\hat{b}^{}_{} e^{+ j \delta^{}_{\rm A B} t}_{} + \hat{a}^{}_{}
\hat{b}^{\dag}_{} e^{- j \delta^{}_{\rm A B} t}_{} \right) \, ,
 \label{H:op:gen:eff}
\end{eqnarray}
where $\Delta^{}_{\rm A} {} \equiv {} \Omega^{}_{\rm Q} -
\omega^{}_{\rm A}$, $\Delta^{}_{\rm B} {} \equiv {} \Omega^{}_{\rm
Q} - \omega^{}_{\rm B}$, and $\delta^{}_{\rm A B} {} \equiv {}
\omega^{}_{\rm A} - \omega^{}_{\rm B}$. From
Eq.~(\ref{H:op:gen:eff}), we can deduce the generalized coupling
coefficient of the switch, $g^{\left| {\rm g} \right\rangle ,
\left| {\rm e} \right\rangle}_{\rm sw} {} \equiv {} g^{}_{\rm A B}
\mp g^{}_{\rm A} g^{}_{\rm B} \sin^2_{} \theta \left( 1 / 2
\Delta^{}_{\rm A} + 1 / 2 \Delta^{}_{\rm B} \right)$ for the qubit
groundstate $\left| {\rm g} \right\rangle$ or excited state
$\left| {\rm e} \right\rangle$, respectively. As a consequence,
the \textit{generalized switch setting condition} becomes
\begin{equation}
\left| \frac{g^{}_{\rm A} g^{}_{\rm B} \sin^2_{} \theta}{2} \left(
\frac{1}{\Delta^{}_{\rm A}} + \frac{1}{\Delta^{}_{\rm B}} \right)
\right| {} = {} \left| g^{}_{\rm A B} \right| \, .
 \label{gen:sw:sett:cond}
\end{equation}
This condition is displayed in
Fig.~\ref{QS:Figure:5:Matteo:Mariantoni:2008} [dashed green (light
grey) line] as a function of the external flux. Here, we assume
two resonators with center frequencies $f^{}_{\rm A} {} = {}
3.5$\,GHz and $f^{}_{\rm B} {} = {} 3.5\,{\rm GHz} + 35\,{\rm
MHz}$. This corresponds to a relatively large center frequency
spread~\cite{ThomasNiemczyk:PrivateComm:2008:a} of $1 \%$. In
addition, we choose the two coupling coefficients $g^{}_{\rm A} /
2 \pi {} = {} 472$\,MHz and $g^{}_{\rm B} / 2 \pi {} = {}
549$\,MHz to differ by approximately $15 \%$. It is remarkable
that, also in this more general case, the switch setting condition
can be fulfilled easily. We confirm this result by means of
numerical simulations [solid blue (dark grey) line in
Fig.~\ref{QS:Figure:5:Matteo:Mariantoni:2008}] of the full
Hamiltonian of Eq.~(\ref{H:op:prime}), assuming fabrication
imperfections. Interestingly, in contrast to the case where
$\Delta^{}_{\rm A} {} = {} \Delta^{}_{\rm B}$ and $g^{}_{\rm A} {}
= {} g^{}_{\rm B}$, we observe a different behavior of the
analytical and numerical curves of
Fig.~\ref{QS:Figure:5:Matteo:Mariantoni:2008} when moving far away
from the qubit degeneracy point. The reasons behind this fact rely
on the conditions used to obtain the second-order Hamiltonian of
Eq.~(\ref{H:op:gen:eff}). If $\delta^{}_{\rm A B} {} \gtrsim {}
\max \left\{ g^{}_{\rm A} g^{}_{\rm B} \sin^2_{} \theta / 2
\Delta^{}_{\rm A} , g^{}_{\rm A} g^{}_{\rm B} \sin^2_{} \theta / 2
\Delta^{}_{\rm B} , g^{}_{\rm A B} \right\}$, as for the
parameters chosen here, this Hamiltonian does not represent an
accurate approximation anymore. In this case, as expected, only a
partial agreement between analytics and numerics is found.
Nevertheless, a clear switch setting condition is obtained in both
cases. We notice that the numerical switch setting condition is
shifted towards the degeneracy point with respect to the
analytical solution. This is due to the detuning $\delta^{}_{\rm A
B}$ present in Eq.~(\ref{H:op:gen:eff}), which is not accounted
for when plotting the analytical solution.

All the above considerations clearly show that the requirements on
the sample fabrication are substantially relaxed.

\subsection{A quantum switch protocol}
 \label{subsection:a:quantum:switch:protocol}

We now propose a possible switching protocol based on three steps
and discuss two different variants to shift from the zero-coupling
to a finite-coupling condition characterized by a coupling
coefficient $g^{\rm on}_{\rm sw}$. It is important to stress that
this protocol is independent of the specific implementation
(capacitive or inductive) of the switch. For definiteness, we
choose a quantum switch based on a flux qubit in the following.

\textit{(i)} - First, we initialize the qubit in the
groundstate~$\left| {\rm g} \right\rangle$.

\textit{(ii)} - Second, in order to fulfill the switch setting
condition, we choose the appropriate detuning $\Delta$ by changing
the quasi-static bias of the qubit. For the switch operation to be
practical, we assume $\Delta {} = {} \Delta^{}_1 > 0$. In this
way, the sign of the coefficient in front of the
$\hat{\sigma}^{}_z$-operator of Eq.~(\ref{H:op:eff}) remains
positive and, as a consequence, the switch is off in the
groundstate $\left| {\rm g} \right\rangle$, i.e., $g^{\left| {\rm
g} \right\rangle}_{\rm sw} = 0$.

\textit{(iii)} - Third, the state of the quantum switch can now be
changed from off to on in two different ways, \mbox{\textit{(a)}
or \textit{(b)}}.

\textit{(a) - Population-inversion}. The qubit is maintained at
the bias point preset in \textit{(ii)}. Its population is then
inverted from $\left| {\rm g} \right\rangle$ to $\left| {\rm e}
\right\rangle$, e.g., applying a Rabi $\pi$-pulse of duration
$t^{}_{\pi}$. Such a pulse effectively changes the switch to the
on-state, $g^{\left| {\rm e} \right\rangle}_{\rm sw} {} = {} 2
g^{}_{\rm A B}$. In this case, $g^{\rm on}_{\rm sw} {} = {} 2
g^{}_{\rm A B}$. Under these conditions, the two resonators are
effectively coupled and the A-to-B transfer time is $t {} = {} \pi
/ 2 g^{\rm on}_{\rm sw}$, which also constitutes the required
time-scale for most of the operations to be discussed in
Subsec.~\ref{subsection:advanced:applications:nonclassical:states:and:entanglement}.

\textit{(b) - Adiabatic-shift pulse}. We opportunely change the
quasi-static bias of the qubit by applying an adiabatic-shift
pulse.~\cite{FDeppe:RGross:PhysRevB:2007:a} In this way, the qubit
transition frequency becomes effectively modified. As a
consequence, the detuning $\Delta$ is changed from $\Delta^{}_1$
to $\Delta^{}_2$ such that $g^{\left| {\rm g} \right\rangle}_{\rm
sw} {} = {} \widetilde{g}^{}_{\rm sw} {} = {} g^{}_{\rm A B} -
g^2_{} \sin^2_{} \theta / \Delta^{}_2 {} \neq 0$. In other words,
the geometric and dynamic coupling coefficients are not balanced
against each other anymore and the switch is set to the on-state.
In this case, $g^{\rm on}_{\rm sw} {} = {} \widetilde{g}^{}_{\rm
sw}$. The rise time $t^{}_{\rm rise}$ of the shift pulse has to
fulfill the
condition~\cite{JJohansson:HTakayanagi:PhysRevLett:2006:a,
FDeppe:RGross:PhysRevB:2007:a} $2 \pi / g^{\rm on}_{\rm sw} {}
\gtrsim {} t^{}_{\rm rise} {} \gtrsim {} \max \left\{ 2 \pi /
\delta^{}_{\rm Q} , 2 \pi / \omega \right\}$.

Variant \textit{(b)} strongly benefits from the dependence of
$\widetilde{g}^{}_{\rm sw}$ on the external quasi-static bias of
the qubit [see
Figs.~\ref{QS:Figure:3:abcd:Matteo:Mariantoni:2008}(c) and
\ref{QS:Figure:3:abcd:Matteo:Mariantoni:2008}(d)]. We can
distinguish between two possible regimes. The first regime is for
a flux bias close to the qubit degeneracy point, where the
qubit-resonator detuning is reduced and, thus, $\Delta^{}_2 {} <
{} \Delta^{}_1$. In this case, the dynamic contribution to
$\widetilde{g}^{}_{\rm sw}$ dominates over the geometric one. This
enables us to achieve very large resonator-resonator coupling
strengths, which is a highly desirable condition to perform fast
quantum operations (e.g.,
cf.~Sec.~\ref{subsection:advanced:applications:nonclassical:states:and:entanglement}).
The second regime is for a flux bias far away from the qubit
degeneracy point, where the qubit-resonator detuning is increased
and, thus, $\Delta^{}_2 {} > {} \Delta^{}_1$. In this case, the
geometric contribution to $\widetilde{g}^{}_{\rm sw}$ dominates
over the dynamic one. Since very far away from the qubit
degeneracy point $\widetilde{g}^{}_{\rm sw}{} \rightarrow {}
\left| 2 g^{}_{\rm A B} \right|$
[cf.~Subsec.~\ref{subsection:balancing:the:geometric:and:dynamic:coupling}
and Figs.~\ref{QS:Figure:4:ab:insab:Matteo:Mariantoni:2008}(a) and
\ref{QS:Figure:4:ab:insab:Matteo:Mariantoni:2008}(b)], operating
the system in the second regime allows us to probe the pure
geometric coupling between A and B. This would constitute a direct
measurement of the geometric second-order coupling when $M^{}_{\rm
A Q} M^{}_{\rm Q B} / M^{}_{\rm Q Q} {} \gg {} m$.

\subsection{Advanced applications: Nonclassical states \\ and entanglement}
 \label{subsection:advanced:applications:nonclassical:states:and:entanglement}

We now provide a few examples showing how the quantum switch
architecture can be exploited to create nonclassical states of the
microwave radiation as well as entanglement of the resonators and
qubit degrees of freedom. In this subsection, when we discuss
about \textit{the} qubit we refer to the one used for the quantum
switch operation. If the presence of another qubit is required, we
refer to it as the auxiliary qubit.

\textit{Fock state transfer and entanglement between the
resonators.} First, we assume the quantum switch to be turned off,
e.g., following the protocol outlined in
Subsec.~\ref{subsection:a:quantum:switch:protocol} with the qubit
in the groundstate $\left| {\rm g} \right\rangle$. In addition, we
assume resonator A to be initially prepared in a Fock state
$\left| 1 \right\rangle^{}_{\rm A}$, while cavity B remains in the
vacuum state $\left| 0 \right\rangle^{}_{\rm B}$. Following the
lines of
Ref.~\onlinecite{AAHouck:DISchuster:RJSchoelkopf:NatureLett:2007:a},
for example, a Fock state $\left| 1 \right\rangle^{}_{\rm A}$ can
be created in A by means of an auxiliary qubit coupled to it. A
population inversion of the auxiliary qubit (via a $\pi$-pulse)
and its subsequent relaxation suffice to achieve this purpose.
Then, we turn on the quantum switch for a certain time $t$
following either one of the two variants \textit{(a)} or
\textit{(b)} introduced in
Subsec.~\ref{subsection:a:quantum:switch:protocol}. The initial
states are $\left| {\rm e} \right\rangle \left| 1
\right\rangle_{\rm A}^{} \left| 0 \right\rangle_{\rm B}$ and
$\left| {\rm g} \right\rangle \left| 1 \right\rangle_{\rm A}^{}
\left| 0 \right\rangle_{\rm B}$ for \textit{(a)} and \textit{(b)},
respectively. The quantum switch is now characterized by an
effective coupling $g^{\rm on}_{\rm sw}$ and the dynamics
associated with the Hamiltonian of Eq.~(\ref{H:op:eff}) is
activated for the time $t$. In this manner, a coherent linear
superposition of bipartite states containing a Fock state single
photon~\cite{Yu-xiLiu:FrancoNori:EurophysLett:2004:a,
MMariantoni:ESolano:arXiv:cond-mat:2005:a,
JJohansson:HTakayanagi:PhysRevLett:2006:a,
AAHouck:DISchuster:RJSchoelkopf:NatureLett:2007:a,
MaxHofheinz:JohnMMartinis:ANCleland:NatureLett:2008:a} can be
created
\begin{equation}
\cos \left( g^{{\rm on}}_{\rm sw} t \right) \left| 1
\right\rangle^{}_{\rm A} \left| 0 \right\rangle^{}_{\rm B} + e^{j
\pi / 2}_{} \sin \left( g^{{\rm on}}_{\rm sw} t \right) \left| 0
\right\rangle^{}_{\rm A} \left| 1 \right\rangle^{}_{\rm B} \, ,
 \label{Fock:states:coher:superpos}
\end{equation}
where the qubit state does not change and qubit and resonators
remain disentangled. If we choose to wait for a time $t {} = {}
\pi / 2 g^{{\rm on}}_{\rm sw}$, we can exploit
Eq.~(\ref{Fock:states:coher:superpos}) as a mechanism for the
transferring of a Fock state from resonator A to resonator B,
$\left| 1 \right\rangle^{}_{\rm A} \left| 0 \right\rangle^{}_{\rm
B} {} \rightarrow {} \left| 0 \right\rangle^{}_{\rm A} \left| 1
\right\rangle^{}_{\rm B}$. In this case, also the resonators
remain disentangled. It is noteworthy to mention that the
controlled transfer of a Fock state between two remote locations
constitutes the basis of several quantum information
devices.~\cite{NKiesel:HWeinfurter:PhysRevLett:2007:a} If we
choose to wait for a time $t {} = {} \pi / 4 g^{{\rm on}}_{\rm
sw}$ instead, we can achieve maximal entanglement between the two
remote resonators. This goes beyond the results obtained in atomic
systems, where two nondegenerate orthogonally polarized modes of
the same cavity have been used to create mode
entanglement.~\cite{ARauschenbeutel:SHaroche:PhysRevARap:2001:a}

\textit{Tripartite entanglement and GHZ states.} We follow a
modified version of variant \textit{(a)} of the switching
protocol. We start from the same initial conditions as in the
previous example. Resonator A is in $\left| 1
\right\rangle^{}_{\rm A}$ and resonator B is in $\left| 0
\right\rangle^{}_{\rm B}$. The qubit is in $\left| {\rm g}
\right\rangle$ and the switch setting condition is fulfilled,
i.e., the switch is off. We then apply a $\pi / 2$-pulse to the
qubit bringing it into the symmetric
superposition~\cite{AWallraff:RJSchoelkopf:PhysRevLett:2005:a}
$\left( \left| {\rm g} \right\rangle + \left| {\rm e}
\right\rangle \right) / \sqrt{2}$. Then, the state of the system
is still disentangled and can be written as
\begin{equation}
\frac{ \left| {\rm g} \right\rangle \left| 1 \right\rangle^{}_{\rm
A} \left| 0 \right\rangle^{}_{\rm B} + \left| {\rm e}
\right\rangle \left| 1 \right\rangle^{}_{\rm A} \left| 0
\right\rangle^{}_{\rm B}}{\sqrt{2}} \, .
 \label{GHZ:initial:state}
\end{equation}
Now, the Hamiltonian of Eq.~(\ref{H:op:eff}) yields the time
evolution
 \newlength{\myGHZ}
 \settowidth\myGHZ{${}\dfrac{1}{\sqrt{2}} \big({}$}
\begin{eqnarray}
{} & {} & \dfrac{1}{\sqrt{2}} \big( \left| {\rm g} \right\rangle
\left| 1 \right\rangle^{}_{\rm A} \left| 0 \right\rangle^{}_{\rm
B}+ \cos \left( g^{{\rm on}}_{\rm sw} t \right) \left| {\rm e}
\right\rangle \left| 1 \right\rangle^{}_{\rm A} \left| 0
\right\rangle^{}_{\rm B} \nonumber\\[1.5mm]
& & {} \hspace{\myGHZ} + e^{j \pi / 2}_{} \sin \left( g^{{\rm
on}}_{\rm sw} t \right) \left| {\rm e} \right\rangle \left| 0
\right\rangle^{}_{\rm A} \left| 1 \right\rangle^{}_{\rm B} \big)
 \label{GHZ:time:evol}
\end{eqnarray}
for the state of the quantum switch. Under these conditions, the
dynamics displayed in Eq.~(\ref{GHZ:time:evol}) is characterized
by two distinct processes. The first one acts on the $\left| {\rm
g} \right\rangle \left| 1 \right\rangle^{}_{\rm A} \left| 0
\right\rangle^{}_{\rm B}$ part of the initial state of
Eq.~(\ref{GHZ:initial:state}). This process is actually frozen
because the quantum switch is turned off when the qubit is in
$\left| {\rm g} \right\rangle$. The second process, instead, acts
on the $\left| {\rm e} \right\rangle \left| 1
\right\rangle^{}_{\rm A} \left| 0 \right\rangle^{}_{\rm B}$ part
of the initial state, starting the transfer of a single photon
from resonator A to resonator B and vice versa. If during such
evolution we wait for a time $t {} = {} \pi / 2 g^{\rm on}_{\rm
sw}$, a tripartite entangled state
\begin{equation}
\frac{ \left| {\rm g} \right\rangle \left| 1 \right\rangle^{}_{\rm
A} \left| 0 \right\rangle^{}_{\rm B} + e^{j \pi / 2}_{} \left|
{\rm e} \right\rangle \left| 0 \right\rangle^{}_{\rm A} \left| 1
\right\rangle^{}_{\rm B}}{\sqrt{2}}
 \label{GHZ:state}
\end{equation}
of the GHZ class~\cite{LFWei:FrancoNori:PhysRevLett:2006:a} is
generated. Here, the two resonators can be interpreted as photonic
qubits because only the Fock states $\left| 0 \right\rangle_{\rm A
, B}^{}$ and $\left| 1 \right\rangle_{\rm A , B}^{}$ are involved.
Hence, Eq.~(\ref{GHZ:state}) represents a state containing maximal
entanglement for a three-qubit system, which consists of two
photonic qubits and one superconducting (charge or flux) qubit.
The generation of GHZ states is important for the study of the
properties of genuine multipartite entanglement. Interestingly,
the quantum nature of our switch is embodied in the linear
superposition of $\left| {\rm g} \right\rangle \left| 1
\right\rangle^{}_{\rm A} \left| 0 \right\rangle^{}_{\rm B}$ and
$\left| {\rm e} \right\rangle \left| 1 \right\rangle^{}_{\rm A}
\left| 0 \right\rangle^{}_{\rm B}$ of the initial state of
Eq.~(\ref{GHZ:initial:state}).

\textit{Entanglement of coherent states.} Finally, we show how to
produce entangled coherent states of the intracavity microwave
fields of the two resonators. These are prototypical examples of
the vast class of states referred to as Schr\"odinger cat
states.~\cite{LDavidovich:SHaroche:PhysRevLett:1993:a,
ESolano:NZagury:PhysRevLett:2001:a,
ESolano:NZagury:JOptBQuantumSemiclassOpt:2002:a,
Yu-xiLiu:FNori:PhysRevA:2005:a,
MJStorcz:MMariantoni:ESolano:arXiv:cond-mat:2007:a} This time, we
start with cavity A populated by a coherent state $\left| \alpha
\right\rangle^{}_{\rm A}$ instead of a Fock state $\left| 1
\right\rangle^{}_{\rm A}$. Again, cavity B is in the vacuum state
$\left| 0 \right\rangle^{}_{\rm B}$ and the qubit in the symmetric
superposition state $\left( \left| {\rm g} \right\rangle + \left|
{\rm e} \right\rangle \right) / \sqrt{2}$, i.e., a modified
version of variant \textit{(a)} of the switching protocol is again
employed. The total disentangled initial state can be written as
\begin{equation}
\frac{\left| {\rm g} \right\rangle \left| \alpha
\right\rangle^{}_{\rm A} \left| 0 \right\rangle^{}_{\rm B} +
\left| {\rm e} \right\rangle \left| \alpha \right\rangle^{}_{\rm
A} \left| 0 \right\rangle^{}_{\rm B}}{\sqrt{2}} \, .
 \label{initial:cat:state}
\end{equation}
The resulting dynamics associated with the Hamiltonian of
Eq.~(\ref{H:op:eff}) yields a time evolution similar to that shown
for Fock states in Eq.~(\ref{GHZ:time:evol}). In this case, the
part of the evolution involving $\left| {\rm e} \right\rangle$ can
be calculated either quantum-mechanically or employing a
semi-classical model. In both cases, after a waiting time $t {} =
{} \pi / 2 g^{\rm on}_{\rm sw}$, resonator B is in the state
$\left| \alpha \right\rangle^{}_{\rm B}$ and A in the vacuum state
$\left| 0 \right\rangle^{}_{\rm A}$. However,
Eq.~(\ref{initial:cat:state}) contains an initial linear
superposition of $\left| {\rm g} \right\rangle \left| \alpha
\right\rangle^{}_{\rm A} \left| 0 \right\rangle^{}_{\rm B}$ and
$\left| {\rm e} \right\rangle \left| \alpha \right\rangle^{}_{\rm
A} \left| 0 \right\rangle^{}_{\rm B}$, requiring a
quantum-mechanical treatment of the time evolution. From this, one
finds that, after the waiting time $t {} = {} \pi / 2 g^{\rm
on}_{\rm sw}$ the quantum switch operation creates the tripartite
GHZ-type entangled state
\begin{equation}
\frac{\left| {\rm g} \right\rangle \left| \alpha
\right\rangle^{}_{\rm A} \left| 0 \right\rangle^{}_{\rm B} + e^{j
\varphi}_{} \left| {\rm e} \right\rangle \left| 0
\right\rangle^{}_{\rm A} \left| \alpha \right\rangle^{}_{\rm
B}}{\sqrt{2}} \, ,
 \label{final:cat:state}
\end{equation}
where $\varphi$ is an arbitrary phase. Again, the creation of such
states clearly reveals the quantum nature of our switch, showing a
departure from standard classical
switches.~\cite{LeonOChua:CharlesADesoer:ErnestSKuh:Book:1987:a}
Remarkably, the state of Eq.~(\ref{final:cat:state}) describes the
entanglement of coherent (``classical'') states in both
resonators. This feature is peculiar to our quantum switch and
cannot easily be reproduced in atomic
systems.~\cite{ARauschenbeutel:SHaroche:PhysRevARap:2001:a} In
principle, in absence of dissipation the quantum switch dynamics
continues transferring back the coherent state to cavity A. In
order to stop this evolution, an ultimate measurement of the qubit
along the $x$-axis of the Bloch
sphere~\cite{MatthiasSteffen:JohnMMartinis:PhysRevLett:2006:a,
AAHouck:DISchuster:RJSchoelkopf:NatureLett:2007:a} is necessary.
This corresponds to a projection associated with the Pauli
operator $\hat{\sigma}^{}_x$, which creates the two-resonator
entangled state
\begin{equation}
\frac{\left| \alpha \right\rangle^{}_{\rm A} \left| 0
\right\rangle^{}_{\rm B} + e^{j \varphi}_{} \left| 0
\right\rangle^{}_{\rm A} \left| \alpha \right\rangle^{}_{\rm
B}}{\sqrt{2}} \, .
 \label{cat:state:after:measurement}
\end{equation}
This state is decoupled from the qubit degree of freedom.

Obviously, all the protocols discussed above need suitable
measurement schemes to be implemented in reality. For instance, it
is desirable to measure the transmitted microwave field through
both resonators and, eventually, opportune cross-correlations
between them by means of a double heterodyne detection apparatus
similar to that of
Ref.~\onlinecite{MMariantoni:ESolano:arXiv:cond-mat:2005:a}. In
addition, a direct measurement of the qubit state, e.g., by means
of a DC~SQUID coupled to
it~\cite{JJohansson:HTakayanagi:PhysRevLett:2006:a,
FrankDeppe:MatteoMariantoni:RGross:NaturePhysLett:2008:a} would
allow for the full characterization of the quantum switch device.

In summary, we show that a rich landscape of nonclassical and
multipartite entangled states can be created and measured by means
of our quantum switch in two-resonator circuit~QED.

\section{TREATMENT OF DECOHERENCE}
 \label{section:treatment:of:decoherence}

The discussion in the previous sections implicitly assumes pure
quantum states. In reality, however, a quantum system gradually
decays into an incoherent mixed state during its time evolution.
This process, called decoherence, is due to the entanglement of
the system with its environment and it is known to be a critical
issue for solid-state quantum circuits. Since it is difficult to
decouple these circuits from the large number of environmental
degrees of freedom to which they are exposed, their typical
decoherence rates cannot be easily minimized. Usually, they are in
the range from $1$\,MHz to $1$\,GHz, depending on the specific
implementation. In this section, we discuss the impact of the
three most relevant decoherence mechanisms on the quantum switch
architecture. These are: First, the population decay of resonators
A and B with rates $\kappa^{}_{\rm A}$ and $\kappa^{}_{\rm B}$,
respectively; second, the qubit relaxation from the energy excited
state to the groundstate at a rate $\gamma^{}_{\rm r}$ due to
high-frequency noise; third, the qubit dephasing (loss of phase
coherence) at a pure dephasing rate $\gamma^{}_{\varphi}$ due to
low-frequency noise. We show by means of detailed analytical
derivations that, despite decoherence mechanisms, a working
quantum switch can be realized with readily available
superconducting qubits and resonators.

Decoherence processes are most naturally described in the qubit
energy eigenbasis $\left\{ \left| {\rm g} \right\rangle , \left|
{\rm e} \right\rangle \right\}$. Under the Markov approximation,
the time evolution of the density matrix of the quantum switch
Hamiltonian $\widehat{H}$ of Eq.~(\ref{H:op}) is described by the
master equation
\begin{equation}
\dot{\hat{\rho}}^{}_{} {} = {} \frac{1}{j \hbar} \left(
\widehat{H}^{}_{} \hat{\rho}^{}_{} - \hat{\rho}^{}_{}
\widehat{H}^{}_{} \right) + \sum_{n = 1}^4 \hat{\mathcal{L}}^{}_n
\hat{\rho}^{}_{} \, .
 \label{rho:op:dot:tot}
\end{equation}
Here, $\hat{\mathcal{L}}^{}_n$ is the Lindblad superoperator
defined as $\hat{\mathcal{L}}^{}_n \hat{\rho}^{}_{} \equiv
\gamma^{}_n \left( \hat{X}^{}_n \hat{\rho}^{}_{} \hat{X}^{\dag}_n
- \hat{X}^{\dag}_n \hat{X}^{}_n \hat{\rho}^{}_{} / 2 -
\hat{\rho}^{}_{} \hat{X}^{\dag}_n \hat{X}^{}_n / 2 \right)$. The
indices $n = 1 , 2 , 3 , 4$ label the decay of resonator A, the
decay of resonator B, qubit relaxation, and qubit dephasing,
respectively. Consequently, the generating operators are
$\hat{X}^{}_1 \equiv \hat{a}^{}_{}$, $\hat{X}^{\dag}_1 \equiv
\hat{a}^{\dag}_{}$, $\hat{X}^{}_2 \equiv \hat{b}^{}_{}$,
$\hat{X}^{\dag}_2 \equiv \hat{b}^{\dag}_{}$, $\hat{X}^{}_3 \equiv
\hat{\sigma}^-_{}$, $\hat{X}^{\dag}_3 \equiv \hat{\sigma}^+_{}$,
and $\hat{X}^{}_4 = \hat{X}^{\dag}_4 \equiv \hat{\sigma}^{}_z$.
The corresponding decoherence rates are $\gamma^{}_1 \equiv
\kappa^{}_{\rm A}$, $\gamma^{}_2 \equiv \kappa^{}_{\rm B}$,
$\gamma^{}_3 \equiv \gamma^{}_{\rm r}$, and $\gamma^{}_4 \equiv
\gamma^{}_{\varphi} / 2$. For the resonators, $\kappa^{}_{\rm A}$
and $\kappa^{}_{\rm B}$ are often expressed in terms of the
corresponding loaded quality factors $Q^{}_{\rm A}$ and $Q^{}_{\rm
B}$, $\kappa^{}_{\rm A} \equiv \omega^{}_{\rm A} / Q^{}_{\rm A}$
and $\kappa^{}_{\rm B} \equiv \omega^{}_{\rm B} / Q^{}_{\rm B}$,
respectively. Although in general all four processes coexist, in
most experimental situations one of them dominates over the
others. In fact, it is a common experimental scenario that
$\gamma^{}_{\varphi} \gg \gamma^{}_{\rm r}$, for example in the
special case of a flux qubit operated away from the degeneracy
point (see, e.g.,
Ref.~\onlinecite{FDeppe:RGross:PhysRevB:2007:a}). In this
situation, we can extract pessimistic relaxation and dephasing
rates from the
literature,~\cite{PBertet:JEMooij:PhysRevLett:2005:a,
FYoshihara:JSTsai:PhysRevLett:2006:a,
KKakuyanagi:AShnirman:PhysRevLett:2007:a,
FDeppe:RGross:PhysRevB:2007:a} $\gamma^{}_{\rm r} {} \simeq {}
1$\,MHz and $\gamma^{}_{\varphi} {} \simeq {} 200$\,MHz. In other
words, dephasing is the dominating source of qubit
decoherence.~\cite{qubit:energy:relaxation} The decay rates of the
resonators can be engineered such
that~\cite{ThomasNiemczyk:PrivateComm:2008:a} $\kappa^{}_{\rm A} ,
\kappa^{}_{\rm B} {} \lesssim {} \gamma^{}_{\rm r} {} \ll {}
\gamma^{}_{\varphi}$. For these reasons, hereafter we focus on
dephasing mechanisms only. Hence, we analyze the following
simplified master equation
\begin{equation}
\dot{\hat{\rho}}^{}_{} {} = {} \frac{1}{j \hbar} \big(
\hat{H}^{}_{} \hat{\rho}^{}_{} - \hat{\rho}^{}_{} \hat{H}^{}_{}
\big) + \hat{\mathcal{L}}^{}_{\varphi} \hat{\rho}^{}_{} \, ,
 \label{rho:op:dot:deph}
\end{equation}
where $\hat{\mathcal{L}}^{}_{\varphi} \hat{\rho}^{}_{} {} \equiv
{} \hat{\mathcal{L}}^{}_4 \hat{\rho}^{}_{} = ( \gamma^{}_{\varphi}
/ 2) ( \sigma^{}_z \hat{\rho}^{}_{} \hat{\sigma}^{}_z -
\hat{\rho}^{}_{} )$.

The impact of qubit dephasing on the switch operation depends on
the chosen protocol
(cf.~Subsec.~\ref{subsection:a:quantum:switch:protocol}). When
employing the population-inversion protocol, qubit dephasing
occurs within the duration time $t^{}_{\pi}$ of the control
$\pi$-pulses. The time $t^{}_{\pi}$ coincides with the inverse of
the qubit Rabi frequency and can be reduced to less than $1$\,ns
using high driving
power.~\cite{SSaito:HTakayanagi:PhysRev:Lett:2006:a} In this way,
the time window during which the qubit is sensitive to dephasing
is substantially shortened. However, it is more favorable to
resort to the adiabatic-shift pulse protocol. In this case, the
qubit always remains in $\left| {\rm g} \right\rangle$ resulting
in a complete elimination of pulse-induced dephasing. The relevant
time scale during which dephasing occurs is therefore set by the
operation time of the switch between two on-off events. Naturally,
this time should be as long as possible if we want to perform many
operations.

The effect of dephasing during the switch operation time is better
understood by inspecting the effective quantum switch Hamiltonian
$\widehat{H}^{}_{\rm eff}$ of Eq.~(\ref{H:op:eff}). In
Subsec.~\ref{subsection:balancing:the:geometric:and:dynamic:coupling},
we deduce this effective Hamiltonian by means of a Dyson series
expansion. This approach is very powerful and compact when dealing
with the analysis of a unitary evolution. However, when treating
master equations, we prefer to utilize a variant of the well-known
Schrieffer-Wolff unitary
transformation,~\cite{AlexandreBlais:RJSchoelkopf:PhysRevA:2004:a,
JulianHauss:GerdSchoen:PhysRevLett:2008:a} $\widehat{U}^{}_{}
\widehat{H}^{}_{} \widehat{U}^{\dag}_{}$, where
\newlength{\myspacetwo}
\settowidth\myspacetwo{$\exp \bigg[$}
\newlength{\myspacethree}
\settowidth\myspacethree{$\bigg]$}
\begin{eqnarray}
\widehat{U}^{}_{} & {} \equiv {} & \exp \bigg[ \frac{g^{}_{\rm A}
\sin \theta}{\Delta} \Big( \hat{\sigma}^-_{} \hat{a}^{\dag}_{} -
\hat{\sigma}^+_{} \hat{a}^{}_{} \Big) \nonumber\\[1.5mm]
& & {} \hspace{\myspacetwo} + \frac{g^{}_{\rm B} \sin
\theta}{\Delta} \Big( \hat{\sigma}^-_{} \hat{b}^{\dag}_{} -
\hat{\sigma}^+_{} \hat{b}^{}_{} \Big) \bigg]
 \label{U:op}
\end{eqnarray}
and $\widehat{U}^{\dag}_{}$ is its Hermitian conjugate. In the
large-detuning regime, $g^{}_{\rm A} \sin \theta , g^{}_{\rm B}
\sin \theta {} \ll {} \Delta$, we can neglect all terms of orders
$\left( g^{}_{\rm A} \sin \theta / \Delta \right)^2_{}$, $\left(
g^{}_{\rm B} \sin \theta / \Delta \right)^2_{}$, $g^{}_{\rm A}
g^{}_{\rm B} \sin^2_{} \theta / \Delta^2_{}$, or higher. After a
transformation into an interaction picture with respect to the
qubit and both resonators
(cf.~Subsec.~\ref{subsection:balancing:the:geometric:and:dynamic:coupling})
and performing opportune RWAs, we obtain again
$\widehat{H}^{}_{\rm eff}$ of Eq.~(\ref{H:op:eff}). The master
equation governing the time evolution of the effective density
matrix $\hat{\rho}^{\rm eff}_{} {} \equiv {} \widehat{U}^{}_{}
\hat{\rho}^{}_{} \widehat{U}^{\dag}_{}$ then becomes
\begin{equation}
\dot{\hat{\rho}}^{\rm eff}_{} {} = {} \frac{1}{j \hbar} \left(
\widehat{H}^{}_{\rm eff} \hat{\rho}^{\rm eff}_{} - \hat{\rho}^{\rm
eff}_{} \widehat{H}^{}_{\rm eff} \right) + \hat{\mathcal{L}}^{\rm
eff}_{\varphi} \hat{\rho}^{\rm eff}_{} \, .
 \label{rho:op:dot:eff}
\end{equation}
The analysis is complicated by the fact that also the Lindblad
superoperator $\hat{\mathcal{L}}^{}_{} \hat{\rho}$ has to be
transformed. For the sake of simplicity, we can assume $g^{}_{\rm
A} {} = {} g^{}_{\rm B} {} \equiv {} g^{}_{}$ and find
\begin{equation}
\hat{\mathcal{L}}^{\rm eff}_{\varphi} \hat{\rho}^{\rm eff}_{} {}
\approx {} \hat{\mathcal{L}}^{}_{\varphi} \hat{\rho}^{\rm eff}_{}
+ 2 \gamma^{}_{\varphi} \times \mathcal{O} \left[ \left( \frac{g
\sin \theta}{\Delta} \right)^{\!2}_{} \right] \, .
   \label{L:op:varphi:eff:rho:op:eff}
\end{equation}
When deriving this expression, all terms of $\mathcal{O} \left( g
\sin \theta / \Delta \right)$ are explicitly neglected by a RWA.
This approximation relies on the condition $\left(
\gamma^{}_{\varphi} / \Delta \right) g \sin \theta \ll \Delta$,
which is well satisfied in the large-detuning regime as long as
$\gamma^{}_{\varphi} {} \lesssim {} \Delta$. The latter
requirement can easily be met by most types of existing
superconducting qubits. In the frame of $\widehat{H}^{}_{\rm
eff}$, $\hat{\mathcal{L}}^{}_{\varphi} \hat{\rho}^{\rm eff}_{}$
has the standard Lindblad dephasing structure and the qubit
appears only in the form of $\hat{\sigma}^{}_z$-operators. Since
the initial state of the switch operation is characterized by
either no (adiabatic-shift pulse protocol) or only very small
(population-inversion protocol) qubit coherences, the effect of
$\hat{\mathcal{L}}^{}_{\varphi} \hat{\rho}^{\rm eff}_{}$ on the
time evolution of the system is negligible. All other nonvanishing
terms are comprised in the expression $2 \gamma^{}_{\varphi}
\times \mathcal{O} [ ( g \sin \theta / \Delta )^{\!2}_{} ]$ of
Eq.~(\ref{L:op:varphi:eff:rho:op:eff}) and scale with a factor
smaller than $\gamma^{\rm eff}_{\varphi} {} \equiv {} 2
\gamma^{}_{\varphi} \left( g \sin \theta / \Delta \right)^2_{}$.
Hence, the operation of the quantum switch is robust to qubit
dephasing on a characteristic time scale $T^{\rm eff}_{\varphi} {}
= {} 1 / \gamma^{\rm eff}_{\varphi} {} \gg {} 1 /
\gamma^{}_{\varphi}$. For completeness, it is important to mention
that the higher-order terms of
Eq.~(\ref{L:op:varphi:eff:rho:op:eff}) can contain combinations of
operators such as $\hat{a}^{\dag}_{} \hat{a}^{}_{}$ and
$\hat{b}^{\dag}_{} \hat{b}^{}_{}$. In this case, $T^{\rm
eff}_{\varphi}$ would be reduced for a large number of photons
populating the resonators. Fortunately, this does not constitute a
major issue since the most interesting applications of a quantum
switch require that the number of photons in the resonators is of
the order of one.

In summary, we show that for suitably engineered cavities the
quantum switch operation time for the adiabatic-shift pulse
protocol is limited only by the effective qubit dephasing time
$T^{\rm eff}_{\varphi}$. The latter is strongly enhanced with
respect to the intrinsic dephasing time $T^{}_{\varphi} {} \equiv
{} 1 / \gamma^{}_{\varphi}$. In this sense, the quantum switch is
superior to the dual setup, where two qubits are dispersively
coupled via one cavity
bus.~\cite{JMajer:JMChow:RJSchoelkopf:NatureLett:2007:a} Moreover,
the intrinsic dephasing time $T^{}_{\varphi}$ and, consequently,
$T^{\rm eff}_{\varphi}$ are further enhanced by choosing a shift
pulse which sets the on-state bias near the qubit degeneracy
point.~\cite{FDeppe:RGross:PhysRevB:2007:a} As explained in
Subsec.~\ref{subsection:a:quantum:switch:protocol}, this regime
takes place for a qubit-resonator detuning $\Delta^{}_2 <
\Delta^{}_1$. In this case, the switch coupling coefficient is
also substantially increased because of a dominating dynamic
interaction. As a consequence, this option is particularly
appealing in the context of the advanced applications discussed in
Subsec.~\ref{subsection:advanced:applications:nonclassical:states:and:entanglement}.
Finally, we notice that for the population-inversion protocol the
switch operation time could be limited by the qubit relaxation
time $T^{}_{\rm r} {} \equiv {} 1 / \gamma^{}_{\rm r}$. However,
the switch setting condition is typically fulfilled for a bias
away from the qubit degeneracy point. There, $T^{}_{\rm r}$ is
considerably enhanced by both a
reduced~\cite{FDeppe:RGross:PhysRevB:2007:a} $\sin \theta$ and by
the Purcell effect of the
cavities.~\cite{AlexandreBlais:RJSchoelkopf:PhysRevA:2004:a}

\section{AN EXAMPLE OF TWO-RESONATOR CIRCUIT~QED WITH A FLUX QUBIT}
 \label{section:an:example:of:two:resonator:circuit:qed:with:a:flux:qubit}

\begin{figure*}[t!]
\centering{%
 \includegraphics[width=0.75\textwidth]{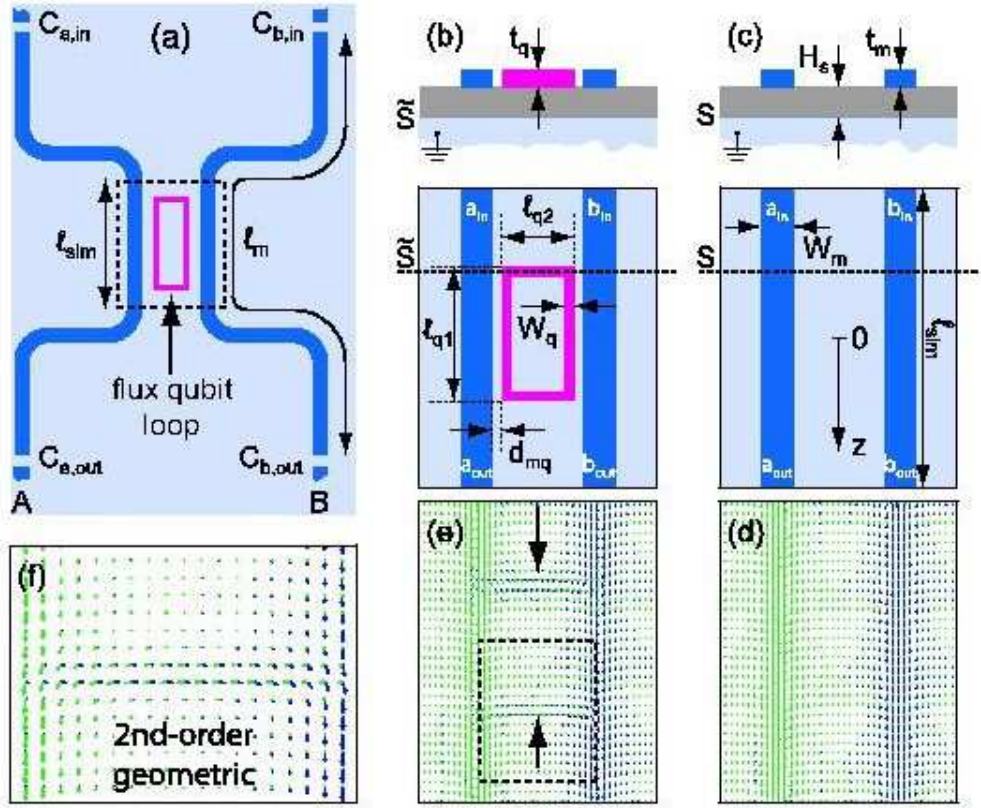}}
\caption{(Color online) A possible setup for two-resonator
circuit~QED with a flux qubit. (a)~Overall structure (dimensions
not in scale). Two microstrip resonators A and B (thick blue
lines) of length $\ell^{}_{\rm m}$ simultaneously coupled to a
flux qubit loop [magenta (middle grey) rectangle]. $C^{}_{\rm a ,
in}$, $C^{}_{\rm b , in}$, $C^{}_{\rm a , out}$, and $C^{}_{\rm b
, out}$: Input and output capacitors for A and B. The dashed black
box indicates the region of the close-up shown in (b).
$\ell^{}_{\rm sim}$: Length of the region used for the
FASTHENRY~\cite{FASTHENRY:MattanKamon:JakobKWhite:IEEEMicrowave:1994:a:two:links}
simulations. (b)~Close-up of the region which contains the flux
qubit loop in (a). $\ell^{}_{{\rm q} 1}$ and $\ell^{}_{{\rm q}
2}$: Qubit loop lateral dimensions. $W^{}_{\rm q}$: Width of the
qubit lines. $d^{}_{\rm m q}$: Distance between the qubit and each
resonator. The dashed black line denominated as $\widetilde{S}$
marks the cross-section reported on the top part of the panel.
$t^{}_{\rm q}$: Thickness of the qubit loop lines. (c) As in (b),
but without the qubit loop. $W^{}_{\rm m}$ and $t^{}_{\rm m}$:
Width and thickness (see cross-section S) of the two microstrip
resonators. $H^{}_{\rm s}$: Height of the dielectric substrate.
The reference axis $0 z$ is also indicated
(cf.~Appendix~\ref{appendix:b:details:of:the:fasthenry:simulations}).
Both in (b) and (c), ${\rm a^{}_{in}}$, ${\rm a^{}_{out}}$, ${\rm
b^{}_{in}}$, and ${\rm b^{}_{out}}$ represents the input and
output probing ports used in the simulations. (d)~Current density
distribution at high frequency ($5$\,GHz) for the structures drawn
in (c). The currents are represented by small arrows, green (light
grey) for resonator A and blue (dark grey) for resonator B.
(e)~Current density distribution at high frequency ($5$\,GHz) for
the structures drawn in (b). The two black arrows indicate two
high-current-density channels between the two resonators. The
dashed black box marks the close-up area shown in (f).
(f)~Close-up of one of the two geometric second-order interaction
channels.}
 \label{QS:Figure:6:abcdef:Matteo:Mariantoni:2008}
\end{figure*}

In this section, we focus on the geometry sketched in
Fig.~\ref{QS:Figure:1:abcde:Matteo:Mariantoni:2008}(c) and present
one specific implementation of two-resonator circuit~QED. As a
particular case, the described setup can be operated as a
superconducting quantum switch. In this example, we consider
microstrip resonators. Coplanar wave guide resonators can also be
used without significantly affecting our main results. In
addition, we choose a flux qubit because this is our main topic of
research.~\cite{MMariantoni:ESolano:arXiv:cond-mat:2005:a,
KKakuyanagi:AShnirman:PhysRevLett:2007:a,
FDeppe:RGross:PhysRevB:2007:a,
FrankDeppe:MatteoMariantoni:RGross:NaturePhysLett:2008:a}
Moreover, as shown in
Subsec.~\ref{subsection:balancing:the:geometric:and:dynamic:coupling},
the dynamic properties of the quantum switch are independent of
specific implementations. As a consequence, in this section we
concentrate on its geometric properties only. It is worth
mentioning again that such properties are inherent to circuit~QED
architectures and constitute a fundamental departure from quantum
optical systems.

In Figs.~\ref{QS:Figure:6:abcdef:Matteo:Mariantoni:2008}(a) and
\ref{QS:Figure:6:abcdef:Matteo:Mariantoni:2008}(b), the design of
a possible two-resonator circuit~QED setup is shown. The overall
structure is composed of two superconducting microstrip
transmission lines, which are bounded by input and output
capacitors, $C^{}_{\rm a , in}$, $C^{}_{\rm b , in}$, $C^{}_{\rm a
, out}$, and $C^{}_{\rm b , out}$. This geometrical configuration
forms the two resonators A and B. The input and output capacitors
also determine the loaded or external quality factors $Q^{}_{\rm
A}$ and $Q^{}_{\rm B}$ of the two
resonators.~\cite{LFrunzio:RSchoelkopf:IEEEApplSupercond:2005:a}
Both A and B are characterized by a length $\ell^{}_{\rm m}$,
which defines their center frequencies $f^{}_{\rm A}$ and
$f^{}_{\rm B}$. We choose $\ell^{}_{\rm m} = \lambda^{}_{\rm m} /
2 = 12$\,mm, where $\lambda^{}_{\rm m} \equiv \lambda^{}_{\rm A} =
\lambda^{}_{\rm B}$ is the full wavelength of the standing waves
on the resonators. The superconducting loop of the flux qubit
circuit is positioned at the current antinode of the two
$\lambda^{}_{\rm m} / 2$ resonators.

In Fig.~\ref{QS:Figure:6:abcdef:Matteo:Mariantoni:2008}(c), only
the two microstrip resonators A and B are considered. They are
chosen to have a width $W^{}_{\rm m} = 10$\,\micro m and a
thickness $t^{}_{\rm m} = 100$\,nm. The height of the dielectric
substrate between each microstrip and the corresponding
groundplane is $H^{}_{\rm s} = 12.3$\,\micro m. The substrate can
opportunely be made of different materials, for example silicon,
sapphire, amorphous hydrogenated silicon, or silicon oxide,
depending on the experimental necessities. A detailed study on the
properties of a variety of dielectrics and on the dissipation
processes of superconducting on-chip resonators can be found in
Refs.~\onlinecite{JohnMMartinis:ClareCYu:PhysRevLett:2005:a,
ADOapConnell:ErikLucero:ANCleland:JMMartinis:ApplPhysLett:2008:a,
JiansongGao:HenryGLeduc:ApplPhysLett:2008:a,
JiansongGao:JohnMMartinis:HenryGLeduc:ApplPhysLett:2008:a} and
\onlinecite{RBarends:TMKlapwijk:ApplPhysLett:2008:a}. The aspect
ratio $W^{}_{\rm m} / H^{}_{\rm s}$ is engineered to guarantee a
line characteristic impedance $Z^{}_{\rm c} \simeq 50\,\Omega$,
even if this is not a strict requirement for the resonators to
function properly.~\cite{on-chip:transmission:lines:matching} The
remaining dimensions of our system are shown in
Fig.~\ref{QS:Figure:6:abcdef:Matteo:Mariantoni:2008}(b): The
lateral dimensions $\ell^{}_{{\rm q} 1} = 200$\,\micro m and
$\ell^{}_{{\rm q} 2} = 87$\,\micro m of the qubit loop, the width
$W^{}_{\rm q} = 1$\,\micro m of each line forming the qubit loop,
the interspace $d^{}_{\rm m q} = 1$\,\micro m between qubit and
resonators, and the thickness $t^{}_{\rm q} ( = t^{}_{\rm m} ) =
100$\,nm of the qubit lines. The dimensions of the qubit loop are
chosen to optimize the qubit-resonator coupling strengths. This
geometry results in a relatively large inductance $L^{}_{\rm q}
\simeq 780$\,pH (number obtained from FASTHENRY
simulations;~\cite{FASTHENRY:MattanKamon:JakobKWhite:IEEEMicrowave:1994:a:two:links}
cf.~Table~\ref{QS:Table:1:Matteo:Mariantoni:2008}). Despite the
large self-inductance $L^{}_{\rm q}$, reasonable qubit coherence
times are expected (see, e.g.,
Refs.~\onlinecite{RHKoch:DPDiVincenzo:PhysRevLett:2006:a} and
\onlinecite{BLTPlourde:JohnClarke:PhysRevBRap:2005:a}). Moreover,
in the light of Sec.~\ref{section:treatment:of:decoherence} these
coherence times easily suffice for the operation of a
superconducting quantum switch, where the qubit acts as a mere
mediator for the exchange of virtual excitations.

\begin{table*}[ht]
\caption{Relevant parameters for a possible two-resonator
circuit~QED setup based on a superconducting flux qubit. The
various constants are described in
Subsec.~\ref{section:an:example:of:two:resonator:circuit:qed:with:a:flux:qubit}
and reported in
Fig.~\ref{QS:Figure:2:abcdef:Matteo:Mariantoni:2008}. All
inductances are simulated using the version of FASTHENRY for
superconducting
materials.~\cite{FASTHENRY:MattanKamon:JakobKWhite:IEEEMicrowave:1994:a:two:links}
The capacitances are calculated analytically. All geometric
second-order inductances are computed analytically and numerically
and then compared to each other for consistency. We find an
excellent agreement between the estimates $M^{}_{\rm q a}
M^{}_{\rm q b} / L^{}_{\rm q}$ and $\widetilde{m} - m$ for the
second-order mutual inductance. Also, the shift inductances
$M^2_{\rm q a} / L^{}_{\rm q}$ and $M^2_{\rm q b} / L^{}_{\rm q}$
coincide with their counterparts $L^{}_{\rm s a} {} \equiv {}
L^{\ast}_{\rm r a} - \widetilde{L}^{\ast}_{\rm r a}$ and
$L^{}_{\rm s b} {} \equiv {} L^{\ast}_{\rm r b} -
\widetilde{L}^{\ast}_{\rm r b}$, respectively. These parameters
are suitable for the implementation of a superconducting quantum
switch.}
 \vspace{3.0pt}
\begin{ruledtabular}
 \begin{tabular}{c c c c c c c c c c c c}
  \vspace{-8.5pt}
   \\
     & $L^{\ast}_{\rm r a}$ & $L^{}_{\rm r a}$ & $C^{}_{\rm r a}$ & $\lambda^{}_{\rm A}$
     & $i^{}_{{\rm A} 0} {} \equiv {} \sqrt{\dfrac{\hbar \omega^{}_{\rm A}}{2 L^{}_{\rm r a}}}$
     & $L^{}_{\rm q}$ & $M^{}_{\rm q a}$ & $\dfrac{M^{}_{\rm q a} M^{}_{\rm q b}}{L^{}_{\rm q}}$
     & $+ L^{}_{\rm s a} {} \equiv {} \dfrac{M^2_{\rm q a}}{L^{}_{\rm q}}$
     & $+ L^{}_{\rm s b} {} \equiv {} \dfrac{M^2_{\rm q b}}{L^{}_{\rm q}}$
   \\
     & \footnotesize{(pH)} & \footnotesize{(nH)} & \footnotesize{(pF)} & \footnotesize{(mm)}
     & \footnotesize{(nA)}
     & \footnotesize{(pH)} & \footnotesize{(pH)} & \footnotesize{(pH)}
     & \footnotesize{(pH)}
     & \footnotesize{(pH)}
   \\
     \vspace{-7.0pt}
   \\
     \hline
     \vspace{-7.0pt}
   \\
     & $252.781$ & $6.06697$ & $3.36369$ & $24$
     & $13.8249$
     & $784.228$ & $61.2387$ & $4.78192$
     & $+ 4.78200$
     & $+ 4.78184$
   \\
     \vspace{-7.0pt}
   \\
     \hline
     \vspace{-7.0pt}
   \\
     & $L^{\ast}_{\rm r b}$ & $L^{}_{\rm r b}$ & $C^{}_{\rm r b}$ & $\lambda^{}_{\rm B}$
     & $i^{}_{{\rm B} 0} {} \equiv {} \sqrt{\dfrac{\hbar \omega^{}_{\rm B}}{2 L^{}_{\rm r b}}}$
     & $m$ & $M^{}_{\rm q b}$ & $\widetilde{m} \! - \! m$
     & \raisebox{0.0pt}[0.0pt][0.0pt]{$\begin{array}{@{}c@{}} - L^{}_{\rm s a} = {}
       \\ \widetilde{L}^{\ast}_{\rm r a} - L^{\ast}_{\rm r a}
       \end{array}$}
     & \raisebox{0.0pt}[0.0pt][0.0pt]{$\begin{array}{@{}c@{}} - L^{}_{\rm s b} = {}
       \\ \widetilde{L}^{\ast}_{\rm r b} - L^{\ast}_{\rm r b}
       \end{array}$}
   \\
     & \footnotesize{(pH)} & \footnotesize{(nH)} & \footnotesize{(pF)}& \footnotesize{(mm)}
     & \footnotesize{(nA)}
     & \footnotesize{(pH)} & \footnotesize{(pH)} & \footnotesize{(pH)}
     & \footnotesize{(pH)}
     & \footnotesize{(pH)}
   \\
     \vspace{-7.0pt}
   \\
     \hline
     \vspace{-7.0pt}
   \\
     & $252.778$ & $6.06693$ & $3.36369$ & $24$
     & $13.8249$
     & $2.90130$ & $61.2377$ & $4.78192$
     & $-4.78100$
     & $-4.78100$
   \\
     \vspace{-8.5pt}
   \\
 \end{tabular}
\end{ruledtabular}
 \label{QS:Table:1:Matteo:Mariantoni:2008}
\end{table*}
In our numerical simulations
(cf.~Appendix~\ref{appendix:b:details:of:the:fasthenry:simulations}),
we restrict ourselves to the region indicated by the black dashed
box in Fig.~\ref{QS:Figure:6:abcdef:Matteo:Mariantoni:2008}(a),
the close-up of which is shown in
Fig.~\ref{QS:Figure:6:abcdef:Matteo:Mariantoni:2008}(b) and, in
the absence of the flux qubit loop, in
Fig.~\ref{QS:Figure:6:abcdef:Matteo:Mariantoni:2008}(c). This
region is characterized by a length $\ell^{}_{\rm sim} {} = {}
500$\,\micro m of the resonators and is centered at a position
where the magnetic field reaches a maximum (antinode) and the
electric field reaches a minimum (node). We notice that magnetic
and electric fields can equivalently be expressed in terms of
currents and voltages, respectively. There are two main hypotheses
behind the validity of our simulation results for the entire
two-resonator-qubit system. These are the uniformity of the
electromagnetic field (voltage and current) in the simulated
region and the abruptly~\cite{abruplty} increasing geometric
distance between resonators A and B outside of it [see sketch of
Fig.~\ref{QS:Figure:6:abcdef:Matteo:Mariantoni:2008}(a)]. The
three main implications of the above assumptions are explained in
the following. First, all coupling strengths are dominated by
inductive interactions and there are no appreciable capacitive
ones. Inside the simulated region, in fact, the voltage is
practically characterized by a node, which results in a vanishing
coupling coefficient. Outside the simulated region, the effective
distance $d^{}_{\rm eff}$ between the cavities strongly increases
together with the geometric one.~\cite{RoberECollin:Book:2000:a,
DavidMPozar:Book:2005:a} As a consequence, the geometric
first-order capacitance $c \propto 1 / d^{}_{\rm eff}$ becomes
negligible. Second, the coupling coefficients between qubit and
resonators can be obtained without integrating over the spatial
distribution of the mode. This is because of the uniformity of the
field, which, for all practical purposes, is constant over the
restricted simulated region. Third, the geometric first-order
coupling between the two resonators, which is proportional to
their mutual inductance $m$, is still accurately determined. In
fact, outside the simulated region any additional contribution to
$m$ becomes negligible. For all the reasons mentioned above, we
are allowed to use the
FASTHENRY~\cite{FASTHENRY:MattanKamon:JakobKWhite:IEEEMicrowave:1994:a:two:links}
calculation software for our simulations. In this section, we
utilize two different versions of FASTHENRY, one for
superconducting materials and one for almost perfect conducting
ones. We use the second version only when we want to obtain
current density distributions or the frequency dependence of an
inductance. In these cases, due to technical limitations of the
software, we cannot use the version valid for
superconductors.~\cite{FASTHENRY:MattanKamon:JakobKWhite:IEEEMicrowave:1994:a:two:links}

Figures~\ref{QS:Figure:6:abcdef:Matteo:Mariantoni:2008}(d) and
\ref{QS:Figure:6:abcdef:Matteo:Mariantoni:2008}(e) display the
simulated current density distributions at a probing frequency of
$5$\,GHz (high-frequency regime) for the different structures
drawn in Fig.~\ref{QS:Figure:6:abcdef:Matteo:Mariantoni:2008}(c)
and \ref{QS:Figure:6:abcdef:Matteo:Mariantoni:2008}(b),
respectively. Similar results can be found in a range between
$1$\,GHz and $10$\,GHz (data not shown). Without loss of
generality, these simulations are performed for almost perfect
conductors using a FASTHENRY version which does not support
superconductivity. The results of
Fig.~\ref{QS:Figure:6:abcdef:Matteo:Mariantoni:2008}(d) clearly
show that the two microstrip lines are regions characterized by a
high current density separated by a region with a low current
density in absence of the flux qubit loop. In this case, the
geometric interaction between resonators A and B is reduced to a
bare first-order coupling, which turns out to be very weak. On the
contrary, in
Fig.~\ref{QS:Figure:6:abcdef:Matteo:Mariantoni:2008}(e) the
presence of the qubit loop clearly opens two new current channels
between A and B. These are located at the position of the upper
and lower qubit loop segments of
Fig.~\ref{QS:Figure:6:abcdef:Matteo:Mariantoni:2008}(b). For
clarity, the close-up of one of these channels is shown in
Fig.~\ref{QS:Figure:6:abcdef:Matteo:Mariantoni:2008}(f). Notably,
the two additional current channels of
Fig.~\ref{QS:Figure:6:abcdef:Matteo:Mariantoni:2008}(e) represent
the geometric second-order coupling.

We now study in more detail the relationship between geometric
first- and second-order inductances for the structures of
Fig.~\ref{QS:Figure:6:abcdef:Matteo:Mariantoni:2008}(b) and
\ref{QS:Figure:6:abcdef:Matteo:Mariantoni:2008}(c). The notation
is that of
Subsec.~\ref{subsection:the:role:of:circuit:topology:two:examples}
and Figs.~\ref{QS:Figure:2:abcdef:Matteo:Mariantoni:2008}(d)-(f).
All quantities are computed numerically with the aid of FASTHENRY
for superconducting
materials~\cite{FASTHENRY:MattanKamon:JakobKWhite:IEEEMicrowave:1994:a:two:links}
assuming a London penetration depth $\lambda^{}_{\rm L} {} = {}
180$\,nm. We notice that, in this case, the simulated inductances
are independent of the probing frequency. In a first run of
simulations, we calculate pure first-order inductances only
(cf.~Appendix~\ref{appendix:b:details:of:the:fasthenry:simulations}).
These are the simulated test inductances $L^{\ast}_{\rm r a}$ and
$L^{\ast}_{\rm r b}$ from which we obtain the self-inductances
$L^{}_{\rm r a}$ and $L^{}_{\rm r b}$ of resonators A and B (more
details in the next paragraph), the first-order mutual inductance
$m$ between the two resonators, the self-inductance $L^{}_{\rm q}$
of the qubit loop, and the mutual inductances $M^{}_{\rm q a}$ and
$M^{}_{\rm q b}$ between qubit and resonators. In a second run of
simulations, we calculate directly
(cf.~Appendix~\ref{appendix:b:details:of:the:fasthenry:simulations})
the sum of first- and second-order inductances. These are the
renormalized test inductances $\widetilde{L}^{\ast}_{\rm r a}$ and
$\widetilde{L}^{\ast}_{\rm r b}$ of the portions of resonators A
and B shown in
Fig.~\ref{QS:Figure:6:abcdef:Matteo:Mariantoni:2008}(b) and the
total mutual inductance $\widetilde{m}^{}_{}$ between the two
resonators. The difference $\widetilde{m}^{}_{} - m =
4.78192$\,pH, i.e., the geometric second-order coupling, coincides
up to the sixth significant digit with the quantity $M^{}_{\rm q
a} M^{}_{\rm q b} / L^{}_{\rm q}$ expected from our general
three-node network approach of
Subsec.~\ref{subsection:the:capacitance:and:inductance:matrices:up:to:second:order}
and, equivalently, from the three-circuit theory of
Subsec.~\ref{subsection:the:role:of:circuit:topology:two:examples}
(cf.~Table~\ref{QS:Table:1:Matteo:Mariantoni:2008}). We also find
that the dominating geometric coupling between A and B is not the
first-order inductance $m = 2.90130$\,pH, but the second-order
one. The ratio between second-order and first-order inductances is
$( \widetilde{m} - m ) / m \simeq 1.6$. In addition, the numerical
simulations yield the two shift test inductances $|
\widetilde{L}^{\ast}_{\rm r a} - L^{\ast}_{\rm r a} | = |
\widetilde{L}^{\ast}_{\rm r b} - L^{\ast}_{\rm r b} | =
4.78100$\,pH (cf.~also next paragraph). These shifts renormalize
the bare center frequencies $f^{}_{\rm A}$ and $f^{}_{\rm B}$ of
resonators A and B, respectively, and are found to be in very good
agreement up to several decimal digits with their analytical
estimates $L^{}_{\rm s a} \equiv M^2_{\rm q a} / L^{}_{\rm q}$ and
$L^{}_{\rm s b} \equiv M^2_{\rm q b} / L^{}_{\rm q}$ of our
three-circuit theory of
Subsec.~\ref{subsection:the:role:of:circuit:topology:two:examples}
(cf.~Table~\ref{QS:Table:1:Matteo:Mariantoni:2008}). We point out
that, in our definition, the quantities $L^{}_{\rm s a}$ and
$L^{}_{\rm s b}$ are strictly positive. Remarkably, our
simulations reveal that $L^{\ast}_{\rm r a} >
\widetilde{L}^{\ast}_{\rm r a}$ and $L^{\ast}_{\rm r b} >
\widetilde{L}^{\ast}_{\rm r b}$, reproducing the minus sign in the
expressions $L^{}_{\rm r a} - L^{}_{\rm s a}$ and $L^{}_{\rm r b}
- L^{}_{\rm s b}$ of
Fig.~\ref{QS:Figure:2:abcdef:Matteo:Mariantoni:2008}(f). These
findings confirm the superiority of the three-circuit theory of
Subsec.~\ref{subsection:the:role:of:circuit:topology:two:examples}
over the simple model which results in the Hamiltonian of
Eq.~(\ref{H:op:f}). In the case of purely inductive interactions,
the resonators suffer a small blue-shift of their center
frequencies, i.e., a shift towards higher values. This is opposite
to the redshift, i.e., towards lower frequencies, experienced by
the resonators for a pure capacitive coupling
[cf.~Subsec.~\ref{subsection:the:role:of:circuit:topology:two:examples}
and Fig.~\ref{QS:Figure:2:abcdef:Matteo:Mariantoni:2008}(c)].

The numerical values of all parameters discussed above are listed
in Table~\ref{QS:Table:1:Matteo:Mariantoni:2008}. The values of
the bare self-inductances of the two resonators are first
evaluated for the test length $\ell^{}_{\rm sim}$ in absence of
the qubit loop. This yields the simulated test inductances
$L^{\ast}_{\rm r a}$ and $L^{\ast}_{\rm r b}$. Then,
$L^{\ast}_{\rm r a}$ and $L^{\ast}_{\rm r b}$ are extrapolated to
the full length $\ell^{}_{\rm m}$ of each microstrip resonator to
obtain $L^{}_{\rm r a}$ and $L^{}_{\rm r b}$, respectively. In the
presence of the qubit loop, the simulated test inductances
$\widetilde{L}^{\ast}_{\rm r a}$ and $\widetilde{L}^{\ast}_{\rm r
b}$ can also be found. The resonator capacitances per unit length,
$c^{}_{\rm r a}$ and $c^{}_{\rm r b}$, are calculated analytically
by means of a conformal mapping
technique:~\cite{RoberECollin:Book:2000:a}
\begin{equation}
c^{}_{\rm r a} {} = {} c^{}_{\rm r b} {} = {} 2 \pi \epsilon^{}_0
\epsilon^{}_r \ln \left( \frac{8 H^{}_{\rm s}}{W^{\rm eff}_{\rm
m}} + \frac{W^{\rm eff}_{\rm m}}{4 H^{}_{\rm s}} \right) \, .
 \label{c:ra:c:rb}
\end{equation}
Here, $\epsilon^{}_0 = 8.854 \times 10^{- 12}_{}\,{\rm F / m}$ is
the permittivity of \mbox{vacuum} (electric
constant),~\cite{NIST:fundamental:physical:constants}
$\epsilon^{}_{\rm r} = 11.5$ the relative dielectric constant of
the substrate (in our example, sapphire or silicon; other
dielectrics could be used), and $W^{\rm eff}_{\rm m} \equiv
W^{}_{\rm m} + 0.398 t^{}_{\rm m} [ 1 + \ln ( 2 H^{}_{\rm s} /
t^{}_{\rm m} ) ]$ the effective width of the
resonators.~\cite{RoberECollin:Book:2000:a} As a consequence, the
resonator capacitances are $C^{}_{\rm r a} = \ell^{}_{\rm m}
c^{}_{\rm r a}$ and $C^{}_{\rm r b} = \ell^{}_{\rm m} c^{}_{\rm r
b}$. Finally, from the knowledge of the velocity of the
electromagnetic waves inside the two resonators, $\bar{c}^{}_{\rm
A} \equiv \ell^{}_{\rm m} / \sqrt{ ( L^{}_{\rm r a} C^{}_{\rm r a}
) }$ and $\bar{c}^{}_{\rm b} \equiv \ell^{}_{\rm m} / \sqrt{ (
L^{}_{\rm r b} C^{}_{\rm r b} ) }$, one can find the full
wavelengths $\lambda^{}_{\rm A} = \bar{c}^{}_{\rm A} / f^{}_{\rm
A}$ and $\lambda^{}_{\rm B} = \bar{c}^{}_{\rm B} / f^{}_{\rm B}$
of the two resonators. As before, all these results are summarized
in Table~\ref{QS:Table:1:Matteo:Mariantoni:2008}.

\begin{figure*}[t!]
\centering{%
 \includegraphics[width=0.85\textwidth]{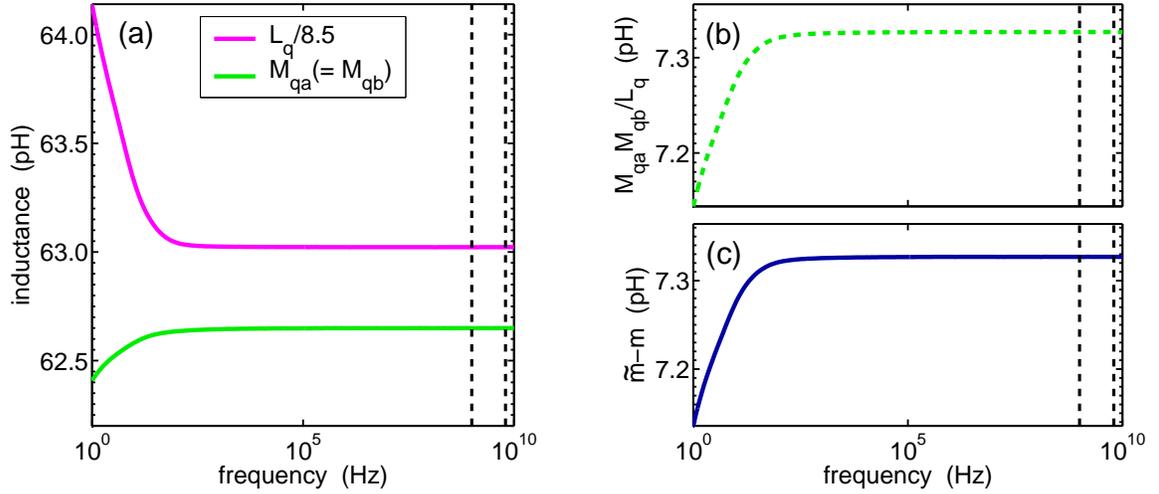}}
\caption{(Color online) FASTHENRY simulation results for the
frequency dependence of some relevant first- and second-order
inductances relative to our example of two-resonator cicruit~QED.
Vertical dashed black lines: Frequency region of interest for the
operation of the quantum switch from $1$\,GHz to $6$\,GHz.
(a)~Magenta (middle grey) line: Qubit loop self-inductance
$L^{}_{\rm q}$ renormalized by a factor of $8.5$ for clarity.
Green (light grey) line: Mutual inductance $M^{}_{\rm q a} ( {} =
{} M^{}_{\rm q b} )$ between the qubit and resonator A (or B).
(b)~Bare second-order mutual inductance between the two resonators
calculated with the results from (a) using the expression
$M^{}_{\rm q a} M^{}_{\rm q b} / L^{}_{\rm q}$. (c)~Bare
second-order mutual inductance between the two resonators
$\widetilde{m} - m$. The agreement with (b) is excellent.}
 \label{QS:Figure:7:abc:Matteo:Mariantoni:2008}
\end{figure*}
\begin{figure}[t!]
\centering{%
 \includegraphics[width=0.90\columnwidth,clip=]{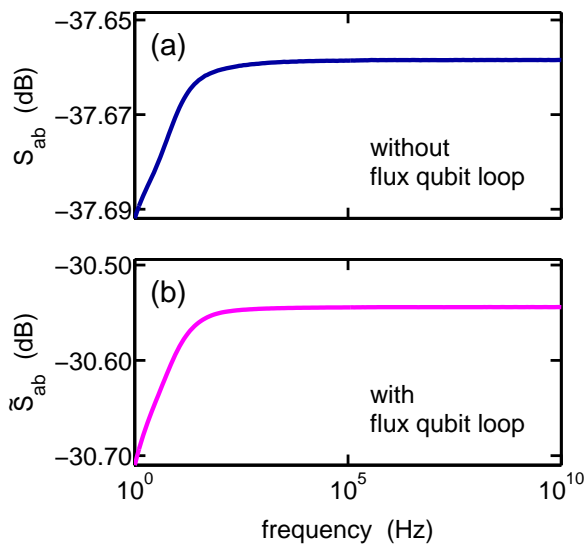}}
\caption{(Color online) FASTHENRY simulation results for the
frequency dependence of the scattering matrix elements between
resonators A and B considering almost perfect conducting
structures. Frequency span: From $1$\,Hz to $10$\,GHz.
(a)~Scattering matrix element $S^{}_{\rm a b}$ in absence of the
flux qubit loop. (b)~Scattering matrix element
$\widetilde{S}^{}_{\rm a b}$ in presence of the flux qubit loop.
Owing to the significant second-order mutual inductance between A
and B, we find $| \widetilde{S}^{}_{\rm a b} | {} < {} | S^{}_{\rm
a b} |$.}
 \label{QS:Figure:8:ab:Matteo:Mariantoni:2008}
\end{figure}

We now analyze the frequency dependence of the geometric first-
and second-order coupling coefficients, i.e., the first- and
second-oder mutual inductances, for a broad frequency span between
$1$\,Hz and $10$\,GHz. Again, we assume almost perfectly
conducting structures and use the FASTHENRY version which does not
support superconductivity. The results are plotted in
Figs.~\ref{QS:Figure:7:abc:Matteo:Mariantoni:2008} and
\ref{QS:Figure:8:ab:Matteo:Mariantoni:2008}. In
Fig.~\ref{QS:Figure:7:abc:Matteo:Mariantoni:2008}(a), we plot the
frequency dependence of the simulated inductances $L^{}_{\rm q}$
(which is renormalized by a factor of $8.5$ for clarity) and
$M^{}_{\rm q a} = M^{}_{\rm q b}$. From these, we then compute the
expression $M^{}_{\rm q a} M^{}_{\rm q b} / L^{}_{\rm q}$ for the
second-order mutual inductance as derived in
Subsecs.~\ref{subsection:the:capacitance:and:inductance:matrices:up:to:second:order}
and \ref{subsection:the:role:of:circuit:topology:two:examples}.
This expression is plotted in
Fig.~\ref{QS:Figure:7:abc:Matteo:Mariantoni:2008}(b). In
Fig.~\ref{QS:Figure:7:abc:Matteo:Mariantoni:2008}(c), we plot the
bare second-order mutual inductance $\widetilde{m} - m$ as a
function of frequency. Remarkably, comparing
Fig.~\ref{QS:Figure:7:abc:Matteo:Mariantoni:2008}(b) to
Fig.~\ref{QS:Figure:7:abc:Matteo:Mariantoni:2008}(c), we find
$M^{}_{\rm q a} M^{}_{\rm q b} / L^{}_{\rm q} {} = {}
\widetilde{m} - m$ with very high accuracy over the entire
frequency range. In the frequency region of interest for the
operation of a quantum switch, i.e., from approximately $1$\,GHz
to $6$\,GHz, we find $L^{}_{\rm q} {} \simeq {} 63.02$\,pH,
$M^{}_{\rm q a} {} = {} M^{}_{\rm q b} {} \simeq {} 7.37$\,pH,
and, consequently, $M^{}_{\rm q a} M^{}_{\rm q b} / L^{}_{\rm q}
{} = {} \widetilde{m} - m {} \simeq {} 7.33$\,pH. All these
results prove again the general validity of the derivations of
Subsecs.~\ref{subsection:the:capacitance:and:inductance:matrices:up:to:second:order}
and \ref{subsection:the:role:of:circuit:topology:two:examples}.

Finally, we study the scattering matrix elements between
resonators A and B both without and with flux qubit loop. In
absolute value, these elements correspond to the isolation
coefficients between A and B. As before, the FASTHENRY simulations
are performed within the regions of
Figs.~\ref{QS:Figure:6:abcdef:Matteo:Mariantoni:2008}(c) and
\ref{QS:Figure:6:abcdef:Matteo:Mariantoni:2008}(b), respectively.
In these figures, we also define the input and output probing
ports used in the simulations as ${\rm a^{}_{in}}$, ${\rm
a^{}_{out}}$, ${\rm b^{}_{in}}$, and ${\rm b^{}_{out}}$,
respectively. Under these assumptions, the scattering matrix
element $S^{}_{\rm a b} {} = {} S^{}_{\rm b a}$ in absence of the
flux qubit loop is given
by~\cite{RoberECollin:Book:2000:a,DavidMPozar:Book:2005:a}
\begin{equation}
S^{}_{\rm a b} {} \equiv {} 20 \log^{}_{} \left| \frac{-
I^{\,-}_{\rm a^{}_{in}}}{I^{+}_{\rm b^{}_{in}}}
\right|^{}_{I^{+}_{} = 0} {} = {} 20 \log^{}_{}
\frac{m}{L^{\ast}_{\rm r a}} \, ,
 \label{S:ab}
\end{equation}
where $I^{+}_{\rm b^{}_{in}}$ is a test current wave incident on
the input probing port ${\rm b^{}_{in}}$ of resonator B. The
current $- I^{\,-}_{\rm a^{}_{in}}$ corresponds to the outgoing
wave from the input probing port ${\rm a^{}_{in}}$ of resonator A.
The remaining current waves incident on the ports of the two
resonators are $I^{+}_{} {} \equiv {} \left\{ I^{+}_{\rm
a^{}_{in}} , I^{+}_{\rm a^{}_{out}} , I^{+}_{\rm b^{}_{out}}
\right\}$. In a similar way, the scattering matrix element
$\widetilde{S}^{}_{\rm a b} {} = {} \widetilde{S}^{}_{\rm b a}$ in
presence of the flux qubit loop is given by
\begin{equation}
\widetilde{S}^{}_{\rm a b} {} = {} 20 \log^{}_{}
\frac{\widetilde{m}}{\widetilde{L}^{\ast}_{\rm r a}} \, .
 \label{S:tilde:ab}
\end{equation}
We note that the same results as in Eqs.~(\ref{S:ab}) and
(\ref{S:tilde:ab}) are obtained replacing the input probing port
${\rm b^{}_{in}}$ with the output probing port ${\rm b^{}_{out}}$
for the incident wave. In this case, the associated current
$I^{+}_{\rm b^{}_{in}}$ has to be exchanged with $I^{+}_{\rm
b^{}_{out}}$. Similar substitutions apply for the probing port and
associated current of the outgoing waves. In the
literature,~\cite{RoberECollin:Book:2000:a} the outgoing waves are
often denominated as reflected waves. Equation~(\ref{S:ab}) can be
straightforwardly found via the definitions of mutual and
self-inductance, $m I^{+}_{\rm b^{}_{in}} {} = {} \Phi^{}_{\rm b
a} {} = {} L^{\ast}_{\rm r a} I^{\,-}_{\rm a^{}_{in}}$. There,
$\Phi^{}_{\rm b a}$ is the flux generated in the portion of
resonator A by the current flowing in the portion of resonator B
of Fig.~\ref{QS:Figure:6:abcdef:Matteo:Mariantoni:2008}(c).
Similar arguments lead to Eq.~(\ref{S:tilde:ab}). When considering
superconducting materials, the scattering matrix elements between
A and B without and with flux qubit loop can be evaluated
inserting the opportune numbers reported in
Table~\ref{QS:Table:1:Matteo:Mariantoni:2008} into
Eqs.~(\ref{S:ab}) and (\ref{S:tilde:ab}). This yields $S^{}_{\rm a
b} {} \simeq {} - 38.80$\,dB and $\widetilde{S}^{}_{\rm a b} {}
\simeq {} - 30.18$\,dB. If we want to calculate the scattering
matrix elements between A and B over a broad frequency span (e.g.,
from $1$\,Hz to $10$\,GHz), we can consider again almost perfect
conducting structures. In this case, the results are plotted in
Figs.~\ref{QS:Figure:8:ab:Matteo:Mariantoni:2008}(a) and
\ref{QS:Figure:8:ab:Matteo:Mariantoni:2008}(b). In the high
frequency region from $1$\,GHz to $6$\,GHz, we find $S^{}_{\rm a
b} {} \simeq {} - 37.66$\,dB and $\widetilde{S}^{}_{\rm a b} {}
\simeq {} - 30.54$\,dB. These numbers are in good agreement with
the results obtained for superconducting materials. In addition,
it is noteworthy to mention that the scattering matrix elements
between A and B calculated here with FASTHENRY for almost perfect
conducting structures are in excellent agreement with those
evaluated for similar structures by means of more advanced
software based on the method of
moments.~\cite{RoberECollin:Book:2000:a,
MatteoMariantoni:PhDThesis:2008:a}

In conclusion, we study a detailed setup of two-resonator
circuit~QED based on a superconducting flux qubit. In this case,
we prove that the geometric second-order inductance found with our
three-node network approach agrees well with that obtained from
numerical simulations. Moreover, we give a set of parameters (many
sets can easily be found) for which the second-order inductance
dominates over the first-order one.

\section{SUMMARY AND CONCLUSIONS}
 \label{section:summary:and:conclusions}

In this work, we first introduce the formalism of two-resonator
circuit~QED, i.e., the interaction between two on-chip microwave
cavities and a superconducting qubit circuit. Starting from the
Hamiltonian of a generic three-node network, we show that the
qubit circuit mediates a geometric second-order coupling between
the two resonators. For the case of strong qubit-resonator
coupling, the geometric second-order interaction is a fundamental
property of the system. In contrast to the geometric first-order
coupling between the two resonators, the second-order one cannot
be arbitrarily reduced by means of proper engineering.

With the aid of two prototypical examples, we then highlight the
important role played by circuit topology in two-resonator
circuit~QED. Our models reveal a clear departure from a less
detailed theory based on the Hamiltonian of a charge quantum
circuit (e.g., a Cooper-pair box or a transmon) or a flux quantum
circuit (e.g., a superconducting one- or three-Josephson-junction
loop) coupled to multiple quantized microwave fields. We
demonstrate that this simplified approach easily produces
artifacts. We also show that our three-node network approach
suffices to obtain correct results when including topological
details appropriately into the definitions of the nodes.

We subsequently demonstrate the possibility of balancing a
geometric coupling against a dynamic second-order one. In this
way, the effective interaction between the two resonators can be
controlled by means of an external bias. Based on this mechanism,
we propose possible protocols for the implementation of a quantum
switch and outline other advanced applications, which exploit the
presence of the qubit.

Remarkably, we find that the quantum switch operation is robust to
decoherence processes. In fact, we show that the qubit acts as a
mere mediator of virtual excitations between the two resonators, a
condition which substantially relaxes the requirements on the
qubit performances.

Finally, we give detailed parameters for a specific setup of
two-resonator circuit~QED based on a superconducting flux qubit.
We perform numerical simulations of the geometric coupling
coefficients and find an excellent agreement with our analytical
predictions. In particular, we confirm the existence of a regime
where the geometric second-order coupling dominates over the
first-order one.

In conclusion, our findings show that, in circuit~QED, the circuit
properties of the system are crucial to provide a correct picture
of the problem and also constitute a major difference with respect
to atomic systems. This peculiar aspect of ciruit~QED makes it a
very rich environment for the prediction and experimental
implementation of unprecedented phenomena.

\section{ACKNOWLEDGEMENTS}
 \label{section:acknowledgements}

Financial support by the German Science Foundation through SFB 631
and the Excellence Cluster `Nanosystems Initiative Munich (NIM)'
is gratefully acknowledged. This project is also partially funded
by the EU EuroSQIP project, the Basque Foundation for Science
(Ikerbasque), the UPV-EHU Grant GIU07/40, and NSERC. We thank
Elisabeth~Hoffmann, Edwin~P.~Menzel, Henning~Christ,
Markus~J.~Storcz, David~Zueco L\'{a}inez, Sigmund~Kohler,
Florian~Marquardt, Jens~Siewert, Martijn~Wubs, William~D.~Oliver,
Johannes~B.~Mejer, Yu-xi~Liu, Franco~Nori, and Peter~Zoller for
fruitful discussions.

\appendix

\section{HIGHER-ORDER CORRECTIONS TO THE CAPACITANCE AND INDUCTANCE MATRICES}
 \label{appendix:a:higher:order:corrections:to:the:capacitance:and:inductance:matrices}

In
Subsec.~\ref{subsection:the:capacitance:and:inductance:matrices:up:to:second:order},
we account for corrections up to second-order capacitive and
inductive interactions between the elements of a three-node
network. Throughout this work, we show that for a three-node
network the geometric second-order coupling coefficients can
dominate over the first-order ones. For this reason, in the
following we can safely assume vanishing first-order coupling
coefficients, $c = m = 0$. Nevertheless, we notice that our
results would not be qualitatively affected even in the presence
of appreciable first-order couplings.

In this appendix, we demonstrate that third- and fourth-order
capacitances and inductances are negligible. We start with the
case of third-order corrections. There are two possible excitation
pathways giving rise to third-order coupling coefficients. These
pathways are between resonator A and qubit Q, A-Q-B-Q, or between
resonator B and qubit Q, B-Q-A-Q. Assuming the two resonators to
have identical properties, we only study the A-Q-B-Q pathway. In
this case, from the knowledge of the electromagnetic energy we can
derive
\begin{eqnarray}
\widehat{H}^{\left( 3 \right)}_{\rm A Q} & {} = {} &
\hat{V}^{}_{\rm A} C^{}_{\rm A Q} \hat{V}^{}_{\rm Q} +
\hat{I}^{}_{\rm A} M^{}_{\rm A Q} \hat{I}^{}_{\rm Q} \nonumber\\
& & {} + \hat{V}^{}_{\rm A} C^{}_{\rm A Q} \frac{1}{C^{}_{\rm Q
Q}} C^{}_{\rm Q B} \frac{1}{C^{}_{\rm B B}} C^{}_{\rm B Q} \hat{V}^{}_{\rm Q} \nonumber\\
& & {} + \hat{I}^{}_{\rm A} M^{}_{\rm A Q} \frac{1}{M^{}_{\rm Q
Q}} M^{}_{\rm Q B} \frac{1}{M^{}_{\rm B B}} M^{}_{\rm B Q}
\hat{I}^{}_{\rm Q} \, ,
 \label{H:op:3:A:Q}
\end{eqnarray}
where the inverse paths Q-A and Q-B-Q-A are already included. In
the equation above, resonator B is only virtually excited. In the
same equation, we identify the capacitance and inductance matrix
elements up to third order
\begin{eqnarray}
C^{\left( 3 \right)}_{\rm A Q} & {} \equiv {} & C^{}_{\rm A Q} \;
\left( 1 + \frac{C^2_{\rm Q B}}{C^{}_{\rm Q Q} C^{}_{\rm B B}} \right) \label{C:3:A:Q} \\[1.5mm]
& & \rm{and} \nonumber\\[1.5mm]
M^{\left( 3 \right)}_{\rm A Q} & {} \equiv {} & M^{}_{\rm A Q}
\left( 1 + \frac{M^2_{\rm Q B}}{M^{}_{\rm Q Q} M^{}_{\rm B B}}
\right) \, .
 \label{M:3:A:Q}
\end{eqnarray}
In circuit theory, it is well-known that the squares of the
electromagnetic coupling coefficients
are~\cite{LeonOChua:CharlesADesoer:ErnestSKuh:Book:1987:a}
$C^2_{\rm Q B} / C^{}_{\rm Q Q} C^{}_{\rm B B} < 1$ and $M^2_{\rm
Q B} / M^{}_{\rm Q Q} M^{}_{\rm B B} < 1$. This implies that the
pure third-order capacitance and inductance are always smaller
than the first-order ones, $C^{\left( 3 \right)}_{\rm A Q} -
C^{}_{\rm A Q} < C^{}_{\rm A Q}$ and $M^{\left( 3 \right)}_{\rm A
Q} - M^{}_{\rm A Q} < M^{}_{\rm A Q}$. For typical experimental
parameters, we find third-order processes to be negligible,
$C^{\left( 3 \right)}_{\rm A Q} - C^{}_{\rm A Q} \ll C^{}_{\rm A
Q}$ and $M^{\left( 3 \right)}_{\rm A Q} - M^{}_{\rm A Q} \ll
M^{}_{\rm A Q}$. For example, using the parameters given in
Sec.~\ref{section:an:example:of:two:resonator:circuit:qed:with:a:flux:qubit}
yields $M^2_{\rm Q B} / M^{}_{\rm Q Q} M^{}_{\rm B B} \simeq 7.88
\times 10^{- 4}_{} \ll 1$.

In a similar way, the fourth-order coupling coefficients for the
excitation pathways A-Q-B-Q-B and, equivalently, B-Q-A-Q-A can
easily be found. In this case, it is the qubit to be only
virtually excited. The capacitance and inductance matrix elements
up to fourth order become
\begin{eqnarray}
C^{\left( 4 \right)}_{\rm A B} & {} \equiv {} & \frac{C^{}_{\rm A
Q} C^{}_{\rm Q B}}{C^{}_{\rm Q Q}} \;
\left( 1 + \frac{C^2_{\rm Q B}}{C^{}_{\rm Q Q} C^{}_{\rm B B}} \right) \\[1.5mm]
& & \rm{and} \nonumber\\[1.5mm]
M^{\left( 4 \right)}_{\rm A Q} & {} \equiv {} & \frac{M^{}_{\rm A
Q} M^{}_{\rm Q B}}{M^{}_{\rm Q Q}} \; \left( 1 + \frac{M^2_{\rm Q
B}}{M^{}_{\rm Q Q} M^{}_{\rm B B}} \right) \, .
 \label{C:4:A:B}
\end{eqnarray}
When comparing the above equations to Eqs.~(\ref{C:3:A:Q}) and
(\ref{M:3:A:Q}), respectively, we find that fourth-order processes
are negligible for typical experimental parameters.

In the light of all these considerations, all higher-order
coupling coefficients can safely be ignored within the scope of
this work.

\section{DETAILS OF THE FASTHENRY SIMULATIONS}
 \label{appendix:b:details:of:the:fasthenry:simulations}

In this appendix, we discuss the details of our FASTHENRY
simulations.~\cite{FASTHENRY:MattanKamon:JakobKWhite:IEEEMicrowave:1994:a:two:links}
First, we verify our hypothesis on the uniformity of the AC
currents (corresponding to the magnetic fields) flowing on the
resonators in the regions of
Figs.~\ref{QS:Figure:6:abcdef:Matteo:Mariantoni:2008}(b) and
\ref{QS:Figure:6:abcdef:Matteo:Mariantoni:2008}(c). To this end,
we derive the quantized current on one of the two resonators
(e.g., A) following similar calculations as in
Refs.~\onlinecite{AlexandreBlais:RJSchoelkopf:PhysRevA:2004:a} and
\onlinecite{MMariantoni:ESolano:arXiv:cond-mat:2005:a}
\begin{equation}
\hat{I}^{}_{\rm r a} \left( z^{}_{} , t \right) {} \equiv {}
i^{}_{{\rm A} 0} \cos \left( \frac{\pi z^{}_{}}{\ell^{}_{\rm m}}
\right) \ j \left[ \hat{a}^{\dag}_{} \left( t \right) -
\hat{a}^{}_{} \left( t \right) \right] \, ,
 \label{I:op:ra:z:t}
\end{equation}
where $z^{}_{}$ represents a coordinate along the longitudinal
direction of the resonator [see
Fig.~\ref{QS:Figure:6:abcdef:Matteo:Mariantoni:2008}(c)] and $t$
is the time. In Eq.~(\ref{I:op:ra:z:t}), the bosonic field
operators are expressed in the Heisenberg picture. We notice that
Eq.~(\ref{I:op:ra:z:t}) is valid for the first mode of the
$\lambda / 2$-resonator(s) considered in our example. The
contribution from the second mode is negligible for two main
reasons. First, the current is characterized by a node at the flux
qubit loop position chosen here. Second, the qubit-resonator
detuning becomes substantially larger, hence resulting in a
correspondingly reduced coupling. The contribution form higher
modes can also be neglected because of the increasing detuning.

Substituting the numbers of
Table~\ref{QS:Table:1:Matteo:Mariantoni:2008} into
Eq.~(\ref{I:op:ra:z:t}) and setting $z^{}_{} = \mp \ell^{}_{\rm
sim} / 2$, we find that the two currents at the boundaries $\mp
\ell^{}_{\rm sim} / 2$ are about $0.998 i^{}_{{\rm A} 0}$, where
$i^{}_{{\rm A} 0}$ is the maximum amplitude of the quantized
current in Eq.~(\ref{I:op:ra:z:t}). This maximum is obtained at
the position $z = 0$ of the mode antinode. The main implications
of current uniformity over $\ell^{}_{\rm sim}$ are explained in
detail in
Sec.~\ref{section:an:example:of:two:resonator:circuit:qed:with:a:flux:qubit}.
In a similar way, we can also estimate the voltage contribution
for the first mode at the boundaries $\mp \ell^{}_{\rm sim} / 2$.
In this case, we must replace the cosine function of
Eq.~(\ref{I:op:ra:z:t}) with a sine function, owing to the
conjugation of quantized currents and voltages. The maximum vacuum
voltage of, e.g., resonator A is given by $v^{}_{{\rm A} 0} {}
\equiv {} \sqrt{\hbar \omega^{}_{\rm A} / 2 C^{}_{\rm r a}} {}
\simeq {} 0.5871$\,\micro V for the realistic parameters of
Table~\ref{QS:Table:1:Matteo:Mariantoni:2008}. At $\mp
\ell^{}_{\rm sim} / 2$, we then obtain the maximum voltages in the
simulated regions, which are approximately $\mp 0.065 v^{}_{{\rm
A} 0}$. Towards the center of the simulated regions the voltage
reduces to zero because its corresponding first mode is
characterized by a node. Also, higher modes do not contribute for
the same detuning arguments outlined above. Therefore, we can
safely neglect all capacitive couplings in our simulations.

Second, we notice that $\ell^{}_{\rm sim} = 500$\,\micro m is
chosen to be large enough compared to the lateral dimension
$l^{}_{{\rm q} 1} = 200$\,\micro m of the flux qubit loop [see
Figs.~\ref{QS:Figure:6:abcdef:Matteo:Mariantoni:2008}(a)-(c)].
This avoids errors due to fringing effects when simulating the
coupling coefficients between qubit and resonators. For
consistency, we have also performed several simulations assuming
larger values of $\ell^{}_{\rm sim}$, up to $1$-$1.5$\,mm (data
not shown).~\cite{MatteoMariantoni:PhDThesis:2008:a} We have not
found any appreciable deviation in the resulting inductances.

Third, we stress that special care has to be taken when using
FASTHENRY to simulate the second-order inductances of our
three-circuit network. In order to compute the inductance matrix,
test currents must be applied to the involved structures at
specific probing ports. However, when applying test currents to
all three circuits simultaneously, only first-order inductances
are calculated. This is due to the boundary conditions that must
be fulfilled at the probing ports. This fact has important
implications for the calculation of the mutual inductance
$\widetilde{m}$, which is the sum of first- and second-order
mutual inductances between resonators A and B. In this case, it is
crucial to apply test currents only to the two resonators, but not
to the qubit circuit. On the contrary, the pure first-order mutual
inductance $m$ between A and B can be simulated in two equivalent
ways: Either the qubit circuit is completely removed from the
network or test currents are applied to all three structures. We
do not notice any difference between these two approaches. The
above arguments also apply to the calculation of the renormalized
self-inductances $\widetilde{L}^{\ast}_{\rm r a}$ and
$\widetilde{L}^{\ast}_{\rm r b}$ of the two resonators and their
pure counterparts $L^{\ast}_{\rm r a}$ and $L^{\ast}_{\rm r b}$.


\begin{thebibliography}{99}

\bibitem{AWallraff:RJSchoelkopf:NatureLett:2004:a}
A.~Wallraff, D.~I.~Schuster, A.~Blais, L.~Frunzio, R.-S.~Huang,
J.~Majer, S.~Kumar, S.~M.~Girvin, and R.~J.~Schoelkopf, Nature
(London) \textbf{431}, 162 (2004).

\bibitem{IChiorescu:JEMooij:NatureLett:2004:a}
I.~Chiorescu, P.~Bertet, K.~Semba, Y.~Nakamura,
C.~J.~P.~M.~Harmans, and J.~E.~Mooij, Nature (London)
\textbf{431}, 159 (2004).

\bibitem{JJohansson:HTakayanagi:PhysRevLett:2006:a}
J.~Johansson, S.~Saito, T.~Meno, H.~Nakano, M.~Ueda, K.~Semba, and
H.~Takayanagi, Phys. Rev. Lett. \textbf{96}, 127006 (2006).

\bibitem{RJSchoelkopf:SMGirvin:NatureHorizons:2008:a}
R.~J.~Schoelkopf, S.~M.~Girvin, Nature (London) \textbf{451}, 664
(2008).

\bibitem{MikaASillanpaeae:RaymondWSimmonds:NatureLett:2007:a}
M.~A.~Sillanp\"{a}\"{a}, J.~I.~Park, and R.~W.~Simmonds, Nature
(London) \textbf{449}, 438 (2007).

\bibitem{JMajer:JMChow:RJSchoelkopf:NatureLett:2007:a}
J.~Majer, J.~M.~Chow, J.~M.~Gambetta, J.~Koch, B.~R.~Johnson,
J.~A.~Schreier, L.~Frunzio, D.~I.~Schuster, A.~A.~Houck,
A.~Wallraff, A.~Blais, M.~H.~Devoret, S.~M.~Girvin, and
R.~J.~Schoelkopf, Nature (London) \textbf{449}, 443 (2007).

\bibitem{AAHouck:DISchuster:RJSchoelkopf:NatureLett:2007:a}
A.~A.~Houck, D.~I.~Schuster, J.~M.~Gambetta, J.~A.~Schreier,
B.~R.~Johnson, J.~M.~Chow, L.~Frunzio, J.~Majer, M.~H.~Devoret,
S.~M.~Girvin, and R.~J.~Schoelkopf, Nature (London) \textbf{449},
328 (2007).

\bibitem{MaxHofheinz:JohnMMartinis:ANCleland:NatureLett:2008:a}
M.~Hofheinz, E.~M.~Weig, M.~Ansmann, R.~C.~Bialczak, E.~Lucero,
M.~Neeley, A.~D.~O'Connell, H.~Wang, J.~M.~Martinis, and
A.~N.~Cleland, Nature (London) \textbf{454}, 310 (2008).

\bibitem{OAstafiev:JSTsai:NatureLett:2007:a}
O.~Astafiev, K.~Inomata, A.~O.~Niskanen, T.~Yamamoto,
Yu.~A.~Pashkin, Y.~Nakamura, and J.~S.~Tsai, Nature (London)
\textbf{449}, 588 (2007).

\bibitem{JMFink:AWallraff:NatureLett:2008:a}
J.~M.~Fink, M.~G\"{o}ppl, M.~Baur, R.~Bianchetti, P.~J.~Leek,
A.~Blais, and A.~Wallraff, Nature (London) \textbf{454}, 315
(2008).

\bibitem{LevSBishop:RJSchoelkopf:arXiv:2008:a}
L.~S.~Bishop, J.~M.~Chow, J.~Koch, A.~A.~Houck, M.~H.~Devoret,
E.~Thuneberg, S.~M.~Girvin, and R.~J.~Schoelkopf, eprint
arXiv:0807.2882 (unpublished).

\bibitem{FrankDeppe:MatteoMariantoni:RGross:NaturePhysLett:2008:a}
F.~Deppe, M.~Mariantoni, E.~P.~Menzel, A.~Marx, S.~Saito,
K.~Kakuyanagi, H.~Tanaka, T.~Meno, K.~Semba, H.~Takayanagi,
E.~Solano, R.~Gross, eprint arXiv:0805.3294 (to appear in Nature
Phys., doi:10.1038/nphys1016).

\bibitem{APalacios-Laloy:DEsteve:JLowTempPhys:2008:a}
A.~Palacios-Laloy, F.~Nguyen, F.~Mallet, P.~Bertet, D.~Vion,
D.~Esteve, J. Low Temp. Phys. \textbf{151}, 1034 (2008).

\bibitem{MSandberg:PDelsing:arXiv:2008:a}
M.~Sandberg, C.~M.~Wilson, F.~Persson, G.~Johansson, V.~Shumeiko,
T.~Duty, and P.~Delsing, Appl. Phys. Lett. \textbf{92}, 203501
(2008).

\bibitem{HMabuchi:ACDoherty:ScienceReview:2002:a}
H.~Mabuchi and A.~C.~Doherty, Science \textbf{298}, 1372 (2002).

\bibitem{SergeHaroche:Jean-MichelRaimond:Book:2006:a}
S.~Haroche and J.-M.~Raimond, \textit{Exploring the Quantum},
(Oxford University Press Inc., New York, 2006).

\bibitem{HerbertWalther:ThomasBecker:RepProgPhys:2006:a}
H.~Walther, B.~T.~H.~Varcoe, B.~G.~Englert, and
T.~Becker, Rep. Prog. Phys. \textbf{69}, 1325 (2006).

\bibitem{MichaelANielsen:IsaacLChuang:Book:2000:a}
M.~A.~Nielsen and I.~L.~Chuang, \textit{Quantum Computation and
Quantum Information}, (Cambridge University Press, Cambridge,
2000).

\bibitem{AlexandreBlais:JayGambetta:RJSchoelkopf:PhysRevA:2007:a}
A.~Blais, J.~Gambetta, A.~Wallraff, D.~I.~Schuster, S.~M.~Girvin,
M.~H.~Devoret, and R.~J.~Schoelkopf, Phys. Rev. A \textbf{75},
032329 (2007).

\bibitem{FerdinandHelmer:FlorianMarquardt:arXiv:2007:a}
F.~Helmer, M.~Mariantoni, A.~G.~Fowler, J.~von Delft, E.~Solano,
F.~Marquardt, eprint arXiv:0706.3625 (unpublished).

\bibitem{MJStorcz:MMariantoni:ESolano:arXiv:cond-mat:2007:a}
M.~J.~Storcz, M.~Mariantoni, H.~Christ, A.~Emmert, A.~Marx,
W.~D.~Oliver, R.~Gross, F.~K.~Wilhelm, and E.~Solano, eprint
arXiv:cond-mat/0612226 (unpublished).

\bibitem{AlexandreBlais:RJSchoelkopf:PhysRevA:2004:a}
A.~Blais, R.-S.~Huang, A.~Wallraff, S.~M.~Girvin, and
R.~J.~Schoelkopf, Phys. Rev. A \textbf{69}, 062320 (2004).

\bibitem{MatteoMariantoni:FrankDeppe:unpublished:2008:a}
M.~Mariantoni and F.~Deppe, unpublished.

\bibitem{DISchuster:AAHouck:RJSchoelkopf:NatureLett:2007:a}
D.~I.~Schuster, A.~A.~Houck, J.~A.~Schreier, A.~Wallraff,
J.~M.~Gambetta, A.~Blais, L.~Frunzio, J.~Majer, B.~Johnson,
M.~H.~Devoret, S.~M.~Girvin, and R.~J.~Schoelkopf, Nature (London)
\textbf{445}, 515 (2007).

\bibitem{RHKoch:DPDiVincenzo:PhysRevLett:2006:a}
R.~H.~Koch, G.~A.~Keefe, F.~P.~Milliken, J.~R.~Rozen, C.~C.~Tsuei,
J.~R.~Kirtley, and D.~P.~DiVincenzo, Phys. Rev. Lett. \textbf{96},
127001 (2006).

\bibitem{YuriyMakhlin:GerdSchoen:RevModPhys:2001:a}
Yu.~Makhlin, G.~Sch\"{o}n, and A.~Shnirman, Rev. Mod. Phys.
\textbf{73}, 357 (2001).

\bibitem{MHDevoret:JMMartinis:arXiv:cond-mat:2004:a}
M.~H.~Devoret, A.~Wallraff, and J.~M.~Martinis, eprint
arXiv:cond-mat/0411174 (unpublished).

\bibitem{JQYou:FrancoNori:PhysToday:2005:a}
J.~Q.~You, Franco~Nori, Phys. Today \textbf{58}, 42 (2005).

\bibitem{GoeranWendin:VitalyShumeiko:BookChapter:2006:a}
G.~Wendin and V.~Shumeiko, in Handbook of Theoretical and
Computational Nanotechnology, edited by M.~Rieth and W.~Schommers
(American Sscientific Publishers, Los Angeles, 2006), Vol.~3,
pp.~223–-309, see also arXiv:cond-mat/0508729.

\bibitem{ARauschenbeutel:SHaroche:PhysRevARap:2001:a}
A.~Rauschenbeutel, P.~Bertet, S.~Osnaghi, G.~Nogues, M.~Brune,
J.~M.~Raimond, and S.~Haroche, Phys. Rev. A \textbf{64}, 050301(R)
(2001).

\bibitem{OlivierBuisson:FrankHekking:BookChapter:2001:a}
O.~Buisson and F.~W.~J.~Hekking, in Macroscopic Quantum Coherence
and Computing, edited by D.~V.~Averin, B.~Ruggiero, and
P.~Silvestrini (Kluwer Academic Publishers, New York, 2001),
pp.~137-146.

\bibitem{Chui-PingYang:SiyuanHan:PhysRevA:2003:a}
C.-P.~Yang, S.-I~Chu, and S.~Han, Phys. Rev. A \textbf{67}, 042311
(2003).

\bibitem{FPlastina:GFalci:PhysRevB:2003:a}
F.~Plastina and G.~Falci, Phys. Rev. B \textbf{67}, 224514 (2003).

\bibitem{JQYou:FrancoNori:PhysRevB:2003:a}
J.~Q.~You and F.~Nori, Phys. Rev. B \textbf{68}, 064509 (2003).

\bibitem{AMessina:ANapoli:JofModernOpt:2003:a}
A.~Messina, S.~Maniscalco, and A.~Napoli, J. of Modern Opt.
\textbf{50}, 1 (2003).

\bibitem{CPSun:FrancoNori:PhysRevA:2005:a}
C.~P.~Sun, L.~F.~Wei, Y.~X.~Liu, F.~Nori, Phys. Rev. A
\textbf{73}, 022318 (2006).

\bibitem{FLSemitao:GJMilburn:arXiv:2008:a}
F.~L.~Semi\~{a}o, K.~Furuya, G.~J.~Milburn, eprint arXiv:0808.0743
(unpublished).

\bibitem{AlexandreBlais:RobertJSchoelkopf:APSDenver:2007:a}
A.~Blais, J.~M.~Gambetta, C.~Cheung, A.~Wallraff, D.~I.~Schuster,
S.~M.~Girvin, R.~J.~Schoelkopf,
http://meetings.aps.org/link/BAPS.2007.MAR.H33.5 (unpublished).

\bibitem{MartijnWubs:PeterHaenggi:PhysicaE:2007:a}
M.~Wubs, S.~Kohler, P.~H\"{a}nggi, Physica E (Amsterdam)
\textbf{40}, 187 (2007).

\bibitem{FerdinandHelmer:FlorianMarquardt:arXiv:2007:b}
F.~Helmer, M.~Mariantoni, E.~Solano, F.~Marquardt, eprint
arXiv:0712.1908 (unpublished).

\bibitem{PengXue:KevinLalumiere:arXiv:2008:a}
P.~Xue, B.~C.~Sanders, A.~Blais, K.~Lalumiere, eprint
arXiv:0802.2750 (unpublished).

\bibitem{LanZhou:FrancoNori:arXiv:2008:a}
L.~Zhou, Z.~R.~Gong, Yu-xi~Liu, C.~P.~Sun, and F.~Nori, eprint
arXiv:0802.4204 (unpublished).

\bibitem{LDavidovich:SHaroche:PhysRevLett:1993:a}
L.~Davidovich, A.~Maali, M.~Brune, J.~M.~Raimond, and S.~Haroche,
Phys. Rev. Lett. \textbf{71}, 2360 (1993).

\bibitem{DVAverin:CBruder:PhysRevLett:2003:a}
D.~V.~Averin and C.~Bruder, Phys. Rev. Lett. \textbf{91}, 057003
(2003).

\bibitem{EIlapichev:AMZagoskin:PhysRevLett:2003:a}
E.~Il'ichev, N.~Oukhanski, A.~Izmalkov, Th.~Wagner, M.~Grajcar,
H.-G.~Meyer, A.~Yu.~Smirnov, Alec~Maassen~van~den~Brink,
M.~H.~Amin, and A.~M.~Zagoskin, Phys. Rev. Lett. \textbf{91},
097906 (2003).

\bibitem{BLTPlourde:JohnClarkePhysRevBRap:2004:a}
B.~L.~T.~Plourde, J.~Zhang, K.~B.~Whaley, F.~K.~Wilhelm,
T.~L.~Robertson, T.~Hime, S.~Linzen, P.~A.~Reichardt, C.-E.~Wu,
and J.~Clarke, Phys. Rev. B \textbf{70}, 140501(R) (2004).

\bibitem{ALupacscu:JEMooij:PhysRevLett:2004:a}
A.~Lupa\c{s}cu, C.~J.~M.~Verwijs, R.~N.~Schouten,
C.~J.~P.~M.~Harmans, and J.~E.~Mooij, Phys. Rev. Lett.
\textbf{93}, 177006 (2004).

\bibitem{MASillanpaeae:PJHakonen:PhysRevLett:2005:a}
M.~A.~Sillanp\"{a}\"{a}, T.~Lehtinen, A.~Paila, Yu.~Makhlin,
L.~Roschier, and P.~J.~Hakonen, Phys. Rev. Lett. \textbf{95},
206806 (2005).

\bibitem{TDuty:PDelsing:PhysRevLett:2005:a}
T.~Duty, G.~Johansson, K.~Bladh, D.~Gunnarsson, C.~Wilson, and
P.~Delsing, Phys. Rev. Lett. \textbf{95}, 206807 (2005).

\bibitem{GJohansson:GWendin:JPhysCondensMatter:2006:a}
G.~Johansson, L.~Tornberg, V.~S.~Shumeiko, and G.~Wendin, J.
Phys.: Condens. Matter \textbf{18}, 901 (2006).

\bibitem{JKoenemann:ABZorin:PhysRevB:2007:a}
J.~K\"{o}nemann, H.~Zangerle, B.~Mackrodt, R.~Dolata, and
A.~B.~Zorin, Phys. Rev. B \textbf{76}, 134507 (2007).

\bibitem{BernardYurke:JohnSDenker:PhysRevA:1984:a}
B.~Yurke and J.~S.~Denker, Phys. Rev. A \textbf{29}, 1419 (1984).

\bibitem{GuidoBurkard:DavidPDiVincenzo:PhysRevB:2004:a}
G.~Burkard, R.~H.~Koch, and D.~P.~DiVincenzo, Phys. Rev. B
\textbf{69}, 064503 (2004).

\bibitem{GuidoBurkard:PhysRevB:2005:a}
G.~Burkard, Phys. Rev. B \textbf{71}, 144511 (2005).

\bibitem{GuidoBurkard:BookChapter:2005:a}
G.~Burkard, in Handbook of Theoretical and Computational
Nanotechnology, edited by M.~Rieth and W.~Schommers (American
Sscientific Publishers, New York, 2005), see cond-mat/0409626
(unpublished).

\bibitem{GWendin:VSShumeiko:LowTempPhys:2007:a}
G.~Wendin, V.~S.~Shumeiko, Low. Temp. Phys. \textbf{33}, 724
(2007).

\bibitem{LeonOChua:CharlesADesoer:ErnestSKuh:Book:1987:a}
L.~O.~Chua, C.~A.~Desoer, and E.~S.~Kuh, \textit{Linear and
Nonlinear Circuits}, (McGraw-Hill C., New York, 1987).

\bibitem{FDeppe:RGross:PhysRevB:2007:a}
F.~Deppe, M.~Mariantoni, E.~P.~Menzel, S.~Saito, K.~Kakuyanagi,
H.~Tanaka, T.~Meno, K.~Semba, H.~Takayanagi, and R.~Gross, Phys.
Rev. B \textbf{76}, 214503 (2007).

\bibitem{capacitances:values}
In two of our previous
works,~\cite{MJStorcz:MMariantoni:ESolano:arXiv:cond-mat:2007:a,
FerdinandHelmer:FlorianMarquardt:arXiv:2007:a} we have studied
charge qubits in multi-resonator systems. In that case, we have
performed numerical simulations to study the geomteric first-order
capicitance between two resonators [in a configuration similar to
that of Fig.~\ref{QS:Figure:1:abcde:Matteo:Mariantoni:2008}(e)].
We have found a scattering matrix element $S^{}_{\rm a b} {}
\simeq {} - 50$\,dB at approximately $4$\,GHz, which gives
$C^{}_{\rm A B} {} = {} c {} = {} 3 \times 10^{- 3}_{} \ C^{}_{\rm
B B}$. Assuming $C^{}_{\rm A Q} {} \approx {} C^{}_{\rm Q B}$, we
can easily compare $c$ to the second-order cross-capacitance
between A and B and obtain $c / \left( C^2_{\rm A Q} / C^{}_{\rm Q
Q} \right) {} \simeq {} 0.9$. For these calculations, we use
$C^{}_{\rm B B} {} \simeq {} 2.2$\,pF, $C^{}_{\rm A Q} {} \simeq
{} 22$\,fF, and $C^{}_{\rm Q Q} {} \simeq {} 67$\,fF from
Ref.~\onlinecite{JMajer:JMChow:RJSchoelkopf:NatureLett:2007:a}. In
particular, we find $c {} \simeq {} 7$\,fF and $C^{}_{\rm A Q}
C^{}_{\rm Q B} / C^{}_{\rm Q Q} {} \simeq {} 7.5$\,fF. In this
case, an example of third-order cross-capacitance is $C^{}_{\rm A
Q} C^{}_{\rm Q B} C^{}_{\rm B Q} / C^{}_{\rm Q Q} C^{}_{\rm B B}
{} \simeq {} 76$\,aF.

\bibitem{MMariantoni:ESolano:arXiv:cond-mat:2005:a}
M.~Mariantoni, M.~J.~Storcz, F.~K.~Wilhelm, W.~D.~Oliver,
A.~Emmert, A.~Marx, R.~Gross, H.~Christ, and E.~Solano, eprint
arXiv:cond-mat/0509737 (unpublished).

\bibitem{JensKoch:TerriMYu:JayGambetta:AlexandreBlais:RJSchoelkopf:PhysRevA:2007:a}
J.~Koch, T.~M.~Yu, J.~Gambetta, A.~A.~Houck, D.~I.~Schuster,
J.~Majer, A.~Blais, M.~H.~Devoret, S.~M.~Girvin, and
R.~J.~Schoelkopf, Phys. Rev. A \textbf{76}, 042319 (2007).

\bibitem{JASchreier:JensKoch:RJSchoelkopf:PhysRevBRap:2008:a}
J.~A.~Schreier, A.~A.~Houck, J.~Koch, D.~I.~Schuster,
B.~R.~Johnson, J.~M.~Chow, J.~M.~Gambetta, J.~Majer, L.~Frunzio,
M.~H.~Devoret, S.~M.~Girvin, and R.~J.~Schoelkopf, Phys. Rev. B
\textbf{77}, 180502(R) (2008).

\bibitem{AAHouck:JensKoch:RJSchoelkopf:arXiv:2008:a}
A.~A.~Houck, J.~A.~Schreier, B.~R.~Johnson, J.~M.~Chow, J.~Koch,
J.~M.~Gambetta, D.~I.~Schuster, L.~Frunzio, M.~H.~Devoret,
S.~M.~Girvin, R.~J.~Schoelkopf, eprint arXiv:0803.4490
(unpublished).

\bibitem{TLindstroem:AYaTzalenchuk:JPhysConfSer:2008:a}
T.~Lindstr\"{o}m, C.~H.~Webster, J.~E.~Healey, M.~S.~Colclough,
C.~M.~Muirhead, A.~Ya.~Tzalenchuk, J. Phys.: Conf. Ser.
\textbf{97} 012319 (2008).

\bibitem{single:Cooper-pair:box:self-capacitance}
In the definition of self-capacitance, we neglect the capacitance
of the island itself because it is small compared to the other
capacitances, $C^{}_{\rm isl} \ll \min \{ C^{}_{\rm g a} , 2
C^{}_{\rm J} , C^{}_{\rm g b} \}$.

\bibitem{displacement-type:operators}
The remaining two interaction terms, when quantizing the AC
excitations, result in a displacement-type of
operator,~\cite{AlexandreBlais:RJSchoelkopf:PhysRevA:2004:a} which
act on the two resonators coordinates. These operators reduce to
zero right at the charge degeneracy point (i.e., for $n^{\rm
DC}_{\rm g} {} = {} 1 / 2$). In general, these terms are small and
can thus be
neglected.~\cite{AlexandreBlais:RJSchoelkopf:PhysRevA:2004:a}

\bibitem{TPOrlando:JEMooij:PhysRevB:1999:a}
T.~P.~Orlando, J.~E.~Mooij, L.~Tian, C.~H.~van~der~Wal,
L.~S.~Levitov, S.~Lloyd, J.~J.~Mazo, Phys. Rev. B \textbf{60},
15398 (1999).

\bibitem{MatteoMariantoni:PhDThesis:2008:a}
M.~Mariantoni, Ph.~D. Thesis, in preparation.

\bibitem{CWildfeuer:DHSchiller:PhysRevA:2003:a}
C.~Wildfeuer and D.~H.~Schiller, Phys. Rev. A \textbf{67}, 053801
(2003).

\bibitem{ThomasNiemczyk:PrivateComm:2008:a}
T.~Niemczyk, private communications (2008).

\bibitem{Yu-xiLiu:FrancoNori:EurophysLett:2004:a}
Yu-xi~Liu, L.~F.~Wei, and F.~Nori, Europhys. Lett. \textbf{67},
941 (2004).

\bibitem{NKiesel:HWeinfurter:PhysRevLett:2007:a}
N.~Kiesel, C.~Schmid, G.~T\'{o}th, E.~Solano, and H.~Weinfurter,
Phys. Rev. Lett. \textbf{98}, 063604 (2007).

\bibitem{AWallraff:RJSchoelkopf:PhysRevLett:2005:a}
A.~Wallraff, D.~I.~Schuster, A.~Blais, L.~Frunzio, J.~Majer,
M.~H.~Devoret, S.~M.~Girvin, and R.~J.~Schoelkopf, Phys. Rev.
Lett. \textbf{95}, 060501 (2005).

\bibitem{LFWei:FrancoNori:PhysRevLett:2006:a}
L.~F.~Wei, Yu-Xi Liu, and F.~Nori, Phys. Rev. Lett. \textbf{96},
246803 (2006).

\bibitem{Yu-xiLiu:FNori:PhysRevA:2005:a}
Yu-xi~Liu, L.~F.~Wei, and F.~Nori, Phys. Rev. A \textbf{71},
063820 (2005).

\bibitem{ESolano:NZagury:PhysRevLett:2001:a}
E.~Solano, R.~L.~de~Matos~Filho, and N.~Zagury, Phys. Rev. Lett.
\textbf{87}, 060402 (2001).

\bibitem{ESolano:NZagury:JOptBQuantumSemiclassOpt:2002:a}
E.~Solano, R.~L.~de~Matos~Filho, and N.~Zagury, J. Opt. B: Quantum
Semiclass. Opt. \textbf{4} S324 (2002).

\bibitem{MatthiasSteffen:JohnMMartinis:PhysRevLett:2006:a}
M.~Steffen, M.~Ansmann, R.~McDermott, N.~Katz, R.~C.~Bialczak,
E.~Lucero, M.~Neeley, E.~M.~Weig, A.~N.~Cleland, and
J.~M.~Martinis, Phys. Rev. Lett. \textbf{97}, 050502 (2006).

\bibitem{PBertet:JEMooij:PhysRevLett:2005:a}
P.~Bertet, I.~Chiorescu, G.~Burkard, K.~Semba,
C.~J.~P.~M.~Harmans, D.~P.~DiVincenzo, and J.~E.~Mooij, Phys. Rev.
Lett. \textbf{95}, 257002 (2005).

\bibitem{FYoshihara:JSTsai:PhysRevLett:2006:a}
F.~Yoshihara, K.~Harrabi, A.~O.~Niskanen, Y.~Nakamura, and
J.~S.~Tsai, Phys. Rev. Lett. \textbf{97}, 167001 (2006).

\bibitem{KKakuyanagi:AShnirman:PhysRevLett:2007:a}
K.~Kakuyanagi, T.~Meno, S.~Saito, H.~Nakano, K.~Semba,
H.~Takayanagi, F.~Deppe, and A.~Shnirman, Phys. Rev. Lett.
\textbf{98}, 047004 (2007).

\bibitem{qubit:energy:relaxation}
At the qubit degeracy point, the scenario can be quite different
and energy relaxation can become the dominating source of
decoherence. However, qubit energy relaxation rates are typically
$\lesssim {} 10$\,MHz in this situation and, thus, a reasonable
operating time for the quantum switch is guaranteed.

\bibitem{SSaito:HTakayanagi:PhysRev:Lett:2006:a}
S.~Saito, T.~Meno, M.~Ueda, H.~Tanaka, K.~Semba, and H.~Takayanagi
Phys. Rev. Lett. \textbf{96}, 107001 (2006).

\bibitem{JulianHauss:GerdSchoen:PhysRevLett:2008:a}
J.~Hauss, A.~Fedorov, C.~Hutter, A.~Shnirman, and G.~Sch\"{o}n,
Phys. Rev. Lett. \textbf{100}, 037003 (2008).

\bibitem{LFrunzio:RSchoelkopf:IEEEApplSupercond:2005:a}
L.~Frunzio, A.~Wallraff, D.~Schuster, J.~Majer, and R.~Schoelkopf,
IEEE Trans. Applied Supercond. \textbf{15}, 860 (2005).

\bibitem{JohnMMartinis:ClareCYu:PhysRevLett:2005:a}
J.~M.~Martinis, K.~B.~Cooper, R.~McDermott, M.~Steffen,
M.~Ansmann, K.~D.~Osborn, K.~Cicak, S.~Oh, D.~P.~Pappas,
R.~W.~Simmonds, and C.~C.~Yu, Phys. Rev. Lett. \textbf{95}, 210503
(2005).

\bibitem{ADOapConnell:ErikLucero:ANCleland:JMMartinis:ApplPhysLett:2008:a}
A.~D.~O'Connell, M.~Ansmann, R.~C.~Bialczak, M.~Hofheinz, N.~Katz,
E.~Lucero, C.~McKenney, M.~Neeley, H.~Wang, E.~M.~Weig,
A.~N.~Cleland, and J.~M.~Martinis, Appl. Phys. Lett. \textbf{92},
112903 (2008).

\bibitem{JiansongGao:HenryGLeduc:ApplPhysLett:2008:a}
J.~Gao, M.~Daal, A.~Vayonakis, S.~Kumar, J.~Zmuidzinas,
B.~Sadoulet, B.~A.~Mazin, P.~K.~Day, and H.~G.~Leduc, Appl. Phys.
Lett. \textbf{92}, 152505 (2008).

\bibitem{JiansongGao:JohnMMartinis:HenryGLeduc:ApplPhysLett:2008:a}
J.~Gao, M.~Daal, J.~M.~Martinis, A.~Vayonakis, J.~Zmuidzinas,
B.~Sadoulet, B.~A.~Mazin, P.~K.~Day, and H.~G.~Leduc, Appl. Phys.
Lett. \textbf{92}, 212504 (2008).

\bibitem{RBarends:TMKlapwijk:ApplPhysLett:2008:a}
R.~Barends, H.~L.~Hortensius, T.~Zijlstra, J.~J.~A.~Baselmans,
S.~J.~C.~Yates, J.~R.~Gao, and T.~M.~Klapwijk, Appl. Phys. Lett.
\textbf{92}, 223502 (2008).

\bibitem{on-chip:transmission:lines:matching}
However, in order to avoid unwanted reflections all on-chip
transmission lines connected to the resonators via the input and
output capacitors have to be properly engineered to be
$50\,\Omega$-matched.

\bibitem{FASTHENRY:MattanKamon:JakobKWhite:IEEEMicrowave:1994:a:two:links}
FASTHENRY, Inductance Analysis Program, RLE Computational
Prototyping Group, Boston; M.~Kamon, M.~J.~Tsuk, and J.~K.~White,
IEEE Trans. Microwave Theory Tech. \textbf{42}, 1750 (1994); see
also http://www.fastfieldsolvers.com and
http://www.wrcad.com/freestuff.html .

\bibitem{BLTPlourde:JohnClarke:PhysRevBRap:2005:a}
B.~L.~T.~Plourde, T.~L.~Robertson, P.~A.~Reichardt, T.~Hime,
S.~Linzen, C.-E.~Wu, and J.~Clarke, Phys. Rev. B \textbf{72},
060506(R) (2005).

\bibitem{abruplty}
Obviously, this does not mean that sharp edges of the microwave
on-chip structures are needed (which would imply unwanted
radiation effects). It only means that the lines of the two
resonators have to rapidly depart from each other.

\bibitem{RoberECollin:Book:2000:a}
R.~E.~Collin, \textit{Foundations for Microwave Engineering}, 2nd
ed. (Wiley-IEEE Press, New Jersey, 2000).

\bibitem{DavidMPozar:Book:2005:a}
D.~M.~Pozar, \textit{Microwave Engineering}, 3rd ed. (J.~Wiley \&
Sons Inc., New Jersey, 2005).

\bibitem{NIST:fundamental:physical:constants}
In all our calculations and simulations, for the permittivity of
vacuum we use the standard value provided by the National
Institute of Standards and Technology~(NIST), retaining all given
decimal digits. The same applies for all other fundamental
physical constants, which can be found at
http://physics.nist.gov/cuu/Constants/.

\end{thebibliography}
\end{document}